   \documentclass[12pt]{article}
   \usepackage{latexsym}
\else\oddsidemargin25mm\evensidemargin25mm\marginparwidth25mm\fi
\def\theequation{\arabic{section}.\arabic{equation}}

\renewcommand{\theequation}{\thesection.\arabic{equation}}
\begin{document}
\makeatletter \@addtoreset{equation}{section} \makeatother
\renewcommand{\theequation}{\thesection.\arabic{equation}}
\newcommand{\ft}[2]{{\textstyle\frac{#1}{#2}}}
\newcommand{\QED}{{\hspace*{\fill}\rule{2mm}{2mm}\linebreak}}
\def\dop{{\rm d}\hskip -1pt}
\def\bfone{\relax{\rm 1\kern-.35em 1}}
\def\bfzero{\relax{\rm I\kern-.18em 0}}
\def\inbar{\vrule height1.5ex width.4pt depth0pt}
\def\IC{\relax\,\hbox{$\inbar\kern-.3em{\rm C}$}}
\def\ID{\relax{\rm I\kern-.18em D}}
\def\IF{\relax{\rm I\kern-.18em F}}
\def\IK{\relax{\rm I\kern-.18em K}}
\def\IH{\relax{\rm I\kern-.18em H}}
\def\II{\relax{\rm I\kern-.17em I}}
\def\IN{\relax{\rm I\kern-.18em N}}
\def\IP{\relax{\rm I\kern-.18em P}}
\def\IQ{\relax\,\hbox{$\inbar\kern-.3em{\rm Q}$}}
\def\IR{\relax{\rm I\kern-.18em R}}
\def\IG{\relax\,\hbox{$\inbar\kern-.3em{\rm G}$}}
\font\cmss=cmss10 \font\cmsss=cmss10 at 7pt
\def\ZZ{\relax\ifmmode\mathchoice
{\hbox{\cmss Z\kern-.4em Z}}{\hbox{\cmss Z\kern-.4em Z}}
{\lower.9pt\hbox{\cmsss Z\kern-.4em Z}} {\lower1.2pt\hbox{\cmsss
Z\kern .4em Z}}\else{\cmss Z\kern-.4em Z}\fi}
\def\a{\alpha}
\def\b{\beta}
\def\d{\delta}
\def\e{\epsilon}
\def\g{\gamma}
\def\G{\Gamma}
\def\l{\lambda}
\def\L{\Lambda}
\def\s{\sigma}
\def\S{\Sigma}
\def\im {{\rm Im}\cN}
\def\cA{{\cal A}}
\def\cB{{\cal B}}
\def\cC{{\cal C}}
\def\cD{{\cal D}}
    \def\cF{{\cal F}}
    \def\cG{{\cal G}}
\def\cH{{\cal H}}
\def\cI{{\cal I}}
\def\cJ{{\cal J}}
\def\cK{{\cal K}}
\def\cL{{\cal L}}
\def\cM{{\cal M}}
\def\cN{{\cal N}}
\def\cO{{\cal O}}
\def\cP{{\cal P}}
\def\cQ{{\cal Q}}
\def\cR{{\cal R}}
\def\cV{{\cal V}}
\def\cW{{\cal W}}
%
%
%
\def\crr{\crcr\noalign{\vskip {8.3333pt}}}
\def\tilde{\widetilde}
\def\bar{\overline}
\def\us#1{\underline{#1}}
\let\shat=\hat
\def\hat{\widehat}
\def\hyp{\vrule height 2.3pt width 2.5pt depth -1.5pt}
\def\square{\mbox{.08}{.08}}
\def\Coeff#1#2{{#1\over #2}}
\def\Coe#1.#2.{{#1\over #2}}
\def\coeff#1#2{\relax{\textstyle {#1 \over #2}}\displaystyle}
\def\coe#1.#2.{\relax{\textstyle {#1 \over #2}}\displaystyle}
\def\half{{1 \over 2}}
\def\shalf{\relax{\textstyle {1 \over 2}}\displaystyle}
\def\dag#1{#1\!\!\!/\,\,\,}
\def\to{\rightarrow}
\def\notin{\hbox{{$\in$}\kern-.51em\hbox{/}}}
\def\shdot{\!\cdot\!}
\def\ket#1{\,\big|\,#1\,\big>\,}
\def\bra#1{\,\big<\,#1\,\big|\,}
\def\equaltop#1{\mathrel{\mathop=^{#1}}}
\def\Trbel#1{\mathop{{\rm Tr}}_{#1}}
\def\inserteq#1{\noalign{\vskip-.2truecm\hbox{#1\hfil}
\vskip-.2cm}}
\def\attac#1{\Bigl\vert
{\phantom{X}\atop{{\rm\scriptstyle #1}}\phantom{X}}}
\def\exx#1{e^{{\displaystyle #1}}}
\def\del{\partial}
\def\delbar{\bar\partial}
\def\nex#1{$N\!=\!#1$}
\def\dex#1{$d\!=\!#1$}
\def\cex#1{$c\!=\!#1$}
\def\eg{{\it e.g.}} \def\ie{{\it i.e.}}
%
\def\IE{\relax{{\rm I\kern-.18em E}}}
\def\cE{{\cal E}}
\def\cU{{\cal U}}
\def\rt{{\cR^{(3)}}}
\def\IGam{\relax{{\rm I}\kern-.18em \Gamma}}
\def\IGa{\IA}
\def\cV{{\cal V}}
\def\Rt{{\cal R}^{(3)}}
\def\W{{\cal W}}
\def\tft#1{\langle\langle\,#1\,\rangle\rangle}
\def\IA{\relax{\hbox{{\rm A}\kern-.82em {\rm A}}}}
\let\picfuc=\fp
\def\hata{{\shat\a}}
\def\hatb{{\shat\b}}
\def\hatA{{\shat A}}
\def\hatB{{\shat B}}
\def\bv{{\bf V}}
\def\Fb{\overline{F}}
\def\nablab{\overline{\nabla}}
\def\Ub{\overline{U}}
\def\Db{\overline{D}}
\def\zb{\overline{z}}
\def\eb{\overline{e}}
\def\fb{\overline{f}}
\def\tb{\overline{t}}
\def\Xb{\overline{X}}
\def\Vb{\overline{V}}
\def\Cb{\overline{C}}
\def\Sb{\overline{S}}
\def\delb{\overline{\del}}
\def\Gammab{\overline{\Gamma}}
\def\Ab{\overline{A}}
\def\Anh{A^{\rm nh}}
\def\alphab{\bar{\alpha}}
\def\cy{Calabi--Yau}
\def\cabg{C_{\alpha\beta\gamma}}
\def\B{\Sigma}
\def\Bh{\hat \Sigma}
\def\Kh{\hat{K}}
\def\Knh{{\cal K}}
\def\A{\Lambda}
\def\Ah{\hat \Lambda}
\def\R{\hat{R}}
\def\V{{V}}
\def\T{T}
\def\Gammah{\hat{\Gamma}}
\def\twot{$(2,2)$}
\def\K{K\"ahler}
\def\rat{({\theta_2 \over \theta_1})}
\def\lv{{\bf \omega}}
\def\w{w}
\def\CP{C\!P}
\def\o#1#2{{{#1}\over{#2}}}
\newcommand{\be}{\begin{equation}}
\newcommand{\ee}{\end{equation}}
\newcommand{\ba}{\begin{eqnarray}}
\newcommand{\ea}{\end{eqnarray}}
\newtheorem{definizione}{Definition}[section]
\newcommand{\bd}{\begin{definizione}}
\newcommand{\ed}{\end{definizione}}
\newtheorem{teorema}{Theorem}[section]
\newcommand{\bth}{\begin{teorema}}
\newcommand{\eth}{\end{teorema}}
\newtheorem{lemma}{Lemma}[section]
\newcommand{\blem}{\begin{lemma}}
\newcommand{\elem}{\end{lemma}}
\newcommand{\brr}{\begin{array}}
\newcommand{\err}{\end{array}}
\newcommand{\nn}{\nonumber}
\newtheorem{corollario}{Corollary}[section]
\newcommand{\bcorol}{\begin{corollario}}
\newcommand{\ecorol}{\end{corollario}}
\def\twomat#1#2#3#4{\left(\begin{array}{cc}
 {#1}&{#2}\\ {#3}&{#4}\\
\end{array}
\right)}
\def\twovec#1#2{\left(\begin{array}{c}
{#1}\\ {#2}\\
\end{array}
\right)}
\begin{titlepage}
\hskip 5.5cm \hskip 1.5cm
\vbox{\hbox{CERN-TH/2001-300}\hbox{hep-th/0110277}\hbox{October,
2001}} \vfill \vskip 3cm
\begin{center}
{\LARGE {Supersymmetry reduction of $N$-extended
supergravities in four dimensions.}}\\
\vskip 1.5cm
  {\bf Laura Andrianopoli$^1$, Riccardo D'Auria$^1$ and Sergio Ferrara$^{1,2,3,4}$} \\
\vskip 0.5cm {\small $^1$ Dipartimento di Fisica, Politecnico di
Torino,\\
 Corso Duca degli Abruzzi 24, I-10129 Torino\\
and Istituto Nazionale di Fisica Nucleare (INFN) - Sezione di
Torino, Italy}\\
{\small $^2$ CERN Theoretical Division, CH 1211 Geneva 23,
Switzerland} \\
{\small $^3$ Istituto Nazionale di Fisica Nucleare (INFN) -
Laboratori Nazionali di Frascati}\\
{\small $^4$ Department of Physics
 \& Astronomy, University of California, Los Angeles, U.S.A.} \\
 \vspace{6pt}
\end{center}
\vskip 3cm \vfill
 {\bf Abstract}: We consider the possible
consistent truncation of $N$-extended supergravities to lower $N'$
theories. The truncation, unlike the case of $N$-extended rigid
theories, is non trivial and only in some cases it is sufficient
just to delete the extra $N-N'$ gravitino multiplets. We explore
different cases (starting with $N=8$ down to $N'\geq 2$) where the
reduction implies restrictions on the matter sector. We perform a
detailed analysis of the interesting case $N=2\longrightarrow
N=1$. This analysis finds applications in different contexts of
superstring and M-theory dynamics.
 {\small
}
\end{titlepage}

\section{Introduction}
It is well known that for globally supersymmetric theories, with
particle content of spin $0,\half, 1$ any theory with $N$
supersymmetries can be regarded as a particular case of a theory
with a number $N'< N$ of supersymmetries \cite{st}. To prove this
it is sufficient to decompose the $N$ supersymmetry--extended
multiplets into $N'$-multiplets.
\par
Of course $N$-extended supersymmetry is more restrictive than $N'<
N$ supersymmetry implying that the former will only allow some
restricted couplings of the latter. As we are going to show in the
present paper the same argument does not apply to supergravity
theories. Indeed, let us consider a standard $N$-extended
supergravity theory with $N$ gravitini and a given number of
matter multiplets with spin $0,\half, 1$: then the $N'$-extended
supergravity obtained by reduction from the mother theory will no
longer be standard because a certain number $N-N'$ of spin $\frac
{3} {2}$ multiplets  appear in the decomposition. Therefore to
obtain a standard $N'$-extended supergravity one must truncate
out at least the $N-N'$ spin $\frac {3} {2}$ multiplets and all
the non-linear couplings they generate in the supergravity action.
\par
 The most known example is $N=2$ supergravity in presence of
 hypermatter \cite{bw1}, \cite{dlv}, \cite{ivanov}, \cite{abcdffm}.
  The non-linear couplings of the hypermultiplets
 generate what is called a ``quaternionic geometry" \cite{bw1}.
 If we regard
 the $N=2$ hypermultiplets as a pair of $N=1$ Wess-Zumino
 multiplets, what we obtain is incompatible with $N=1$
 supergravity where the non-linear couplings must describe a
 K\"{a}hler-Hodge manifold geometry \cite{bw2}. Therefore, in order to consistently
reduce a $N=2$
 supergravity to a $N=1$ theory, the former theory must have the
 property that a certain submanifold of the original quaternionic
 manifold be a K\"{a}hler-Hodge manifold.\\
Note that in rigid supersymmetry
 hypermultiplet couplings are described by HyperK\"{a}hler
 geometry which is instead compatible with $N=1$ supersymmetry.
\\
As an illustrative example let us consider  maximal $N=8$
supergravity in $D=4$ \cite{cg} truncated to lower $N'$
supergravities. In this situation the consistent truncation
consists in deleting only spin $\frac {3} {2}$ multiplets for
sufficiently high $N'$ ($N'=6,5,4)$, but for $N'\leq 4$, where the
matter sectors begin to appear, the consistent truncation also
requires to delete some matter multiplets. We will illustrate how
this process of reduction can be understood in group-theoretical
and geometrical terms, by requiring that certain geometrical
conditions dictated by supergravity define some submanifold of the
original scalar manifold $E_{7(7)}/SU(8)$ of $N=8$ supergravity.
\par
 Returning to
the $N=2\longrightarrow N=1$ case,
 we can show that this generally demands a reduction of both
 special K\"ahler manifold ($\cM^{SK}(n_V)$) \cite{dlv},\cite{str}, \cite{cadf}, \cite{abcdffm}
  and quaternionic  manifold
 ($\cM^{Q}(n_H)$), where $n_V$ and $n_H$ are the number of
 vector multiplets  and hypermultiplets respectively.
 By equipping these manifolds with complex coordinates $z^{{\mathcal I}} \in
 \cM^{SK}$ (${\mathcal I}=1,\cdots , n_V$)
 and real coordinates $q^u \in \cM^Q$ ($u=1,\cdots , 4n_H$) the
 Riemann tensors are given respectively by:
 \begin{eqnarray}
 R_{{\mathcal I} \bar{\mathcal J}{\mathcal K} \bar{\mathcal L}} &=& g_{  {\mathcal I}
 \bar{\mathcal L}} g_{{\mathcal K}\bar{\mathcal J}} +g_{ {\mathcal K}
\bar{\mathcal L}} g_{{\mathcal I}\bar{\mathcal J}} -\bar
C_{\bar{\mathcal J}\bar{\mathcal L}\bar{\mathcal N}}
C_{ {\mathcal I}{\mathcal K} {\mathcal M}} g^{{\mathcal M} \bar{\mathcal N}}\\
{\cal R}^{uv}_{\phantom{uv}{pq}} {\cal U}^{\alpha A}_u {\cal
U}^{\beta B}_v  & = & -\,{{\rm i}\over 2} \Omega^x_{pq}
 (\sigma_x)^{AB} C^{\alpha \beta}+
 \IR^{\alpha\beta}_{pq}\epsilon^{AB}
\end{eqnarray}
where the $SU(2)$ triplet and singlet parts are the $SU(2)$ and
$Sp(2n_H)$ curvatures respectively.\footnote{Here by $Sp(2n_H)$
we denote the compact form of the symplectic group sometimes
called $USp(2n_H)$ ({\it i.e.} $Sp(2) = SU(2)$).} Here $A,B=1,2;
x=1,2,3$ are indices of the fundamental and adjoint
representation of $SU(2)$ and $\a,\b...=1,\cdots 2 n_H$ are
indices in the fundamental representation of $Sp(2n_H)$ .
\\
The K\"ahler metric $g_{  {\mathcal I}\bar{\mathcal L}}=
\partial_{{\mathcal I}} \partial_{\bar{\mathcal L}}
 \cK$, with $\cK =- {\rm log} [{\rm i} (\bar X^{{\bf \L}}
 F_{{\bf \L}} - \bar F_{{\bf \L}} X^{{\bf \L}})]$, is given in
 terms of
 $(X^{{\bf \L}}, F_{{\bf \L}})$ which are the holomorphic symplectic
 sections of $\cM^{SK}$ (they are related to the covariantly holomrphic symplectic sections
 $(L^{{\bf \L}}, M_{{\bf \L}})$ by
 $(L^{{\bf \L}}, M_{{\bf \L}})=
  e^{\frac{\cK}2 } (X^{{\bf \L}}, F_{{\bf \L}}) $).
The tensor $C_{ {\mathcal I}{\mathcal K} {\mathcal M}}$ is
threefold symmetric and covarianty holomorphic, i.e. $C_{
{\mathcal I}{\mathcal K} {\mathcal M}}= e^{\cK} W_{ {\mathcal
I}{\mathcal K} {\mathcal M}} $ (with $W_{ {\mathcal I}{\mathcal
K} {\mathcal M}} $ holomorphic).
\\
On $\cM^Q$, ${\cal U}^{\alpha A}$ denotes the vielbein 1-form.
 Furthermore, we have:
\begin{eqnarray}
\Omega^x \,& \equiv &\, d \omega^x + {1\over 2} \epsilon^{x y z}
\omega^y \wedge \omega^z = - {\rm i}\,  \IC_{\alpha\beta} (\sigma
_x)_{AB} {\mathcal U}^{\alpha A} \wedge {\mathcal
U}^{\beta B}, \\
\IR^{\alpha}_{\ \beta} &\equiv &d \Delta^{\alpha}_{\ \beta} +
\Delta^{\alpha}_{\ \gamma} \wedge \Delta^{\gamma}_{\ \beta}
\nonumber \\&=&-\epsilon_{AB}{\mathcal U}^{A\alpha}\wedge
{\mathcal U}^{B}_{\beta}+{\mathcal U}^{A\gamma}\wedge {\mathcal
U}^{B\delta}
 \epsilon_{AB}
\IC^{\alpha\rho} \Omega_{\rho\beta\gamma\delta}.
\end{eqnarray}
where $\Omega_{\rho\beta\gamma\delta}$ is completely symmetric in
its four indices.
\par
The $N=2 \to N=1$ reduction imposes a number of conditions on the
above defined structures, which have to be satisfied in order to
have a consistent reduction. In particular, we find that the two
scalar manifolds $\cM^{SK}$ and $\cM^Q$ have to be reduced to the
submanifolds $\cM_R (n_C) \subset \cM^{SK}$ and $\cM^{KH}(n_h)
\subset \cM^Q$, where $n_C \leq n_V - n'_V$, $n_h \leq n_H$ are
the complex dimensions of the two K\"ahler--Hodge manifolds
$\cM_R$ and $\cM^{KH}$ and $n'_V$ is the number of $N=1$ vector
multiplets.
\\
We first discuss the two extreme cases $n'_V =n_V $ ($n_C =0$) and
$n'_V =0$ ( $n_C =n_V$). In the first case no $N=1$ chiral
multiplet coming from $N=2$ vector multiplet is retained and all
$N=1$ vector multiplets may remain. In the second case, all the
$N=1$ vector multiplets are truncated out, and no restrictions
appear on the special-K\"ahler manifold: $\cM_R = \cM^{SK}$.
\\
In the general case, let us decompose the coordinates on
$\cM^{SK}$:
\begin{equation}
z^{{\mathcal I}} \to (z^i, z^\a )
\end{equation}
and those on $\cM^Q$:
\begin{equation}
q^u \to ( w^s, \bar w^{\bar s}, n^t, \bar n^{\bar t} )
\end{equation}
where $z^i $ ($i = 1 ,\cdots , n_C$) and  $w^s$ ($s =1,\cdots
,n_h$) are the holomorphic coordinates in $\cM_R$  and $\cM^{KH}$
respectively, and $z^\a $ ($\a = 1 ,\cdots , n_V' =n_V- n_C$) and
$n^t$ ($t =1,\cdots ,n_H - n_h$) are the holomorphic coordinates
in their orthogonal complements. Splitting furthermore the $N=2$
vector indices ${\bf \L}\to (\L , X )$, where $\L = 1, \cdots ,
n'_V$ and $X =0,1, \cdots , n_C$, we find the following
constraints to be satisfied on $\cM_R \times \cM^{KH}$ from
supersymmetry reduction. On $\cM_R$ we get, for consistent
reduction of the special geometry sector in the ungauged case:
\begin{eqnarray}
&&C_{ij \a}|_{\cM_R} = 0 \, ; \quad C_{\a\b\g}|_{\cM_R} = 0\\
&& L^\L|_{\cM_R} = 0 \, ,\quad f^\L_i |_{\cM_R} \equiv \nabla_i L^\L |_{\cM_R}=0\\
&& f^{X}_\a |_{\cM_R}\equiv \nabla_\a L^{X} |_{\cM_R}=0
\end{eqnarray}
The parent  (non holomorphic) vector kinetic matrix $\cN_{{\bf \L}
{\bf \S}}$ satisfies on $\cM_R$:
\begin{equation}
\cN_{ \L Y}|_{\cM_R}=0,
\end{equation}
Furthermore we obtain that $\bar \cN_{\L \S}|_{\cM_R} \equiv \half
f_{\L\S}$ is holomorphic, while $\cN_{X Y}$ has no restrictions
and gives the period matrix on $\cM_R$, which is indeed a
Special-K\"ahler manifold.
\par
For the hypermultiplet sector, the reduction is more subtle
because we have to reduce the holonomy from $SU(2) \times
Sp(2n_H)$ to $U(1) \times SU(n_h)$ which corresponds to decompose
the $SU(2)$ indices $A,B ,...\to (1,2)$ and the $Sp(2n_H)$ indices
$\a , \b ... \to (I , \dot I)$. The following constraints are
found on the geometrical structure of the manifold
$\cM^{KH}\subset \cM^Q$
\begin{eqnarray}
&& \Omega_{IJK\dot L}|_{\cM^{KH}} =0 \\
&& {\cal U}^{2I}|_{\cM^{KH}} = ({\cal U}^{1\dot
I})^*|_{\cM^{KH}}=0 .
\end{eqnarray}
In particular, the second equation implies that the complex
scalars of the chiral multiplets coming from
 the reduced quaternionic manifold are at most half of the
 quaternionic dimension of the original $N=2$ manifold \cite{am}.
\\
The present investigation concerning the $N=2\rightarrow N=1$
 reduction is further analyzed in the most general case when isometries
 of the scalar manifolds are gauged.
\par
In particular we find that the number of reduced $N=1$ vector
multiplets
  and of $N=1$
 chiral multiplets obtained by truncation of the $N=2$ vector
 multiplets (which are in the adjoint representation of some gauge group $G^{(2)}$)
  depend on the gauge
 group $G^{(1)}$ under which the reduced hypermultiplets are
 charged.
 Indeed, if $Adj(G^{(2)}) \to  Adj(G^{(1)}) + R(G^{(1)})$, then the
 chiral multiplets coming from $N=2$  vector multiplets are in
 $R(G^{(1)})$.\\
The reduction of the gauge group further implies constraints on
the special geometry and quaternionic Killing vectors and
prepotentials \cite{gali},\cite{abcdffm}. For the K\"ahlerian
Killing vectors $k^{{\mathcal I}}_{{\bf \L}}$ and prepotential
$P^0_{{\bf \L}}$ we find:
\begin{eqnarray}
&& k^i_{X} =0 \, , \quad k^\a_\L =0\\
 && k^i_\L = {\rm i}g^{i\bar\jmath} \partial_{\bar\jmath} P^0_\L \neq 0 \\
 && P^0_{X} = 0.
\end{eqnarray}
Furthermore, for the quaternionic Killing vectors $k^u_{{\bf
\L}}$ and $SU(2)$-valued prepotentials $P^x_{{\bf \L}}$, we find:
\begin{eqnarray}
&& k^s_{X} =0 \, , \quad k^t_\L =0 , \\
 && k^s_\L = {\rm i}g^{s\bar s} \partial_{\bar s} P^3_\L \neq 0 , \\
 && P^3_{X} = 0 ,\\
 && P^{\rm i}_{\L} =0 \, , \quad ({\rm i} =1,2).
\end{eqnarray}
The $N=1$ D-term and superpotential are respectively given by
\footnote{Particular cases of these formulae have been obtained
in \cite{ps} - \cite{tatar}.}:
\begin{eqnarray}
D^\L &=&-2 ( {\rm Im} f)^{-1\L\S}(P^0_\S(z,\bar z) + P^3_\S(w,\bar w) )\\
L &=& e^{\frac{\cK_R + \cK_H}{2}}  W(z,w) = \frac{\rm i}2
L^{X}\left(P^1_{X} - {\rm
 i}P^2_{X}\right)
\end{eqnarray}
where $\cK_R$, $\cK_H$ are the K\"ahler potentials on $\cM_R$ and
$\cM^{KH}$ respectively.
\par
 This reduction may find applications and is in fact related to many
 interesting aspects of string theory or $M$ theory compactified on
 a Calabi-Yau threefold. Indeed $M$-theory on a Calabi-Yau
 threefold originates a $N=2$ theory in five dimensions \cite{ccdf}. Trivial
 reduction on $S^1$ would give a $N=2$ theory in $D=4$. However, if
 we reduce on the orbifold $S^1/Z_2$ \cite{hw1} - \cite{aq}, then we obtain a $N=1$
 theory with a particular truncation of the $D=5, N=2$
 supergravity states. Other applications are related to
 brane-dynamics where the theory on the brane has lower
 supersymmetry than the theory on the bulk \cite{bz2}, \cite{bkvp}.
 \par
 A different mechanism is obtained by considering Type IIB theory
 on a Calaby-Yau threefold in presence of $H$-fluxes \cite{ps}
 -
  \cite{tatar} where also
 $N=1$ (or $N=0$) supersymmetric vacua can be studied.
 \par
 A related issue is the partial supersymmetry breaking of $N=2$
 down to $N=1$ through a superHiggs mechanism \cite{fgp}, \cite{fgpt}. If one integrates
 out the massive gravitino, then the theory should become a $N=1$
 theory. In this case to ``integrate out" is in principle different
 from truncating unless very special situations occur. However in
 the minimal model studied in reference \cite{fgp}, the resulting
 $N=1$ Lagrangian is a particular case of the general case studied
 here.
 \par
 The paper is organized as follows:
\\
 In section 2 we study the decomposition of the $N=8$ supergravity
 multiplet into $N'<8$ supermultiplets and infer the reduced theories from
 group-theoretical arguments.
\\
 In section 3 we extend the analysis to three, five and six
 dimensional maximal
 supergravities  reduced to eight supercharges.
 \\
In section 4
 we give the interpretation of the reduction
 procedure in a geometrical setting which will be useful to
apply our results to the specific problem of the
 $N=2\longrightarrow N=1$ reduction.
\\
 In section 5 we discuss the constraints coming
 from supersymmetry when the reduction procedure is applied to ungauged theories.
\\
 In section 6, which is the heart of the paper, we give the analysis of the
 $N=2\longrightarrow N=1$ reduction in detail, also in presence of gauging,
  both in the vector multiplet and hypermultiplet
  sectors. At the beginning of the section we discuss the
  constraints coming from the gravitino truncation, while in
  section (6.1) and (6.2) we study the reduction of the $N=2$
  vector multiplet sector.
Subsection (6.3) is devoted to the truncation of the
hypermultiplets sector, while subsection (6.4) discusses further
consequences of the gauging. In subsection (6.5) the computation
of the reduction of the scalar potential is given, and finally in
subsection (6.6) we give examples  of supergravity models which
realize this consistent truncation.
\\
 The Appendices include some technical details related to the
  reduction. In particular, in Appendices A and B we show the
  consistency of the $N=8 \to N=N'$ truncation in the superspace
  Bianchi identities formalism and we apply it to the $N=2 \to
  N=1$ reduction of gauged supergravity. In Appendix C we prove a
  formula valid for the $N=2$ vector multiplets which is
  useful for the truncation. Appendix D refers to the reduction of the
  special-K\"ahler
  manifolds with special coordinates; Appendix E contains the reduction of
  an important relation valid on quaternionic geometry in presence
  of isometries and Appendix F shows the consistency of the
  reduction of the $N=2$ scalar potential to $N=1$ and exploits some magic properties of
  the supersymmetry Ward identities. Finally,
  Appendix G
  contains the explicit form of the
  $N=2$ and $N=1$ lagrangians which are left invariant under the
  supersymmetry transformation laws given in the text.
\section{$N=8\longrightarrow N'$ reduction without gauging}
Reduction of $N=8$ supergravity to $2\leq N'\leq 6$ offers
interesting examples of consistent truncations of standard
supergravity \cite{cre},\cite{ju}.
\par
\noindent
We restrict our analysis to theories whose
$\sigma$-models are given by symmetric spaces $G/H$. This
includes all the theories with
 $N'\geq 3$ and a subset of the $N=2$ theories. The analysis turns out to be
particularly simple in all these cases.
 \par
 Let us first consider
$N'=5,6$ where the reduction only involves the graviton multiplet
and $N-N'$ spin $\frac {3} {2}$ multiplets.
\\
 In the $N=6$ case the $N=8$ R-symmetry group $SU(8)$ decomposes
as:
\begin{equation}\label{holdec1}
SU\left(8\right) \rightarrow  SU\left(6\right)\times
U\left(1\right) \times SU\left(2\right)
\end{equation}
where $SU(2)$ is the group commuting with the $N'=6$ \,
$R$-symmetry $U(6)$. Correspondingly the $N=8$ graviton multiplet
decomposes into $N'=6$ spin $2$ and  $\frac {3} {2}$ multiplets,
as follows:
\begin{eqnarray*}\label{8to6}
\left[{(\bf 2)},8{(\bf \frac{3} {2})},28{(\bf 1)},56{(\bf
\half)},70{(\bf 0)}\right]\longrightarrow
\end{eqnarray*}
\begin{eqnarray}
\left[{(\bf 2)},6{(\bf \frac{3} {2})},(15+1){(\bf 1)},(20+6){(\bf
\half)},(15+\bar 15){(\bf 0)}\right]\oplus 2\left[{(\bf \frac{3}
{2})},6{(\bf 1)},15{(\bf \half)},20{(\bf 0)}\right]
\end{eqnarray}
The hypersurface corresponding to freeze $40$ scalars of the spin
$\frac {3} {2}$ multiplets is precisely the $N'=6$ $\sigma$-model
described by the symmetric space $SO^{\star}(12)/U(6)$. Therefore
by just deleting the two spin $\frac {3} {2}$ multiplets one
obtains standard $N=6$ supergravity.
\par
Let us now consider $N'=5$. In this case the decomposition of the
$N=8$ graviton multiplet into $N'=5$ multiplets, corresponding to
the R-symmetry decomposition
\begin{equation}\label{corro}
SU\left(8\right) \rightarrow  SU\left(5\right) \times
U\left(1\right)\times SU\left(3\right)
\end{equation}
 is:
\begin{eqnarray}\label{8to5}
&&\left[ {(\bf 2)},8{(\bf \frac {3} {2})},28{(\bf 1)},56{(\bf
\half)},70{(\bf 0)}\right]\longrightarrow \nonumber  \left[{(\bf
2)},5{(\bf \frac {3} {2})},10{(\bf 1)},(10+1){(\bf
\half)},(5+\bar 5){(\bf 0)}\right]\nonumber \oplus \\ &&\oplus
\,\,\,3\left[{(\bf \frac {3} {2})},(5+1){(\bf 1)},(10+\bar 5){(\bf
\half)},(10+\bar 10){(\bf 0)}\right]
\end{eqnarray}
If we  delete the three spin $\frac {3} {2}$ multiplets we obtain
standard $N=5$ supergravity, or, geometrically, freezing the 60
scalars inside the spin $\frac {3} {2}$ multiplets corresponds to
single out the manifold $SU(5,1)/U(5)\subset E_{7(7)}/SU(8)$.
\par
When $N'\leq 4$ a new phenomenon appears since in this case also
matter multiplets start to appear in the decomposition of $N=8$
supergravity into $N'$-extended supergravities. Therefore in this
case deleting the spin  $\frac {3} {2}$ multiplets is only a
necessary, but not sufficient condition to obtain a consistent
$N'$-extended supergravity theory.
\par
Let us first start with $N'=4$ (this actually corresponds to
compactify a Type II theory in ten dimensions on $T_2\otimes
T_4/Z_2$). The decomposition of the $N=8$ graviton multiplet into
$N'=4$ multiplets, corresponding to
\begin{equation}\label{corr}
SU\left(8\right) \rightarrow  SU\left(4\right) \times
SU\left(4\right) \times  U\left(1\right)
\end{equation}
 is:
\begin{eqnarray}\label{8to4}
&&\left[ {(\bf 2)},8{(\bf \frac {3} {2})},28{(\bf 1)},56{(\bf
\half)},70{(\bf 0)}\right]\longrightarrow \left[{(\bf 2)},4{(\bf
\frac {3} {2})},6{(\bf 1)},4{(\bf \half)},2{(\bf
0)}\right]\nonumber
\\ &&\oplus\ \ 4\,\,\left[{(\bf \frac {3} {2})},4{(\bf 1)},(6+1){(\bf
\half)},(4+\bar 4){(\bf 0)}\right] \oplus 6\left[{(\bf 1)},4{(\bf
\half)},6{(\bf 0)}\right]
\end{eqnarray}
If we now delete the 4 spin  $\frac {3} {2}$ multiplets this is
equivalent to freeze $32$ scalars. When this occurs the $
E_{7(7)}/SU(8)$ manifold reduces to the submanifold
$\left(SU(1,1)/U(1)\right)\times SO(6,6)/SU(4)\times SU(4)$,
corresponding to the product space of the $N=4$ supergravity
$\sigma$-model and the $\sigma$-model of $6$ vector multiplets. In
this case a standard $N'=4$ supergravity coupled to 6 vector
multiplets corresponds to a consistent truncation since
$E_{7(7)}\supset SU(1,1)\times SO(6,6)$.
\par
Let us now consider $N'=3$. In this case we have the following
decomposition of the $N=8$ R-symmetry group:
\begin{equation}\label{hol8to3}
SU\left(8\right) \rightarrow  SU\left(3\right)  \times
U\left(1\right)\times SU\left(5\right)
\end{equation}
$ SU\left(3\right) \times  U\left(1\right)$ being the R-symmetry
of the $N=3$ theory. Note that this case is dual to the $N'=5$
case with the roles of $SU(N')$ and $SU(N-N')$ exchanged. The
decomposition of the $N=8$ multiplet is now:
\begin{eqnarray}\label{8to3}
&&\left[ {(\bf 2)},8{(\bf 3/2)},28{(\bf 1)},56{(\bf
\half)},70{(\bf 0)}\right]\longrightarrow  \left[{(\bf 2)},3{(\bf
3/2)},3{(\bf 1)},{(\bf \half)}\right]\nonumber
\\ &&\oplus\,\ 5\left[{(\bf 3/2)},3{(\bf 1)},3{(\bf
\half)},2{(\bf 0)}\right]\oplus\,\ 10\,\left[{(\bf 1)},(3+1){(\bf
\half)},(3+\bar 3){(\bf 0)}\right]
\end{eqnarray}
If we now delete the spin  $\frac {3} {2}$ multiplet we freeze the
corresponding $10$ scalars. In this case, however, it is obvious
that we cannot define a submanifold of $E_{7(7)}/SU(8)$: indeed
the standard $N=3$ supergravity coupled to $n$ vector multiplets
\cite{cama} has a non linear $\sigma$-model of the form
$SU(3,n)/SU(3)\times U(1) \times SU(n)$ and , for $n=10$,
$SU(3,10)$ is not a subgroup of $ E_{7(7)}$. Therefore we must
ask the question whether there is some $n$ for which
$SU(3,n)\subset E_{7(7)}$. The answer is $n=4$ since
\begin{equation}\label{cont}
 E_{7(7)}\supset SU(4,4)\supset SU(3,4)\times U(1)
\end{equation}
Therefore the maximal $N'=3$ supergravity contained inside the
$N=8$ theory corresponds to the coupling with $4$ matter
multiplets and the corresponding $\sigma$-model lives in the
submanifold
\begin{equation}\label{sub}
U(3,4)/U(3)\times U(4)\subset E_{7(7)}/SU(8)
\end{equation}
As far as (continuous) duality is concerned, we see that the 3
graviphotons and 4 matter vectors are in the fundamental of
$SU(3,4)$ as required by supersymmetry since
\begin{equation}
56 \to 21 + 21' + 7 + 7'
\end{equation}
This means that the $15 + 6$ vectors coming from the five
gravitino multiplets and six residual matter multiplets should
combine in the antisymmetric of $SU(3,4)$.
\\
We note that if we instead use the chain
\begin{equation}
N=8 \to
N'=4 \to N'=3
\end{equation}
we would only obtain a non-maximal theory with three matter
multiplets since in that case
\begin{equation}
E_{7(7)} \to SU(1,1) \times SO(6,6) \to SU(3,3) \times U(1)
\end{equation}
The latter is a particular case of the more general fact that
$N=4$ with $2n$ vector multiplets can be consistently truncated
to $N=3$ with $n$ vector multiplets\footnote{Note that in string
theory this would imply $n=11$ in agreement with \cite{fkou}}
using the chain\footnote{$N=3$ models based on brane flux
supersymmetry breaking have recently been constructed \cite{fpo}.}
\begin{equation}
SO(6,2n) \supset SU(3,n) \times U(1)
\end{equation}
and
\begin{equation}
\frac{SO(6,2n)}{SO(6) \times SO(2n)} \supset
\frac{SU(3,n)}{SU(3)\times SU(n) \times U(1)}
\end{equation}

\par
The last case we would like to consider is $N'=2$ where there are
two kinds of matter multiplets, namely the vector multiplets and
the hypermultiplets. In the standard $N=2$ theory the
corresponding $\sigma$-model generally is not a coset, but we
limit ourselves to examine this case, namely $\cM =G/H$. The
consistent truncation will now receive severe constraints on the
matter content since the submanifold of the $N=8$ $\sigma$-model
must factorize as:
\begin{equation}\label{facto}
\cM^{SK} (n_V)\times {\cal M}^{Q} (n_H) \subset E_{7(7)}/SU(8)
\end{equation}
where we have denoted  with ${\cal M}^{SK}(n_V)$ and ${\cal M}^Q
(n_H)$ the special-K\"{a}hler and quaternionic manifolds of real
dimensions $2n_V$ and $4n_H$ respectively.
\\
 The decomposition of the $N=8$ graviton multiplet gives now:
\begin{eqnarray}\label{8to2}
&&\left[ {(\bf 2)},8{(\bf 3/2)},28{(\bf 1)},56{(\bf
\half)},70{(\bf 0)}\right]\longrightarrow \left[{(\bf 2)},2{(\bf
3/2)},{(\bf 1)}\right]\nonumber
\\ &&\oplus \,\ 6\left[{(\bf 3/2)},2{(\bf 1)},{(\bf
\half)}\right]\oplus 15\left[{(\bf 1)},2{(\bf \half)},2{(\bf
0)}\right] \oplus 20\left[{(\bf \half)},2{(\bf 0)}\right]
\end{eqnarray}
We immediately see that deleting the spin  $\frac {3} {2}$
multiplets all the scalars survive.  So the question is now, how
many scalars we must delete so that the scalar submanifold enjoys
the  above property of reducing to ${\cal M}^{SK}(n_V)\times {\cal
M}^Q
(n_H)$. \\
 Two immediate solutions are obtained \cite{gst}. For $n_H=0$,
$n_V=15$ we find:
\begin{equation}\label{so12}
{\cal M}^{SK}(n_V=15)=SO^*(12)/U(6)\subset  E_{7(7)}/SU(8)
\end{equation}
which is indeed a  special-K\"{a}hler manifold (coinciding with
the $\sigma$-model of $N=6$ supergravity). The other solution is
$n_V=0,\, n_H=10$ for which
\begin{equation}\label{quater}
{\cal M}^{Q}(n_H=10)\,=\,E_{6(2)}/SU(6)\times SU(2) \subset
E_{7(7)}/SU(8)
\end{equation}
which is indeed a quaternionic space. It corresponds to the
$\sigma $-model obtained by compactification of Type IIB on
$T_6/Z_3$ where only the untwisted states were retained. \\
 By
$c$-map of (\ref{quater}) we obtain another solution with $n_V=9$
and $n_H=1$ corresponding to Type IIA on $T_6/Z_3$ \cite{cfg}:
\begin{equation}\label{9,1}
\frac {SU(3,3)} {SU(3)\times SU(3)\times U(1)}\times \frac
{SU(2,1)} {SU(2)\times U(1)}\subset E_{7(7)}/SU(8)
\end{equation}
\\ If we  look for other maximal subgroups $G_1\times G_2
\subset E_{7(7)}$ we find \cite{Sla} (see Table\begin{equation}
SO(6,2n) \subset SU(3,n) \times U(1)
\end{equation} \ref{groups}):
\begin{equation}\label{product}
G_1\times G_2 = Sp(6,R)\times G_{2(2)}\,\,;SU(1,1)\times
F_{4(4)}\,\,;SU(1,1)\times SO(6,6)\,\,;SU(4,4)
\end{equation}
The first two correspond to $(n_V,n_H)=(6,2)$ and to its $c$-map
image $(1,7)$, namely:
\begin{equation}\label{deco1}
 \frac {Sp(6,R)} {U(3)}\times \frac {G_{2(2)}} {SO(4)} \subset
E_{7(7)}/SU(8)
\end{equation}
\begin{equation}\label{deco2}
\frac {SU(1,1)} {U(1)}\times \frac {F_{4(4)}} {Usp(6)\times
Usp(2)}\subset E_{7(7)}/SU(8)
\end{equation}
From the last two cases we can obtain a $N=2$ truncation of $N=4$
and $N=3$ supergravities with six and four hypermultiplets
respectively:
\begin{equation}\label{trunca4}
\frac {SO(6,6)} {SO(6)\times SO(6)}\longrightarrow \frac {SO(6,4)}
{SO(6)\times SO(4)}
\end{equation}
 \begin{equation}\label{trunca3}
\frac {SU(4,3)} {SU(4)\times SU(3)\times U(1)}\longrightarrow
\frac {SU(4,2)} {SU(4)\times SU(2)\times U(1)}
\end{equation}
with $(n_V,n_H)=(1,6)$ and $(n_V,n_H)=(0,4)$ respectively. \\
 The
first, together with its $c$-map $(n_V,n_H)=(5,2)$
\begin{equation}\label{cmap}
\frac {SU(1,1)} {U(1)}\times \frac {SO(2,4)} {SO(2)\times
SO(4)}\times \frac {SO(4,2)} {SO(4)\times SO(2)}
\end{equation}
corresponds to Type IIB (Type IIA) on $T_6/Z_4$\footnote{Note that
$\frac {SO(4,2)} {SO(4)\times SO(2)} \sim \frac {SU(2,2)}
{SU(2)\times SU(2)\times U(1)}$.}. The last is a truncation of the
$(n_V,n_H)=(0,10)$ case and its $c$-map is
\begin{equation}\label{cmap1}
\frac {SU(3,1)} {SU(3)\times U(1)}\times \frac {SU(2,1)}
{SU(2)\times U(1)}
\end{equation}
with $(n_V,n_H)=(3,1)$ ( This is a truncation of the
$(n_V,n_H)=(9,1)$ case.).

By the decomposition
\begin{equation}\label{ulte}
SO(6,6)\longrightarrow SO(4,6-p)\times SO(2,p)
\end{equation}
we can obtain the additional cases:
\begin{eqnarray}\label{add}
  (n_V=2,n_H=5) & p=1 \\
  (n_V=4,n_H=3) &p=3   \\
  (n_V=3,n_H=4) & p=2  \\
  (n_V=6,n_H=1) & p=5  \\
  (n_V=7,n_H=0) & p=6
\end{eqnarray}
Their $c$-map do not give new models. We note that the case $p=6$
is a truncation of the  $(n_V,n_H)=(15,0)$ case and that the case
$p=5$ is not a subcase of the $(n_V,n_H)=(9,1)$ case because the
corresponding quaternionic manifold is in this case
$\frac{Usp(2,2)}{Usp(2)\times Usp(2)}$ which is not the same of
the $(n_V,n_H)=(9,1)$ case.
\\
 In conclusion we have found eleven ``maximal" cases: the cases $(n_V,n_H)= (15,0),
 (6,1)$ which have no $c$-map counterpart, the case $(n_V,n_H)= (3,4)$
 which is self conjugate under $c$-map and four pairs conjugate under
 $c$- map, namely:
\begin{eqnarray}\label{6cases}
&&(n_V=6,n_H=2)\stackrel {c-map}\longleftrightarrow (n_V=1,n_H=7)\\
&&(n_V=5,n_H=2)\stackrel {c-map}\longleftrightarrow(n_V=1,n_H=6)\\
&&(n_V=0,n_H=10)\stackrel {c-map}\longleftrightarrow
(n_V=9,n_H=1)\\
&&(n_V=4,n_H=3)\stackrel {c-map}\longleftrightarrow (n_V=2,n_H=5)
\end{eqnarray}

Many of these cases can be retrieved from Type II string theories
compactified on $\ZZ_N$ orbifolds which preserve one left and one
right supersymmetry \cite {cfg},\cite {fp}.
\section{$D=3$, $D=5$ and $D=6$ reduction of maximal supergravity to theories with eight supercharges.}
The same analysis can be carried out in $N=2$ theories (eight
supercharges) in $D=3$, $D=5$ (for the cases where the scalars
span a symmetric space) and in  $D=6$.
\par
In $D=5$, $N=8$ supergravity has a non-linear $\sigma$-model
$E_{6(6)}/USp(8)$\cite{cre}. We consider only the $N=8 \to N=2$
case.
\par
The 42 scalars, decomposed with respect to the $N=2$ theory,
consist of $14$ scalars belonging to vector multiplets and
$4\times 7=28$ scalars belonging to quaternionic multiplets,
giving $(n_V=14, n_H=0)$ and $(n_V =0,n_H =7)$ models which
correspond to $\frac{SU^*(6)}{USp(6)} \subset
\frac{E_{6(6)}}{USp(8)}$ and $\frac{F_{4(4)}}{USp(6)\times USp(2)}
\subset \frac{E_{6(6)}}{USp(8)}$ \cite{gst}. For each model in
$D=4$ there is a parent in $D=5$ (the above correspond to the $n_V
\cdot n_H =0$ cases).
\par
If we now look to spaces with isometry groups $G_1 \times G_2
\subset E_{6(6)}$, where $G_1$, $G_2$ correspond to real special
geometry and quaternionic geometry respectively, we find (see
Table \ref{groups}):
\begin{equation}
 G_1 \times G_2 = SL (3,\IC ) \times SU(2,1)
 \end{equation}
 which
give rise to
\begin{equation} \frac{SL(3,\IC )}{SU(3)} \times
\frac{SU(2,1)}{SU(2) \times U(1)} \subset \frac{E_{6(6)}}{USp(8)}
\qquad (n_V=8 \, , \, n_H=1)
\end{equation} and \cite{Sla}
\begin{equation}
G_1 \times G_2 = SL (3,\IR ) \times G_{2(2)}
\end{equation}
 giving
\begin{equation} \frac{SL(3,\IR )}{SO(3)} \times
\frac{G_{2(2)}}{SO(4)} \subset \frac{E_{6(6)}}{USp(8)} \qquad
(n_V=5 \, , \, n_H=2).
\end{equation}
If we go through the $N=4$
theory we also get the series of six cases
\begin{equation} SO(1,1) \times \frac{SO(1,p)}{SO(p)}\times
\frac{SO(4,5-p)}{SO(4)\times SO(5-p)} \qquad (n_V=p+1 \, , \,
n_H=5-p), \quad 0\leq p \leq 5. \label{5dseries}\end{equation} So
we see that there are ten $D=5$ cases with similar types of
quaternionic manifold as in $D=4$ (with the only exception of the
$n_H =10$ case.). \vskip 5mm
\par In $D=6$ the $N=8$ ($(2,2)$ theory) $\sigma$-model is $SO(5,5) / SO(5) \times
SO(5)$. If we decompose the $(2,2)$ theory with respect to the
$(1,0)$ theory we get 5 tensor multiplets and 5 hypermultiplets
corresponding to \begin{equation} \frac{SO(1,5)}{SO(5)} \subset
\frac{SO(5,5)}{SO(5) \times SO(5)} \, , \quad
\frac{SO(4,5)}{SO(4)\times SO(5)} \subset \frac{SO(5,5)}{SO(5)
\times SO(5)}. \label{6d}\end{equation} These are the $n_T \cdot
n_H =0$ cases.
\par
Again we can now look at subgroups $G_1 \times G_2 \subset
SO(5,5)$ where $G_1 = SO(1,n_T)$ and $G_2$ is the isometry group
of a quaternionic manifold.
\\
We find a series analogous to the $D=5$ case (\ref{5dseries}),
with
\begin{equation}
G_1 = SO(1,p) \, , \quad G_2 = SO(4,5-p) \qquad (n_T = p, n_V
=5-p)
\end{equation}
corresponding to the manifolds \begin{equation}
\frac{SO(1,p)}{SO(p)}\times \frac{SO(4,5-p)}{SO(4)\times SO(5-p)}
\qquad (n_T=p \, , \, n_H=5-p), \quad 0\leq p \leq 5
\label{6dseries}\end{equation} which contains also the above
mentioned $n_T \cdot n_H =0$ cases (\ref{6d}).
\vskip 5mm
\par
The reduction of $N=8 \to N=2$ supergravity studied in $D=6,5$
and $4$ finds a further simplification if we look for theories
with eight supercharges in $D=3$, where the R-symmetry is
$SU(2)_1 \times SU(2)_2$.\\
In fact, if we compactify Type II on a Calabi-Yau threefold times
$S_1$, down to $D=3$, then Type IIA and IIB become the same
theory with $1\Leftrightarrow 2$. The $N=4$ $\s$-model is a
product of two quaternionic geometries, where $n_{H_1} =h_{1,1}
+1$, $n_{H_2} =h_{2,1} +1$, the extra quaternion coming from the
graviton and graviphoton degrees of freedom.
\par
More generically, suppose we have a theory which at $D=4$ has a
$\s$-model $\cM^{SK}(n_V) \times \cM^Q(n_H)$, then its dimensional
reduction to $D=3$ will give rise to a $N=4$ $SU(2)_1 \times
SU(2)_2$ theory with $\s$-model $\cM^{Q_1}(n_{H_1}=n_V+1)\times
\cM^{Q_2}(n_{H_2}=n_H)$, where $\cM^{Q_1}$ is the dual
quaternionic manifold of $\cM^{SK}{(n_V)}$.\\
Using the previous recipe, if we look to the $D=4$, $N=2$
theories of section 2 obtained from $N=8$, we can predict $N=4$
theories at $D=3$ which are embedded in the
$\frac{E_{8(8)}}{SO(16)}$ $\s$-model of $D=3$, $N=16$ maximal
supergravity.
\par
From $(n_V =15 , n_H =0)$ and $(n_V =0, n_H =10)$ we respectively
obtain:
\begin{eqnarray}
  (n_{H_1} =16 , n_{H_2} = 0)\, ,  && \frac{E_{7(-5)}}{SO(12)\times SU(2)} \,\subset \, \frac{E_{8(8)}}{SO(16)} \\
  (n_{H_1} =1 , n_{H_2} = 10)\, , && \frac{SU(2,1)}{SU(2)\times U(1)}\times \frac{E_{6(2)}}{SU(6)\times
  SU(2)}\,
  \subset \, \frac{E_{8(8)}}{SO(16)} \\
 (n_{H_1} =2 , n_{H_2} = 7)\, , && \frac{G_{2(2)}}{SO(4)}\times \frac{F_{4(4)}}{USp(6)\times USp(2)} \,
 \subset \, \frac{E_{8(8)}}{SO(16)}.
\end{eqnarray}
Also, by using the embedding \cite{gil} $E_{8(8)} \supset SO(8,8)$
we have the further possibility:
\begin{equation}
\frac{E_{8(8)}}{SO(16)} \, \supset \, \frac{SO(8,8)}{SO(8)\times
SO(8) }\, \supset \, \frac{SO(4,k) \times SO(4, 8-k)}{SO(4) \times
SO(k) \times SO(4) \times SO(8-k)}
\end{equation}
with $n_{H_1} = k$, $n_{H_2} = 8-k$ ($k=0$ is a subcase of the
$(n_{H_1} =16, n_{H_2} =0)$ case, since $SO(4,8) \times SU(2)
\subset E_{7(-5)}$.). They are all dimensional reductions of the
cases previously studied at $D=4$.\\
For the decomposition of the isometry group of maximal $D=3$
supergravity to maximal subgroups, see Table \ref{groups}.
\begin{table}[h]\label{groups}
\centering
  \caption{Decomposition of ``duality'' groups of maximal $D=3,4,5$
  supergravities with respect to maximal subgroups relevant for supergravity reduction.\cite{gil},\cite{Sla}}
---------------------------------------------\\
{\bf Some Maximal subgroups of $E_{8(8)}$:}\\
$SO(16)$\\
$SO(8,8)$\\
$E_{7(-5)} \times SU(2)$ \\
$E_{6(2)} \times SU(2,1)$ \\
$G_{2(2)} \times F_{4(4)}$ \\
---------------------------------------------\\
{\bf Some Maximal subgroups of $E_{7(7)}$:}\\
$SU(8)$\\
$SU(4,4)$\\
$E_{6(2)} \times SO(2)$\\
$SO^*(12) \times SU(2)$ \\
$SO(6,6) \times SU(1,1)$\\
$SU(3,3) \times SU(2,1)$\\
$SU(1,1) \times F_{4(4)}$\\
$Sp(6,\IR )\times G_{2(2)}$ \\
---------------------------------------------\\
{\bf Some Maximal subgroups of $E_{6(6)}$:}\\
$USp(8)$\\
$F_{4(4)}$\\
$SU^*(6) \times SU(2)$\\
$SO(5,5) \times SO(1,1)$\\
$SL(3,\IR ) \times G_{2(2)}$\\
$SL(3,\IC ) \times SU(2,1)$ \\
---------------------------------------------\\
\end{table}
\section{Geometrical interpretation}
It is interesting to analyze the results of the previous section
in geometrical terms, that is to explore the consistency of the
reduction of the $N=8$ $\sigma$-model $E_{7(7)}/SU(8)$ to the
appropriate submanifolds for different values of $N'$. A
consistent truncation of a manifold of dimension $n$ to a
submanifold of dimension $n-k$ can be obtained by considering a
set of $k$ 1-forms $\phi^i$, $i=1,\cdots,k$, which vanish on the
submanifold and such that they are in involution, that is:
\begin{equation}\label{inv}
d\phi^i = \theta ^i_j\wedge \phi^j
\end{equation}
where $\theta ^i_j$ are suitable 1-forms on the manifold.
\par
 To apply this result, known as Frobenius theorem, to our problem we
consider the coset representative $U$ of $E_{7(7)}/SU(8)$ in the
$\bf 56$ fundamental representation of $E_{7(7)}$ and the
corresponding left invariant 1-form \footnote{We use notations as
in ref.\cite{adf}.}:
\begin{equation}
  \label{defgamma}
  \G  \equiv U^{-1} dU =
\pmatrix{\Omega & \bar P \cr P & \bar \Omega \cr }
\end{equation}
satisfying the Cartan-Maurer equation
\begin{equation}\label{carta}
d\G  + \G  \wedge \G  = 0.
\end{equation}
 Here the $28\times 28$
subblocks $ \Omega$ and $P$ embed the $SU(8)$ connection and the
vielbein of $E_{7(7)}/SU(8)$. Introducing indices $A,B=1,\cdots ,
8$ we have explicitly:
\begin{equation}\label{expli}
\Omega \equiv 2\omega^{[A}_{\phantom{[C}[C}
\delta^{B]}_{D]};\,\,\,P\equiv P_{ABCD}
\end{equation}
where $\omega^A_{\ B}$ is the $SU(8)$ connection and $P_{ABCD}$ is
the vielbein of $E_{7(7)}/SU(8)$, antisymmetric in its four
indices and satisfying the reality condition:
\begin{equation}\label{vie}
P_{ABCD}= \frac {1} {24} \epsilon_{ABCDPQRS}\bar P^{PQRS} .
\end{equation}
From the Cartan-Maurer equations one easily finds the two
structure equations:
\begin{eqnarray}
R^A_{\phantom{B}B} &\equiv& d\omega^A_{\phantom{B}B}
+\omega^A_{\phantom{C}C}\wedge \omega^C_{\phantom{B}B}=-\frac {1}
{3} \bar P^{ALMN}\wedge P_{BLMN}\label{stru}\\ \nabla \bar
P^{ABCD}&\equiv& d\bar P^{ABCD}-4 \omega^{[ A}_{\phantom{L}L} \bar
P^{BCD]L}=0 .\label{torsionless}
\end{eqnarray}
Equation (\ref{stru}) gives the  $SU(8)$ Lie algebra valued
curvature $R^A_{\phantom{B}B}$ in terms of the vielbein of the
symmetric coset $E_{7(7)}/SU(8)$ and equation (\ref{torsionless})
expresses the fact that the same manifold is torsionless. Note
that, since the coset is symmetric, the Lie algebra connection
$\omega^A_{\ B}$ is simply related via a structure constant to the
Riemannian spin connection.
\par
Let us now consider how the vielbein $P_{ABCD}$ decomposes under
the holonomy reduction $SU(8)\longrightarrow SU(N')\times
U(1)\times SU(8-N')$. We call $a,b,c,\cdots =1,\cdots ,N'$ the
indices of $SU(N')$ and $i,j,k\cdots = 1,\cdots ,8-N'$ the indices
of $SU(8-N')$. Then the holonomy reduction gives the following
fragments:
\begin{equation}\label{defrag}
P_{ABCD}\longrightarrow P_{abcd}\oplus P_{abci}\oplus
P_{abij}\oplus P_{aijk}\oplus P_{ijkl}
\end{equation}
where actually some of the fragments can be zero if the number of
antisymmetric indices of $SU(N')$ or $SU(8-N')$ exceeds $N'$ or
$8-N'$, respectively. Now we observe that  $P_{abcd}$ satisfies
equation (\ref{torsionless}) which gives for this particular
component:
\begin{equation}\label{redtor}
 d\bar P^{abcd}-4 \omega^{[ a}_{\phantom{[a}\ell} \bar
P^{bcd]\ell} -4 \omega^{[ a}_{\phantom{[a}i} \bar P^{bcd]i}=0
\end{equation}
We see that, in order that equation (\ref{redtor}) describe a
torsionless submanifold with $SU(N')\times U(1)$ holonomy, we must
set $\omega^a_{\phantom{a}i}=0$ and since
\begin{equation}\label{zerocur}
R^a_{\phantom{i}i} \equiv d\omega^a_{\phantom{i}i}
+\omega^a_{\phantom{c}c}\wedge
\omega^c_{\phantom{i}i}+\omega^a_{\phantom{j}j}\wedge
\omega^j_{\phantom{i}i}=-\frac {1} {3} \bar P^{aLMN}\wedge
P_{\phantom{i}iLMN}
\end{equation}
we must also impose that, on the submanifold whose vielbeins are $
P^{abcd}$, the curvature with mixed indices is zero, namely
$R^a_{\phantom{i}i}=-\frac {1} {3} \bar P^{aLMN}\wedge
P_{iLMN}=0$. Using the decomposition (\ref{defrag}), equation
(\ref{zerocur}) can be rewritten as follows
\begin{eqnarray}\label{funda}
d\omega^a_{\phantom{i}i}\,&=&\,-\omega^a_{\phantom{c}c}\wedge
\omega^c_{\phantom{i}i}-\omega^a_{\phantom{j}j}\wedge
\omega^j_{\phantom{i}i}-\frac {1} {3}\bar P^{abcd}\wedge
P_{ibcd}\nonumber \\&-&\frac {1} {3}\bar P^{abcj}\wedge
P_{ibcj}-\frac {1} {3}\bar P^{abjk}\wedge P_{ibjk}-\frac {1}
{3}\bar P^{ajkl}\wedge P_{ijkl}
\end{eqnarray}
On the basis of the Frobenius theorem, each term on the r.h.s. of
(\ref{funda}) must be in involution with
$\omega^a_{\phantom{a}i}$; this is satisfied for the terms
bilinear in the $\omega$-connections, but not for those involving
the vielbein. In order to obtain involution, we must also set to
zero some of the vielbein 1-forms and verify that also these are
actually in involution. Let us see how we can achieve this result
in the various cases.
\par
When $N'=6$, $P_{ijkl}=P_{ibjk}\equiv 0$ because we have 4-fold or
threefold antisymmetrization of the $SU(2)$ indices. Therefore it
is sufficient to set
\begin{equation}\label{vielzero1}
P_{ibcd} \equiv \bar P^{ibcd}=0
\end{equation}
on the submanifold in order to obtain involution, since in this
case equation (\ref{funda}) reduces to
\begin{equation}\label{redu}
d\omega^a_{\phantom{i}i}\,=\,-\omega^a_{\phantom{c}c}\wedge
\omega^c_{\phantom{i}i}-\omega^a_{\phantom{j}j}\wedge
\omega^j_{\phantom{i}i}\longrightarrow R^a_{\phantom{i}i}=0
\end{equation}
We still have to verify that also the vanishing 1-forms $P_{ibcd}$
are in involution with themselves and with
$\omega^a_{\phantom{i}i}$. Indeed, from equation
(\ref{torsionless}), we find:
\begin{equation}\label{verif}
d\bar P^{abci}=3\omega^{[a}_{\phantom{d}d}\bar
P^{bc]di}+3\omega^{[a}_{\phantom{j}j}\bar P^{bc]ji}
+\omega^{i}_{\phantom{d}d}\bar
P^{abcd}+\omega^{i}_{\phantom{j}j}\bar P^{abcj}
\end{equation}
and we see that every term in the r.h.s. contains either $\bar
P^{abcj}$ or $\omega^{a}_{\phantom{i}i}$ so that we get
involution.
\\
We note that condition (\ref{vielzero1}) is equivalent to impose
that the $SU(6)\times U(1)\times SU(2)$ representation ${\bf
(20,0,2)}$ must be absent in the reduction of the scalar vielbein,
and this implies that all the $40$ scalars of the $N'=6$ spin
$\frac {3} {2}$ multiplets must be frozen according to our
analysis in the previous section. In conclusion, setting
$P^{abcj}=0$ and $\omega^{a}_{\phantom{a}i}=0$, we define a
consistent truncation of the $N=8$ theory down to a $N'=6$ theory
since the above conditions define a submanifold of holonomy
$SU(6)\times U(1)\times SU(2)$  whose curvature is easily seen to
be given by
\begin{equation}\label{n6cur}
R^a_{\phantom{b}b} = -\frac {1} {3}\bar P^{almn}\wedge P_{blmn}.
\end{equation}
The corresponding manifold has dimension $30$ and of course
coincides with $SO^*(12)/U(6)$.
\par
The cases $N'=5$ and $N'=4$ can be treated in exactly the same
way. For $N'=5$, equation (\ref{funda}) does not contain
$P_{ijkl}$ and in order to get involution we have to set
\begin{equation}\label{n5tru}
P_{abci}=P_{abij}=0
\end{equation}
which corresponds to delete, in the holonomy reduction
(\ref{corro}), the representations ${\bf(10,1,3)}$ and ${\bf (\bar
10,-1,\bar 3)}$ for the vielbein (because of the reality
condition(\ref{vie})). According to the discussion of the previous
section, this is equivalent to freeze the 60 scalars of the spin
$\frac {3} {2}$ multiplets. Using again equation
(\ref{torsionless}), we can immediately verify that $P_{abci},
P_{abij}$ and $\omega^{a}_{\phantom{a}i}$ are indeed in involution
so that the reduction to the submanifold $SU(1,5)/U(5)$ is indeed
consistent.
\par
For $N'=4$, equation (\ref{funda}) contains all the terms bilinear
in the vielbeins. However it is sufficient to set
\begin{equation}\label{cons4}
P_{abci}=P_{aijk}(\approx \bar P^{abci})=0
\end{equation}
to achieve the vanishing of the r.h.s. of (\ref{funda}) on the
submanifold. This corresponds to delete, in the holonomy reduction
(\ref{8to4}), the representations ${\bf (4,1,4)}$ and ${\bf (\bar
4,-1, \bar 4)}$ in the decomposition of the scalar vielbein, that
is to freeze the 32 scalars appearing in the $N'=4$ spin $\frac
{3} {2}$ multiplets. Again the structure equation
(\ref{torsionless}) can be used to show that $P_{abci}, P_{aijk}$
and $\omega^{a}_{\phantom{a}i}$ are in involution so that we get a
consistent reduction to the $N'=4$ submanifold $SU(1,1)/U(1)\times
SO(6,6)/[SU(4)\times SU(4)]$.
\par

\par
The reduction to the submanifold of the $N'=3$ theory requires a
little more labor. In this case equation (\ref{funda}) does not
contain the term $\bar P^{abcd}\wedge P_{ibcd}$ and if we set
\begin{equation}\label{n3}
 P^{abci} = P_{ijkl}=0
\end{equation}
then
\begin{equation}\label{curn3}
R^a_{\phantom{i}i} = -\frac {1} {3}\bar P^{abjk}\wedge
P_{ibjk}\neq 0
\end{equation}
We could of course set also $\bar P^{abjk}=0$, but then we would
be left with a theory without scalars, that is pure $N'=3$
supergravity theory. \\
In order to obtain a matter coupled $N'=3$ theory, we further
reduce the submanifold holonomy:
\begin{equation}\label{further}
SU(8)\longrightarrow SU(3)\times U(1)\times SU(5)\longrightarrow
SU(3)\times U(1)\times SU(4)
\end{equation}
To see that in this case we obtain a consistent submanifold, we
split the $SU(5)$ indices $i,j,\cdots =1,\cdots ,5$ into $SU(4)$
indices $\alpha,\beta,\cdots =1,\cdots ,4$ and the index 5. Then
we have:
\begin{eqnarray}\label{5to4}
R^a_{\phantom{i}i}&\longrightarrow&
R^a_{\phantom{\alpha}\alpha};\,\,
R^a_{\phantom{5}5}\\
R^a_{\phantom{\alpha}\alpha}&=&-\frac {1} {3} \bar P^{ab\beta
\gamma}\wedge P_{\alpha b \beta \gamma}-\frac {2} {3} \bar
P^{ab\beta 5}\wedge P_{\alpha b \beta 5}\\
R^a_{\phantom{\alpha}5}&=&-\frac {1} {3} \bar P^{ab\beta
\gamma}\wedge P_{5 b \beta \gamma}
\end{eqnarray}
The vielbeins $P_{ab\alpha \beta}$ and $P_{5\beta ab}$ are in the
representations ${\bf (\bar 3,6)}$ and ${\bf (\bar 3,4)}$ of
$SU(3)\times SU(4)$, respectively. Hence if we delete the
representation ${\bf (\bar 3,6)}$, that is if we set
\begin{equation}\label{furtherdel}
P_{ab\alpha \beta}= P_{5 b \beta \gamma}=0
\end{equation}
we get $R^a_{\phantom{\alpha}\alpha}= R^a_{\phantom{5}5}=0$. On
the light of the discussion given in the previous section for the
same case, this corresponds to select, as different from zero on
the submanifold, only the vielbeins with indices in the $({\bf
\bar 3}, {\bf 4})$ rep. of the holonomy group $U(3) \times SU(4)$.
We obtain in this case a consistent reduction to the submanifold
spanned by the vielbeins $P_{ab\gamma 5}$ since it can be easily
verified that $P_{abc\alpha},P_{abc5},P_{\alpha \beta \gamma
\delta},P_{\alpha \beta \gamma
5},\omega^a_{\phantom{\alpha}\alpha},\omega^a_{\phantom{5}5}$ are
all in involution among themselves.
\par
Finally, in the $N'=2$ case, in order to have involution for
$\omega^a_{\ i} =0 $, we must have \begin{equation} R^a_{\ i} =
-\frac {1} {3}\bar P^{abjk}\wedge P_{ibjk}-\frac {1} {3}\bar
P^{ajkl}\wedge P_{ijkl}=0 \end{equation} on the manifold.
\\
If we take $\bar P^{abjk}$ (and its complex conjugate
$P_{ijk\ell}$) or $\bar P^{ajkl}$ vanishing on the submanifold
this corresponds to delete the complex representation ${\bf (\bar
1,-1,15)}$ or the real representation ${\bf (2,0,20)}$ of the
holonomy group $SU(2)\times U(1) \times SU(6)$. We may check
immediately that in both cases the vanishing vielbein are indeed
in involution with $\omega^a_{\ i}$ and with themselves. Indeed:
\begin{eqnarray} d\bar P^{abci}&=& 3\omega^{[a}_{\phantom{d}d}\bar
P^{bc]di}+3\omega^{[a}_{\phantom{j}j}\bar P^{bc]ji}
+\omega^{i}_{\phantom{d}d}\bar
P^{abcd}+\omega^{i}_{\phantom{j}j}\bar P^{abcj}\\
d\bar P^{ajk\ell}&=& \omega^{a}_{\phantom{a}b}\bar
P^{bjk\ell}+\omega^{a}_{\phantom{j}i}\bar P^{ijk\ell}
+3\omega^{[j}_{\phantom{[d}b}\bar P^{k\ell ]
ab}+3\omega^{[j}_{\phantom{[j}i}\bar P^{k\ell ] ia}
\end{eqnarray}
and we see that in both cases the involution condition is
satisfied. Therefore we have found a consistent reduction to the
submanifolds $SO^*(12)/U(6)$ and $E_{6(2)}/SU(6)\times SU(2)$
which are special-K\"{a}hler and quaternionic manifolds
respectively of maximal holonomy.
\\
The other cases treated group theoretically in the previous
section can be handled in an analogous way, provided we reduce the
holonomy of the resulting submanifold in a suitable way. We just
give an example.\\
Consider the manifold given in equation (\ref{9,1}), corresponding
to $(n_V,n_H)=(9,1)$. We decompose the representation ${\bf 6}$ of
$SU(6)$ into the representation  ${\bf ( 3,1)} + {\bf (1,3)} $ of
$SU(3) \times SU(3)$. Correspondingly, the index $i$ in the ${\bf
6}$ of $SU(6)$ is decomposed: \begin{equation} i \to \a , \dot\a
\, , \, \quad (\a , \dot\a =1,2,3) \end{equation} where $\a$ and
$\dot\a$ run on the fundamental rep of the two $SU(3)$ groups.
Then we have: \be R^a_{\ i} \to R^a_{\ \a} \, , \, R^a_{\ \dot\a}
\end{equation} and we find:
 \begin{eqnarray}
  R^a_{\ \a} &=& \stackrel{(1,\bar 3 , 1)}{\bar
 P^{ab\beta\gamma}}\wedge
 \stackrel{(2,1 , 1)}{ P_{\a b\beta\gamma}}+
 \stackrel{(1, 3 , 3)}{\bar
 P^{ab\beta\dot\gamma}}\wedge
 \stackrel{(2,\bar 3 , 3)}{ P_{\a b\beta\dot\gamma}}
+ \stackrel{(1 , 1,\bar 3)}{\bar
 P^{ab\dot\beta\dot\gamma}}\wedge
 \stackrel{(2,3 ,\bar 3)}{ P_{\a b\dot\beta\dot\gamma}}
 \nonumber\\
&+& \stackrel{(2,\bar 3 , 3)}{\bar
 P^{a\beta\gamma\dot\delta}}\wedge
 \stackrel{(1,1 ,3)}{ P_{\a \beta\gamma\dot\delta}}
+ \stackrel{(2 , 3,\bar 3)}{\bar
 P^{a\beta\dot\gamma\dot\delta}}\wedge
 \stackrel{(1,\bar 3 , \bar 3)}{ P_{\a \beta\dot\gamma\dot\delta}}
 +
 \stackrel{(2 , 1,1)}{\bar
 P^{a\dot\beta\dot\gamma\dot\delta}}\wedge
 \stackrel{(1, 3 , 1)}{ P_{\a \dot\beta\dot\gamma\dot\delta}}
\end{eqnarray}
where we have set on the top of each vielbein the rep of $SU(2)
\times SU(3) \times SU(3)$ to which it belongs. We see that
deleting the vielbein in the reps $ ({\bf 1}, {\bf\bar 3 },{\bf
1})$, $({\bf 2} ,{\bf \bar 3} ,{\bf 3})$ and $({\bf 1},{\bf
1},{\bf\bar 3} )$ (and their complex conjugates) we get $R^a_{\
\a} =0$ so that involution is satisfied. An analogous computation
can be done, with the same conclusions, for $R^a_{\ \dot\a}$. Note
that the vielbein which survive, $P_{a\a\beta\gamma} $ and
$P_{ab\beta\dot\gamma}$, in the representations $({\bf 2},{\bf
1},{\bf 1})$ and $({\bf 1},{\bf 3},{\bf 3})$ respectively, do in
fact describe the vielbein system of the given manifold.
\\
The involution of the deleted vielbein is also easily proved.
Indeed:
\begin{equation}
d\bar P^{ab\beta\gamma} = 2 \omega^{[a}_{\phantom{[a} c} \wedge
\bar P^{b]c\beta\gamma} + 2 \omega^{[a}_{\phantom{[a}\a} \wedge
\bar P^{b]\a\beta\gamma} + 2 \omega^{[a}_{\phantom{[a} \dot\a}
\wedge \bar P^{b]\dot\a\beta\gamma} - 2
\omega^{[\beta}_{\phantom{[\b} \delta} \wedge \bar
P^{\gamma]\delta ab} - 2 \omega^{[\beta}_{\phantom{[\b}
\dot\delta} \wedge \bar P^{\gamma]\dot\delta ab}
 \ee
 and we see that each term contains at least a 1-form which is
 zero on the submanifold.
\par
 It is a simple exercise to verify that one can
actually  further reduce the holonomy to all the holonomy
subgroups of the various cases treated in Section 2 and find
consistent reduction to the corresponding special-K\"{a}hler and
quaternionic symmetric coset submanifolds.

\section{Consistency constraints from supersymmetry}
In the previous sections we have analyzed the effects of
truncating out some of the supercharges in the supergravity
theories. In particular, in section 3 we have considered the
effects of the reduction of the holonomy group for the various
supermultiplets at the linearized level, while in section four we
have studied the consequences of such a reduction on the scalar
sectors.
\par
We still have to analyze if the consistency found at the level of
$\sigma$-model in the geometrical analysis can be extended to the
full supersymmetric level.
\\
For this purpose, we analyze the supersymmetry transformation laws
of $N=8$ supergravity, when the R-symmetry gets reduced from
$SU(8)$ to $SU(N')\times U(1)$.  They are, neglecting three
fermions terms:
\begin{eqnarray}
\delta V^a_\mu &=& -{\rm i} \bar \psi^A_\mu \gamma^a \epsilon_A +
h.c.
\\
 \delta \psi_{A\mu} &=& \nabla_\mu \epsilon_A +
T_{AB|\nu\rho}^{- }
\gamma_{\mu}^{\phantom{\mu} \nu}\gamma^{\rho} \epsilon^B \\
\delta \chi_{ABC} &=& P_{ABCD,\alpha} \partial_\mu \phi^\a
\gamma^\mu \epsilon^D + T^-_{[AB|\mu\nu}\gamma^{\mu\nu} \epsilon_{C]} \\
\delta A^{\L\S}_\mu &=& f^{\L\S}_{AB}\left(\bar\psi^A_\mu
\epsilon^B
+ \bar\chi^{ABC} \gamma_\mu\epsilon_C \right) + h.c.\\
\delta \phi^\alpha &=&\bar P^{ABCD,\a} \bar\chi_{ABC}\epsilon_D +
h.c.
\end{eqnarray}
  (the $SU(8)$ indices $A,\cdots $ run from 1 to 8).
We use the same notation as in reference \cite{adf}: we call $U$
the coset representative of $E_{7(7)}/SU(8)$ parametrized as
follows:
\begin{equation}
\label{u} U = \frac{1}{\sqrt{2}} \pmatrix{
  f+{\rm i}h & \bar f+{\rm i}\bar h \cr
  f-{\rm i}h & \bar f-{\rm i}\bar h
}
\end{equation}
 where
$f^{\L\S}_{AB}$ and $h_{\L\S AB}$ ($\L , \S , \cdots =1,\cdots ,
8$) are labelled by couples of antisymmetric indices $\L\S$
and$AB$ with $\L,\S=1\cdots,8$ and $A,B=1\cdots,8$. Therefore they
describe $28 \times 28$ sub-blocks of the $56\times 56$ symplectic
matrix (coinciding with the fundamental ${\bf 56}$ representation
of $E_{7(7)}$). Note that $U$ transforms on the left as the ${\bf
56}$ representation of $E_{7(7)}$ and on the right as the ${\bf
28\oplus \bar{ 28}}$ of $SU(8)$ .
\\
In terms of $f$ and $h$, the $2$-form $T_{AB}$ is given by:
 \be
T_{AB} = -\frac{\rm i}{2} (\bar f^{-1})_{AB\L\S} F^{\L\S} = \half
\left(h_{\L\S AB} F^{\L\S} - f^{\L\S}_{AB} {\cal G}_{\L\S}\right)
\end{equation} where ${\cal G}_{\L\S}$ is the magnetic counterpart of the
field-strength $F^{\L\S}$. The spinor fields $\psi_{A\mu}$ and
$\chi_{ABC}$ are the $N=8$ left-handed gravitinos and dilatinos
respectively. Finally, the covariant derivative acting on the
spinors is defined as follows:
\begin{equation}
\nabla \epsilon_A = {\cal D} \epsilon_A + \omega_A^{\phantom{A}B}
\epsilon_B
\end{equation}
where $\omega_A^{\phantom{A}B}$ is the $SU(8)$ connection and
${\cal D}_\mu$ denotes the Lorentz covariant derivative.
\par
\par
Let us first analyze the gravitino decomposition. We want to
reduce the theory to an $N' \leq 8$ one. Therefore, to reduce the
R-symmetry $SU(8) \to SU(N')\times U(1)$, we decompose the
holonomy indices $A,\cdots \Rightarrow (a,i)$ with $a=1,\cdots ,
N'$ and $i=1,\cdots , 8-N'$. We then have to truncate out (to set
to zero) the $8-N'$ gravitinos $\psi_{i\mu }$ and the
corresponding supersymmetry parameters $\epsilon_i$. We get:
\begin{eqnarray}
\delta \psi_{a\mu} &=& {\cal D}_\mu \epsilon_a +
\omega_a^{\phantom{A}b} \epsilon_b + T_{ab|\nu\rho}^{- }
\gamma_{\mu}^{\phantom{\mu} \nu}\gamma^{\rho} \epsilon^b \label{surv}\\
\delta \psi_{i\mu} &=& \omega_i^{\phantom{A}a} \epsilon_a +
T_{ia|\nu\rho}^{- } \gamma_{\mu}^{\phantom{\mu} \nu}\gamma^{\rho}
\epsilon^a \equiv 0
\end{eqnarray}
The second equation, consistency condition for the truncation,
implies
\begin{equation}
\omega_i^{\ a} = 0 , \quad \quad T_{ia}^{- }=0 \label{condition}
\end{equation}
The first condition in (\ref{condition}) confirms the restriction
of the scalar $\sigma$-models found in the previous section from
the geometrical analysis, while the second one kills the vector
superpartners of
the erased gravitinos at the full interaction level. \\
Then what is left, equation (\ref{surv}), is the correct
transformation law for the survived gravitini, provided
$T_{ab}=-\frac{\rm i}{2} ({\bar f}^{\ -1})_{ab\L\S} F^{\L\S}$ (and
$T_{ij}=-\frac{\rm i}{2} ({\bar f}^{\ -1})_{ij\L\S} F^{\L\S}$ for
$N'=6$) describe the correct expression for the (dressed)
graviphotons in the reduced theory, $T_{ab} = -\frac{\rm i}{2}
({\bar f}^{\ -1})_{ab{\bf \L}} F^{{\bf \L}}$, with ${\bf \L}$
running on the appropriate representation of the U duality group
of the reduced theory\footnote{With abuse of language, we call U
duality group the continuous group whose restriction to the
integers is the U duality group of the theory.
}. \\
 To this aim, let us first recall that, in all
$N$-extended theories, the electric and magnetic field-strengths
transform in a representation of the U duality group whose
dimension is the same as the fundamental representation  of the
embedding symplectic group $Sp(2n_v)$ \cite{gz} ($n_v$ is the
total number of vectors). Let us consider separately the cases
$N'=5,6$, where all the vectors are graviphotons, from the $N'
\leq 4$ cases, where matter vectors are present.
\\
In the former cases, note that $E_{7(7)}$ (the isometry group of
$N=8$ theory) contains, as maximal subgroups: $ SO^*(12) \times
SU(2)$ and $SU(5,1) \times SU(3)$. The duality groups for the
$N'=6,5$ are $SO^*(12)$ and $SU(5,1)$ respectively. The rep ${\bf
56}$, in which the $N=8$ vectors field strengths and their duals
lie, decomposes respectively as follows (see also Table
\ref{dual4}):
\begin{eqnarray}
E_{7(7)} \to SO^*(12) \times SU(2) &&{\bf 56} \to ({\bf 32,1}) +
({\bf 12,2}) \label{e76}\\
E_{7(7)} \to SU(5,1) \times SU(3) &&{\bf 56} \to ({\bf 20,1}) +
({\bf 6,3}) + ({\bf\bar 6 , \bar 3} )\label{e75}
\end{eqnarray}
\begin{table}[h]
  \centering
  \caption{Duality reduction in $D=4$ }\label{dual4}
  \begin{tabular}{|c|c|c|}
\hline $G=E_{7(7)} $ & $G_1 \times G_2$ & $G \to G_1 \times G_2 $
\\ \hline
$N=6$ (\# vect. $=16$) & $SO^*(12) \times SU(2)$ & $ 56
\longrightarrow (32 , 1) + (12,2)$ \\\hline $N=5$ (\# vect. $=10$)
& $SU(5,1) \times SU(3)$&$ 56 \to (10 , 1) + ({\bar {10}},1) + (6
, 3) + ({\bar
{6}},{\bar 3})$\\
\hline $N=4$ (\# vect. $=12$) & $SO(6,6) \times SU(1,1)$ & $ 56
\longrightarrow (12 , 2) +  (32 , 1) $
\\\hline
$N=3$ (\# vect. $=4$) & $SU(3,4) \times U(1)$ & $ 56
\longrightarrow 21 + \bar{21} + 7 + \bar 7 $
\\\hline
$N=2$ (\# vect. $= n_V +1$) &  &
\\
$n_V =0$ & $E_{6(2)} \times U(1)$ & $ 56 \to 1 + 1' + 27 + {\bar{
27}} $ \\
$n_V = 15$ & $SO^*(12) \times SU(2)$ & $ 56 \to (32 , 1) + (12,2)$ \\
$n_V = 9$ & $SU(3,3) \times SU(2,1)$ & $ 56 \to (20 , 1) + (6,3) + ({\bar 6} , {\bar 3})$ \\
$n_V = 6$ & $Sp(6,\IR ) \times G_{2(2)}$ & $ 56 \to (1 , 14) + (6,7) $ \\
$n_V = 2$ & $SU(1,1) \times F_{4(4)}$ & $ 56 \to (4 , 1) + (2,26)$
\\\hline
  \end{tabular}
\end{table}
We note that in each case only a subset of the 56 field-strengths
is transformed only with respect to the (reduced theory) duality
group, while it is a singlet of the $SU(8-N')$ commuting group,
and this immediately identifies the electric and magnetic field
strengths which remain in the gravitational multiplet after
truncation. (Indeed this exactly reproduces the counting at the
linearized level, since we expect to have, in the gravitational
multiplet of the $N'=6$ (respectively $N'=5$) theory, 16
(respectively 10) electric field strengths parametrized by
$T_{ab}, T_{ij}$ (respectively by $T_{ab}$).
\\
 Therefore, in performing the truncation, we also
have to decompose the representations of the $N=8$ U duality group
with respect to its maximal subgroups as in (\ref{e76}),
(\ref{e75}), and to keep only the irrepses, in the decomposition,
which are
singlets under the commuting group, as shown in Table \ref{dual4}.\\
Note that this prescription automatically guarantees the
consistency of the truncation, since the objects to be truncated
out (in particular the $(12,2)$  (respectively $({\bf 6,3}) +
(\bf{\bar 6 , \bar 3 })$) field strengths given by $T_{ai}$ and
their magnetic duals), being in a non trivial representation of
the commuting group $SU(8-N')$, can never mix with those which
have been kept, which are instead singlets.
\\
Let us now consider the matter coupled theories, and in particular
$N'=4$ (the $N'=2$ case is similar). Here the argument is reversed
with respect to the higher $N'$ theories, but with analogous
conclusions. Indeed, the U duality group for the $N'=4$ theory
 is $SU(1,1) \times SO(6,n)$, and, for $n=6$,
  it is indeed a maximal subgroup of the $N=8$
 U duality group, (no commuting subgroup).
 Note that the $U$-duality group is now factorized
into the S-duality group $SU(1,1)$, which mixes electric with
magnetic field strengths, and the electric T-duality group
$SO(6,6)$. We have, for the decomposition of the ${\bf 56}$ of
$E_{7(7)}\to SU(1,1) \times SO(6,6)$:
\begin{equation}
{\bf 56} \to ({\bf 2},{\bf 12}) + ({\bf 1},{\bf 32})
\end{equation}
In this case it is the $({\bf 2},{\bf 12})$ field strengths (given
by the six graviphotons $T_{ab}$ and the six matter vectors
$T_{ij}$, together with their magnetic counterpart) which have to
be retained, since they have the appropriate transformation
property under the U duality group, while the extra 32
field-strengths (given by $T_{ai}$ and its magnetic dual), which
are spinors under $SO(6,6)$, have to be truncated out and do
indeed belong to the extra gravitini multiplets. A similar
argument as given previously still works for the consistency;
indeed the field-strengths in the $({\bf 1}, {\bf 32})$, spinors
under $SO(6,6)$, can be set to zero consistently since they cannot
mix with the other field-strengths which are not in the spinor
representation of $SO^*(12)$.
 As far as the  transformation laws for the vectors, scalars and spin one half fields
 are concerned, one sees
 that the decomposition confirms the results of the analysis at the linearized
 level given in section 3, as summarized in
Table \ref{mult}.
\begin{table}[h]
  \centering
  \caption{Decomposition of $N=8$ into $N=N'$ supergravity multiplet}\label{mult}
  \begin{tabular}{|c|c|c|c|}
\hline
    $N'$ & multiplet & max spin & multiplicity \\\hline
    8 & $(g_{\mu\nu},\psi_{A\mu} T_{AB|\mu\nu}, \chi_{ABC}, P_{ABCD})$& 2
   & 1\\ \hline
    6 & $(g_{\mu\nu},\psi_{a\mu} T_{ab|\mu\nu},T_{ij|\mu\nu}, \chi_{abc}, \chi_{aij},
    P_{abcd}, P_{abij})$ &2 & 1\\
     & $(\psi_{i\mu} T_{ai|\mu\nu}, \chi_{abi},
    P_{abci})$  & $\frac 32$ & 2
    \\\hline
    5 & $(g_{\mu\nu},\psi_{a\mu} T_{ab|\mu\nu}, \chi_{abc}, \chi_{ijk},
    P_{abcd}, P_{aijk})$ &2 & 1\\
& $(\psi_{i\mu} T_{ai|\mu\nu}, T_{ij|\mu\nu },
\chi_{abi},\chi_{aij},
    P_{abci}, P_{abij})$  & $\frac 32$ & 3
    \\\hline
    4 & $(g_{\mu\nu},\psi_{a\mu} T_{ab|\mu\nu}, \chi_{abc},
    P_{abcd}, P_{ijk\ell })$ &2 & 1\\
& $(\psi_{i\mu} T_{ai|\mu\nu}, \chi_{abi},\chi_{ijk},
    P_{abci}, P_{aijk})$  & $\frac 32$ & 4\\
     & $(T_{ij|\mu\nu }, \chi_{aij}, P_{abij})$ & 1& 6 \\\hline
     3 & $(g_{\mu\nu},\psi_{a\mu} T_{ab|\mu\nu}, \chi_{abc})$ &2 & 1\\
& $(\psi_{i\mu} T_{ai|\mu\nu}, \chi_{abi},
    P_{abci}, P_{ijk\ell })$  & $\frac 32$& 5\\
     & $(T_{ij|\mu\nu }, \chi_{aij}, P_{abij}, P_{aijk})$ & 1 & 10\\\hline
     2 & $(g_{\mu\nu},\psi_{a\mu} T_{ab|\mu\nu})$ &2 & 1\\
& $(\psi_{i\mu} T_{ai|\mu\nu}, \chi_{abi})$  & $\frac 32$& 6\\
     & $(T_{ij|\mu\nu }, \chi_{aij}, P_{abij})$ & 1 & 15\\
     & $(\chi_{ijk} , P_{aijk})$& $\half $& 10\\\hline
    1 & $(g_{\mu\nu},\psi_{a\mu} )$ &2 & 1\\
& $(\psi_{i\mu} T_{ai|\mu\nu})$  & $\frac 32$& 7\\
     & $(T_{ij|\mu\nu }, \chi_{aij})$ & 1 & 21\\
     & $(\chi_{ijk} , P_{aijk})$& $\half $& 35\\\hline
  \end{tabular}
\end{table}
\\
For the case $N=8 \to N=2$, we see from Table \ref{dual4} that the
vectors belonging to the six spin $\frac 32$ multiplets and to
those vector multiplets which are truncated out are tied together
by an irrep.  of  $G_1 \times G_2$. This means that to delete only
the spin $\frac 32$ multiplets would be inconsistent.
\par
 The same analysis applies to theories
in higher dimensions and, for the $D=5$ case, the duality
reduction, for some interseting cases, is given in Table
\ref{dual5}.

\begin{table}[h]
  \centering
  \caption{Duality reduction in $D=5$ }\label{dual5}
  \begin{tabular}{|c|c|c|}
\hline $G=E_{6(6)} $ & $G_1 \times G_2$ & $G \to G_1 \times G_2 $
\\ \hline
$N=2$ (\# vect. $= n_V +1$) &  &
\\
$n_V =0$ & $F_{4(4)}$ & $ 27 \to 1 + 26 $ \\
$n_V = 14$ & $SU^*(6) \times SU(2)$ & $ 27 \to (15 , 1) + (6,2)$ \\
$n_V = 8$ & $SL(3,\IC) \times SU(2,1)$ & $ 27 \to (3,3' , 1) + (1,3,3') + (3',1,3)$ \\
$n_V = 5$ & $SL(3,\IR ) \times G_{2(2)}$ & $ 27 \to (6 , 1) + (3,7) $ \\
\hline
  \end{tabular}
\end{table}

\section{$N=2\longrightarrow N=1$ reduction}
This section is devoted to a thorough analysis of the consistent
truncation of $N=2$ supergravity down to $N=1$ in four dimensions.
 The $N=2\longrightarrow N=1$ reduction of the supersymmetry
transformation laws presents different features in the vector
multiplet and in the hypermultiplet sectors. The vector multiplet
case is simpler since the special geometry is already a
K\"{a}hler-Hodge geometry while for hypermultiplets we are
confronted with the more difficult task of reducing a quaternionic
manifold to a K\"{a}hler-Hodge one.\\
Note that, differently from what done in the preceeding sections,
where we discussed only ungauged theories, the present reduction
is given at the level of the complete $N=2$ gauged theory.
\par
 In the first two subsections we begin to analyze the
reduction in the vector multiplet sector, where much of the
special geometry relations are needed. In  subsection 6.3 we
analyze the reduction in the hypermultiplet sector. In both cases
the geometrical approach discussed in section 3 will be essential
for the discussion. The other subsections are devoted to a careful
analysis of the implications of the gauging, to the reduction of
the scalar potential and to the discussion of some explicit
examples.
\par
The reduction is obtained by truncating the spin $3/2$ multiplet
containing the second gravitino $\psi_{\mu 2}$ and the
graviphoton.
\par
Here and in the following we use the notations both for $N=2$ and
$N=1$ supergravity as given in reference \cite{df}, the only
differences being that we use here world indices ${\mathcal
I},\bar{\mathcal I} =1,\cdots ,n_V$ and boldfaced gauge indices
${\bf \L}= 0,1,\cdots ,n_V$ for quantities in the $N=2$ vector
multiplets (since we want to reserve the notation $\L$ and
 $i,\bar \imath$ for the indices of the reduced $N=1$ theory)
 and that the holomorphic matrix appearing in the kinetic
term of the vectors in the $N=1$ theory will be renamed as
follows:
\begin{equation}\label{rename}
  \bar\cN_{ \L\S}(z^{i})\equiv  f_{ \L \S}
  (z^{ i}).
\end{equation}
\par
 Let us write down the supersymmetry
transformation laws of the $N=2$ theory, up to 3-fermions terms
\cite{abcdffm}:
 \vskip 0.5cm
 {\bf Supergravity transformation rules of
the (left--handed) Fermi  fields}:
\begin{eqnarray}
\delta\,\psi _{A \mu} &=& {\hat{\nabla}}_{\mu}\,\epsilon _A\,
 + \left ( {\rm i} \, g \,S_{AB}\eta _{\mu \nu}+
\epsilon_{AB} T^-_{\mu \nu} \right) \gamma^{\nu}\epsilon^B
 \label{gravtrasf} \\
\delta \,\lambda^{{\mathcal I}A}&=&
 {\rm i}\,\nabla _ {\mu}\, z^{{\mathcal I}}
\gamma^{\mu} \epsilon^A +G^{-{\mathcal I}}_{\mu \nu} \gamma^{\mu
\nu} \epsilon _B \epsilon^{AB}\,+\, gW^{{\mathcal I}AB}\epsilon _B
\label{gaugintrasf}\\
 \delta\,\zeta _{\alpha}&=&{\rm i}\,
{\mathcal U}^{B \beta}_{u}\, \nabla _{\mu}\,q^u \,\gamma^{\mu}
\epsilon^A \epsilon _{AB}\,C_{\alpha  \beta} \,+\,g
N_{\alpha}^A\,\epsilon _A \label{iperintrasf}
\end{eqnarray}
where:
\begin{equation}
\hat{\nabla}_{\mu}\,\epsilon _A = {\mathcal D}_\mu \epsilon_A
 + \hat{\omega}_{\mu |
A}^{\phantom{\mu |A}B} \epsilon_B +\hat{\cal Q}_\mu \epsilon_A
\end{equation}
 and the $SU(2)$ and $U(1)$ 1-form ``gauged'' connections are
respectively given by:
\begin{eqnarray}
 \hat{\omega}_{A}^{\phantom{A}B}&=&{\omega}_{A}^{\phantom{A}B} +
g_{({\bf \L})} \, A^{{\bf \L}} \,P^x_{{\bf \L}}\,
(\s^x)_{A}^{\phantom{A}B}\,,\label{oab}\\
\hat{\cal Q} &=& {\cal Q} + g_{({\bf \L})}\, A^{{\bf \L}}
\,P^0_{{\bf \L}} \,,\label{qu}\\
{\cal Q} &=& -\frac{\rm i}2 \left(\partial_{{\mathcal I}} \cK d
z^{{\mathcal I}} - \partial_{\bar{\mathcal I}} \cK d \bar
z^{\bar{\mathcal I}}\right)\label{qu2}
\end{eqnarray}
 ${\omega}_{A}^{\phantom{A}B}$, ${\cal Q}$ are the
 $SU(2)$ and $U(1)$ connections of the ungauged
theory. Moreover we have:
\begin{eqnarray}
\nabla_\mu z^{{\mathcal I}} &=& \partial_\mu z^{{\mathcal I}} +
g_{({\bf \L}
)} A^{{\bf \L}}_\mu k^{{\mathcal I}}_{{\bf \L}}\\
\nabla_\mu q^{u} &=& \partial_\mu q^{u} + g_{({\bf \L})} A^{{\bf
\L}}_\mu k^{u}_{{\bf \L}}
\end{eqnarray}
 \vskip 0.5cm {\bf Supergravity transformation rules of the
Bose fields}:
\begin{eqnarray}
\delta\,V^a_{\mu}&=& -{\rm i}\,\bar {\psi}_{A
\mu}\,\gamma^a\,\epsilon^A -{\rm i}\,\bar {\psi}^A _
\mu\,\gamma^a\,\epsilon_A\\
\delta \,A^{{\bf \L}} _{\mu}&=& 2 \bar L^{{\bf \L}} \bar \psi
_{A\mu} \epsilon _B \epsilon^{AB}\,+\,2L^{{\bf \L}
}\bar\psi^A_{\mu}\epsilon^B \epsilon
_{AB}\nonumber\\
&+&{\rm i} \,f^{{\bf \L}}_{{\mathcal I} }\,\bar
{\lambda}^{{\mathcal I}A} \gamma _{\mu} \epsilon^B \,\epsilon
_{AB} +{\rm i} \, {\bar f}^{{\bf \L}}_{\bar{\mathcal I}}
\,\bar\lambda^{\bar{\mathcal I}}_A
\gamma _{\mu} \epsilon_B \,\epsilon^{AB} \label{gaugtrasf}\\
\delta\,z^{{\mathcal I}} &=& \bar{\lambda}^{{\mathcal I}A}\epsilon _A \label{ztrasf}\\
\delta\,z^{\bar{\mathcal I}}&=& \bar{\lambda}^{\bar {\mathcal
I}}_A \epsilon^A
\label{ztrasfb}\\
  \delta\,q^u &=& {\mathcal U}^u_{\alpha A} \left(\bar {\zeta}^{\alpha}
  \epsilon^A + C^{\alpha  \beta}\epsilon^{AB}\bar {\zeta}_{\beta}
  \epsilon _B \right). \label{quatertrasf}
 \end{eqnarray}
  Here $T^-_{\mu\nu}$ appearing in the supersymmetry
transformation law of the $N=2$ left-handed gravitini is the
``dressed'' graviphoton defined as:
\begin{equation}\label{gravif}
T^-_{\mu\nu} \equiv  2{\rm i} {\rm Im} {\cal N}_{{\bf \L} {\bf
\S}} L^{{\bf \S}} F_{\mu\nu}^{{\bf \L} -}.
\end{equation}
while
\begin{equation}\label{vectors}
G^{{\mathcal I}-}_{\mu\nu} = - g^{{\mathcal I}\bar{\mathcal J}}
 \bar f^{{\bf \G} }_{\bar{\mathcal J}}
{\rm Im } {\cal N}_{{\bf \G} {\bf \L}}
  {F}^{{\bf \L} -}_{\mu\nu}
\end{equation}
are the ``dressed'' field strengths of the vectors inside the
vector multiplets. Moreover the fermionic shifts $S_{AB}$, $
W^{{\mathcal I}\,AB}$ and $ N^A_{\alpha}$ are given in terms of
the prepotentials and Killing vectors of the quaternionic
geometry (suitably dressed with special geometry data) and of the
special geometry Killing vectors, as follows:
\begin{eqnarray}\label{trapsi2}
S_{AB}&=& {\rm i} \frac {1}{2} P_{AB\, {\bf \L}} \,
 L^{{\bf \L}}
 \equiv {\rm i} \frac {1}{2} P^x_{{\bf \L}} \sigma^x_{AB}L^{{\bf \L}} \\
\label{tralam}
 W^{{\mathcal I}\,AB}&=& {\rm {i}} P^{AB}_{
{\bf \L}}\,g^{{\mathcal I}\bar{\mathcal J}} f^{{\bf \L}
}_{\bar{\mathcal J}} + \epsilon^{AB} k^{{\mathcal I}}_{{\bf \L}
}{\overline L}^{{\bf \L}}
\\
N^A_{\alpha}&=& 2\,{\mathcal U}^A_{\alpha u} \,k^u_{{\bf \L}}
{\overline L}^{{\bf \L}} \label{traqua1}\\
N^{\alpha}_A&=& -2 \,{\mathcal U}_{A u}^{\alpha} \,k^u_{{\bf \L}
} L^{{\bf \L}} \label{traqua2}
\end{eqnarray}
We recall that the Killing vectors $k^{{\mathcal I}}_{{\bf \L}}$
and $k^u_{{\bf \L}}$ are related to the prepotentials by:
\begin{eqnarray}
k^{{\mathcal I}}_{{\bf \L}}&=&{\rm i} g^{{\mathcal I}\bar{\mathcal J}}\partial_{\bar{\mathcal J}}P^0_{{\bf \L}}\\
k^u_{{\bf \L}}&=&\frac{1}{6 \l^2} \Omega^{x |
vu}\nabla_{v}P^x_{{\bf \L}} \,; \quad \l =-1
\end{eqnarray}
where $\Omega^{x}_{uv}$ is the $SU(2)$-valued 2-form defined in
Section (6.3) below, and that the prepotential $P^0_{{\bf \L}}$
satisfies:
\begin{equation}
  P^0_{{\bf \L}} L^{{\bf \L}}\,=\, P^0_{{\bf \L}}{\bar L}^{{\bf \L}}\,=\,0
\end{equation}
\par
Since we are going to compare the $N=2$ reduced theory with the
standard $N=1$ supergravity, we also quote the supersymmetry
transformation laws of the latter theory \cite{bw2},\cite{cfgv}.
We have, up to 3-fermions terms:
 \\
{\bf $N=1$ transformation laws}
\begin{eqnarray}
\label{trapsi1}\delta \psi_{\bullet \mu} &=& {\cal D}_{\mu}
\epsilon_{\bullet}+ \hat Q_\mu\epsilon_{\bullet}
+{\rm {i}} L(z, \bar z) \gamma_{\mu} \varepsilon^{\bullet} \\
\label{trachi1} \delta \chi^i &=& {\rm {i}}
\left(\partial_{\mu}z^i + g_{(\L )} A^\L_\mu k^i_\L \right)
\gamma^{\mu} \varepsilon_{\bullet}  + N^{i}\varepsilon_{\bullet}\\
\label{tralamb1}\delta \lambda^{\L }_{\bullet} &=& {\mathcal
{F}}_{\mu \nu}^{(-) \L } \gamma^{\mu \nu} \varepsilon_{\bullet }
+{\rm {i}} D^{\L } \varepsilon_{\bullet} \\
\label{traviel1}\delta V^a_{\mu} &=& -\rm {i}
\psi_{\bullet} \gamma_{\mu} \varepsilon^{\bullet} + h.c.\\
\label{travec1}\delta A^{\L }_{\mu} &=& \rm{i}\frac {1}{2} \bar
\lambda^{\L }_{\bullet} \gamma_{\mu}
\varepsilon^{\bullet} + h.c.\\
 \label{trasca1}\delta z^i  &=& \bar \chi^i
\varepsilon_{\bullet}
\end{eqnarray}
where $\hat \cQ$ is defined in a way analogous to the $N=2$
definition (\ref{qu}) and:
\begin{eqnarray}\label{n1def}
L(z,\bar z)&=& W(z)\,e^{\frac {1}{2} {\mathcal {K}_V(z, \bar z)}} \\
\label{defn} N^i &=& 2\, g^{i\bar\jmath} \,\nabla_{\bar\jmath}\, \bar L \\
 \label{dlambda} D^{\L } &=& - 2 ({\rm {Im}} f_{\L
\S })^{-1} P_{\S }(z,\bar z)
\end{eqnarray}
and $W(z),{\mathcal {K}}_{(1)}(z, \bar z) ,P_{\S }(z,\bar z),
f_{\L\S}(z)$ are the superpotential,  K\"{a}hler potential,
Killing prepotential and  vector kinetic matrix respectively
\cite{cfgv}, \cite{bw2}, \cite{bagger}. Note that for the
gravitino and gaugino fields we have denoted by a lower (upper)
dot left-handed (right-handed) chirality. For the spinors of the
chiral multiplets $\chi$, instead, left-handed (right-handed)
chirality is encoded via an
 upper holomorphic (antiholomorphic) world index
 ($\chi^i, \chi^{\bar\imath}$).\\
 The supersymmetric lagrangians which are left
invariant by these
 transformation laws are given in Appendix G.
\par
 To perform the truncation
we set $A$=1 and 2 successively, putting $\psi_{2\mu}
=\epsilon_2=0$, and we get from equation  (\ref{gravtrasf}):
\begin{equation}\label{reductio}
\delta\,\psi _{1 \mu} = {\mathcal D}_\mu \epsilon_1  -\hat{\cal
Q}_\mu \epsilon_1 - \hat{\omega}_{\mu | 1}^{\phantom{\mu |A}1}
\epsilon_1  \,
 +  {\rm i} \, g \,S_{11}\eta _{\mu \nu}
 \gamma^{\nu}\epsilon^1
\end{equation}
where ${\mathcal D}$ denotes the Lorentz covariant derivative (on
the spinors, ${\mathcal D}_\mu = \partial_\mu -\frac 14
\omega_\mu^{ab}\g_{ab}$), while, for consistency:
\begin{equation}
\delta\,\psi _{2 \mu}\equiv 0 =  - \hat{\omega}_{\mu |
2}^{\phantom{\mu |A}1} \epsilon_1 +
  \left ( {\rm i} \, g \,S_{21}\eta _{\mu \nu}-T^-_{\mu\nu}\right)
 \gamma^{\nu}\epsilon^1
\end{equation}
\par
 For a consistent truncation in the ungauged case we must
set to zero the graviphoton:
\begin{equation} T^-=
T_{{\bf \L}}  F^{-{\bf \L}} = 0 \label{gravifot},
\end{equation}
where
\begin{equation}
T_{{\bf \S}}\equiv 2 {\rm i}{\rm Im} {\cN}_{{\bf \L}{\bf \S}}
L^{{\bf \L}} \label{tlambda}
\end{equation}  is the projector on the graviphoton \cite{cdf},
and the component $\omega_1^{\ 2}$ of the $SU(2)$ connection
1-form:
\begin{equation}\label{ounodue}
  \omega _1^2 = 0
\end{equation}
 In the gauged case we have the further constraints:
\begin{eqnarray} S_{21} &=& \frac {\rm i}2 P^x_{{\bf \L}}
(\s^x)_{12} L^{{\bf \L}} = \frac {\rm i}2 P^3_{{\bf \L}}  L^{{\bf \L}} = 0,\label{first}\\
\hat{\omega}_1^{\phantom{2}2}&=& \omega_1^{\phantom{2}2}
+g_{({\bf \L})}A^{{\bf \L}} P^x_{{\bf \L}}
(\sigma^x)_1^{\phantom{A}2} \equiv g_{({\bf \L})}A^{{\bf \L}}
P^x_{{\bf \L}} (\sigma^x)_1^{\phantom{A}2}= 0\label{second}.
\end{eqnarray}
while no further restriction comes from (\ref{qu}) since the form
of the gauged $U(1)$ connection should not change in the reduced
theory.

 Comparing (\ref{trapsi1}) with (\ref{reductio}), we learn
that we must identify:
\begin{eqnarray}
\psi_{1\mu} &\equiv & \psi_{\bullet \mu} \\
\epsilon_1&\equiv & \epsilon_{\bullet }
\end{eqnarray}
Furthermore,  $g\,S_{11}= \frac{\rm i}{2} g_{({\bf \L})}
P^x_{{\bf \L}} (\s^x)_{11} L^{{\bf \L}}$ must be identified with
the superpotential of the $N=1$ theory, that is to the
covariantly holomorphic section $L$ of the $N=1$ K\"{a}hler-Hodge
manifold. Therefore we have \cite{ps} - \cite{tatar}:
\begin{equation}\label{sup}
  L(q,z,\bar z)=\frac{\rm i}{2} g_{({\bf \L})} P^x_{{\bf \L}}
(\s^x)_{11} L^{{\bf \L}} = \frac{\rm i}{2} g_{({\bf \L})} \left(
P^1_{{\bf \L}} - {\rm i} P^2_{{\bf \L}} \right) L^{{\bf \L}}
\end{equation}
We will show in the following (Section (6.4)) that, after
consistent reduction of the special-K\"ahler manifold $\cM^{SK}$
and of the quaternionic $\s$-model $\cM^Q$, $L$ will in fact
 be a covariantly holomorphic function of
 the K\"ahler coordinates $w^s$ of the reduced manifold
${\cal M}^{KH} \subset {\cal M}^Q$ and of some subset $z^i \in
\cM_R$ of the scalars $z^{{\mathcal I}}$ of the $N=2$
special-K\"ahler manifold $\cM^{SK}$.
\par
The condition on the graviphoton $T^- =0$ will be analyzed in
subsection (6.1), while the condition $\omega_1^{\ 2}  =0$ will be
discussed in section (6.3) and the  constraints appearing in the
gauged theory will be analyzed in section (6.4).
\par
 Here and
in the following we will denote by $\cM^{{SK}}$ and $\cM^Q$ the
special-K\"ahler and quaternionic manifolds of the $N=2$ theory
while the special-K\"ahler and K\"ahler-Hodge manifolds obtained
by reduction of $\cM^{{SK}}$ and $\cM^Q$ will be denoted by
$\cM_R$ and $\cM^{KH}$ respectively.

\subsection{Reduction of the $N=2$ vector multiplet sector}
 Let us now consider the gaugino transformation
laws. When $\epsilon_2 =0$ we get:
\begin{eqnarray}
\delta \,\lambda^{{{\mathcal I}}1}&=&
 {\rm i}\,\nabla _ {\mu}\, z^{{\mathcal I}}
\gamma^{\mu} \epsilon^1 +\, W^{{{\mathcal I}}11}\epsilon _1
\label{chiral} \\
\delta \,\lambda^{{{\mathcal I}}2}&=&
 - G^{-{{\mathcal I}}}_{\mu \nu} \gamma^{\mu \nu}
\epsilon _1\,+\, g \, W^{{{\mathcal I}}21}\epsilon _1
\label{gaugin}
\end{eqnarray}
where, using (\ref{tralam})
\begin{eqnarray}
W^{{{\mathcal I}}21}&=&{\rm {i}} P^3_{ {\bf \L}}\,g^{{\mathcal
I}\bar{\mathcal J}} f^{{\bf \L}}_{\bar{\mathcal J}} -k^{{\mathcal
I}}_{{\bf \L}} \bar L^{{\bf \L}}
 \label{w12}\\
W^{{{\mathcal I}}11} &=& {\rm i} P^{11}_{{\bf \L}} g^{{\mathcal
I}\bar{\mathcal J}}f^{{\bf \L}}_{\bar{\mathcal J}} = \left(
P^2_{{\bf \L} }  - {\rm i}P^1_{{\bf \L}} \right) g^{{{\mathcal
I}}{\bar{\mathcal J}}}f^{{\bf \L}}_{\bar{\mathcal J}}\label{w11}
\end{eqnarray}
From eqs. (\ref{chiral}) and (\ref{gaugin}) we immediately see
that the spinors $\l^{{\mathcal I} 1}$ transform into the scalars
$z^{{\mathcal I}}$ (and should therefore give rise to $N=1$ chiral
multiplets) while the spinors $\l^{{\mathcal I} 2}$ transform
into the matter vectors field strengths $G^{-{{\mathcal I}}}_{\mu
\nu}$ (and should then be identified with the gauginos of the
$N=1$ vector multiplets).
\par
However,
before entering the details of the identification,
 we have to discuss the implications of putting to zero
 the graviphoton $T^-$,
 equation
(\ref{gravifot}). We observe that this condition gives a
constraint on the scalar and vector content of the $N=1$ reduced
theory, that is on the number of chiral and vector multiplets
which are retained after truncation.

Now, since the graviphoton projector $T_{{\bf \L}}$
(\ref{tlambda}) is a scalar field dependent quantity, the request
that  equation  (\ref{gravifot}) is verified all over the manifold
can be trivially realized either by setting to zero all the
scalars $z^{{\mathcal I}}$ and the graviphoton $A^0_\mu$, which
implies on the symplectic section $L^{{\bf \L}} \Rightarrow (L^0
=1 ; L^\L = 0, \L=1,\cdots n_V )$, or, alternatively, by
truncating out all the vectors $A^{{\bf \L}}$, leaving an $N=1$
theory with only chiral matter content.

 There is however a more interesting and non trivial
way to satisfy equation  (\ref{gravifot}), by imposing a suitable
constraint on the set of vectors and of scalar sections which can
be retained in the reduction. Indeed, if we decompose the index
${\bf \L}$ labelling the vectors into two disjoint sets ${\bf \L}
\Rightarrow ( \L , X ), \L =1,\cdots ,n_V'=n_V -n_C;X =
0,1,\cdots ,n_C $, we may satisfy the relation (\ref{gravifot})
as an ``orthogonality relation'' between the subset $ \L$ running
on the retained vectors and the subset $X$ running on the
retained scalar sections. That is we set:
\begin{eqnarray}
F^{X}_{\mu\nu} &=& 0 ;\\
{\rm Im} \cN_{\L{\bf \S}} L^{{\bf \S}} &=& T_\L=0 \label{nl}
\end{eqnarray}
We note that if we delete the electric field strengths $F^{-X}$
we must also delete their magnetic counterpart
\begin{equation}
\cG^-_{X}= \bar \cN_{X Y} F^{-Y} + \bar \cN_{X   \S} F^{- \S}=0
\end{equation}
so that we must also impose
\begin{equation}
  \cN_{X  \S} =0.
\end{equation}
Then, the constraint (\ref{nl}) reduces to
\begin{equation}
{\rm Im} \cN_{ \L \S} L^{ \S} =0\label{nl1}
\end{equation}
which implies
\begin{equation}
 L^{ \S} =0\label{scalred1}
\end{equation}
since the vector-kinetic matrix ${\rm Im} \cN_{ \L \S}$ has to be
invertible.
\\
Note that conditions (\ref{nl1}) and (\ref{scalred1}) imply a
reduction of the $N=2$ scalar manifold $\cM^{SK} \to \cM_R$, since
it says that some coordinate dependent sections on $\cM^{SK}$ have
to be zero in the reduced theory.

Let us decompose the world indices ${{\mathcal I}}$ of the $N=2$
special-K\"ahler $\s$-model as follows: ${{\mathcal I}}
\Rightarrow ( i , \a )$, with $ i =1,\cdots , n_C$, $\a =1,\cdots
, n'_V=n_V-n_C$, where $n_C$ and $n'_V$ are respectively the
number of chiral and vector multiplets of the reduced $N=1$
theory while $n_V$ is the number of $N=2$ vector multiplets.
\\
Then from eq (\ref{dl0}) it follows that the metric on $\cM_R $ is
pulled back to the following form \cite{abcdffm}, \cite{cdf}:
\begin{equation}
g_{  i  {\bar\jmath}} =-2 f^{X}_{  i} {\rm Im}\cN_{X Y}\bar
f^{Y}_{ {\bar\jmath} }
\end{equation}
To examine further the implications of the reduction of the
special-K\"ahler manifold to the submanifold $\cM_R $, it is
convenient to write the special geometry objects using flat
indices. We then define a set of K\"ahlerian vielbeins $P^{\hat
I}= P^{\hat I}_{{\mathcal I}}dz^{{\mathcal I}}$ on $\cM^{SK}$
together with their complex conjugates. Performing the reduction,
they decompose as: $P^{\hat I} \Rightarrow (P^I, P^A )$, where
$I$ and $A$ are flat indices in the submanifold $\cM_R $ and on
its orthogonal complement respectively. By an appropriate choice
of coordinates, we call $z^i$ the coordinates on $\cM_R $, $z^\a$
the coordinates on the orthogonal complement. Then we may set
$P^I_\a =0, P^A_i=0$, so that the metric $g_{{\mathcal I}
\bar{\mathcal J}} = P^{\hat I}_{{\mathcal I}} \bar P^{\hat
I}_{\bar{\mathcal J}}$ has only components $g_{i\bar\jmath},
g_{\a\bar\b}$, while $g_{i\bar\a}=0$.
\par
Then, if we decompose the gauginos $\l^{{\mathcal I} 2}
\Rightarrow (\l^{ i 2}, \l^{\a 2})$, the above truncation
implies, by supersymmetry, $\l^{ i 2}=0$ and, for consistency,
\begin{equation}\label{gau2}
\delta \,\lambda^{{ i}2}=
 - G^{-{ i}}_{\mu \nu} \gamma^{\mu \nu}
\epsilon _1\,+\, g \, W^{{ i}21}\epsilon _1 =0.
\end{equation}
Setting  $G^{-  i}_{\mu\nu}=0$ gives:
\begin{equation}\label{mattervectors}
  G^{-  i}_{\mu\nu} = -g^{ i  \bar{\mathcal J}} \nabla
  _{ \bar{\mathcal J}}\bar L^{{\bf \L}}  {\rm Im} \cN_{{\bf \L}{\bf \S}}
  F^{-{\bf \S}}_{\mu\nu}
=- g^{ i  {\bar\jmath}} \nabla
  _{ {\bar\jmath}}\bar L^{ \L}  {\rm Im} \cN_{ \L \S}
  F^{- \S}_{\mu\nu}=0
\end{equation}
implying
\begin{equation}
\nabla
_{ {\bar\jmath}}\bar L^{ \L} =\bar f
  _{ {\bar\jmath}}^{ \L}=0.
  \label{dl0}
\end{equation}
Moreover, $W^{{ i}21}=0$ implies:
\begin{equation}
P^3_{X} =0 \, , \quad k^i_{X} =0.
\end{equation}
\par
Note that the integrability condition of equation  (\ref{dl0}) is:
\begin{equation}\label{intdl0}
  \nabla_i \nabla_j L^{ \L} = {\rm i} C_{ij{\mathcal K}} g^{{\mathcal K}
  \bar{\mathcal K}}\nabla_{\bar{\mathcal K}} \bar L^{ \L} = {\rm i} C_{ij  k}
  g^{  k{\bar k}}\nabla_{ {\bar k}} \bar L^{ \L}+ {\rm i} C_{ij\a}
   g^{\a
  {\bar\a}}\nabla_{{\bar a}} \bar L^{ \L}=0.
\end{equation}
where $C_{ijk}$ is the 3-index symmetric tensor appearing in the
equations defining the special geometry (see {\it e.g.} ref.
\cite{cdf},\cite{abcdffm}).\\ Since the first term on the r.h.s.
of equation  (\ref{intdl0}) is zero on $\cM_R $ (equation
(\ref{dl0})), equation  (\ref{intdl0}) is satisfied by imposing:
\begin{equation}\label{ccond1}
C_{ij\a}=0
\end{equation}
so that only the $N=2$ special-K\"ahler manifolds satisfying the
constraint (\ref{ccond1}) are suitable for reduction.
\\
Note that, since $C_{ij\a}$ is defined as a symplectic scalar
product \cite{dlv},\cite{str},\cite{cadf},\cite{abcdffm} in terms
of the symplectic section $U=(L^{{\bf \L}}, M_{{\bf \L}})$:
\begin{equation}
  C_{ij\a} = < \nabla _{i} \nabla _{ j} U , \nabla _{ \a} U >,
\end{equation}
it follows that
\begin{equation}
C_{ij\a} = 0 \Rightarrow \nabla _{ \a} U = 0 \Rightarrow \nabla
_{ \a} L^{X}=0 \, ,\, \nabla _{ \a} M_{X}=0.
\end{equation}
\par
The same constraint (\ref{ccond1}) can also be retrieved by
looking at the integrability conditions of the $N=2$ special
geometry as given in \cite{adf}. The relevant ones for our
discussion are the following:
\begin{eqnarray}
\nabla P^{\hat I} &=& dP^{\hat I} + {\rm i} \cQ \wedge P^{\hat I}
+
\omega^{\hat I}_{\phantom{\hat I}\hat J}\wedge P^{\hat J}=0\label{metpost}\\
R^{\hat {\bar J}}_{\phantom{\hat{\bar J}}\hat{\bar I}} &\equiv &
(d\omega + \omega\wedge\omega )^{\hat {\bar J}}_{\phantom{\hat
J}\hat{\bar I}}=P_{\hat{\bar I}}\wedge \bar P^{\hat {\bar J}}
-{\rm i} K \delta^{\hat {\bar J}}_{\hat {\bar I}} - C^{\hat{\bar
J}} _{\phantom{\hat J}\hat L}\wedge \bar C^{\hat L}
_{\phantom{\hat L}\hat{\bar I}}
\end{eqnarray}
where $\cQ$ is the K\"ahler connection 1-form, $K=d\cQ$ is the
K\"ahler 2-form, $\omega^{\hat I}_{\phantom{\hat I}\hat J}$ is
the $SU(n_V)$-Lie algebra valued connection and the 1-form
$C^{\hat{\bar J}} _{\phantom{\hat J}\hat L}$ can be written in
terms of the 3-world indices symmetric tensor $C_{{\mathcal I}
{\mathcal J} {\mathcal K}}$, whose properties are given in ref.
\cite{adf}, via:
\begin{equation}
C^{\hat{\bar J}} _{\phantom{\hat J}\hat L}= P^{\hat {\bar J}
{\mathcal I}}P_{\hat L}^{\phantom{\hat L} {\mathcal J}}
C_{{\mathcal I} {\mathcal J} {\mathcal K}} dz^{{\mathcal K}}.
\end{equation}
Let us restrict the previous equations to the submanifold $\cM_R
$. From the vanishing of the torsion, eq (\ref{metpost}), we find:
\begin{eqnarray}
\nabla P^{ I} &=& dP^{ I} + {\rm i} \cQ \wedge P^{ I} + \omega^{
I}_{\phantom{\hat I} J} \wedge P^{ J}+ \omega^{
I}_{\phantom{\hat I} A} \wedge P^{A} =0\\
\nabla P^{ A} &=& dP^{ A} + {\rm i} \cQ \wedge P^{ A} + \omega^{
A}_{\phantom{\hat I}  J} \wedge P^{  J} + \omega^{
A}_{\phantom{\hat I} B} \wedge P^{B}=0.
\end{eqnarray}
With the same procedure illustrated in the general discussion of
section 4 and in the example of section (6.1), we easily find
that the vanishing of the torsion on $\cM_R $ implies $\omega^{
I}_{\phantom{\hat I} A}|_{\cM_R }=0$, from which it follows,
taking into account the Frobenius theorem and the definition of
$R^{\hat {\bar J}}_{\phantom{\hat{\bar J}}\hat{\bar I}}$:
\begin{equation}
 R^{  {\bar
J}}_{\phantom{\hat{\bar J}} {\bar A}}|_{\cM_R }=P_{ {\bar
A}}\wedge \bar P^{  {\bar J}} - C^{ {\bar J}} _{\phantom{\hat J}
I}\wedge \bar C^{ I} _{\phantom{\hat L} {\bar A}} - C^{ {\bar J}}
_{\phantom{\hat J} B}\wedge \bar C^{B} _{\phantom{\hat L} {\bar
A}} =0.
\end{equation}
Now, expanding the vielbein and the $C$-tensor along the
differentials of the coordinates, we easily find
\begin{equation}
R^{  {\bar J}}_{\phantom{\hat{\bar J}} {\bar A}}|_{\cM_R }= -
\bar P^{ {\bar J}i} P_{\bar A}^{\a} \left( C_{ijk} \bar C^{j}
_{\phantom{j} \a \ell }+  C_{i\b k} \bar C^{\b} _{\phantom{\b} \a
\ell}\right)dz^k \wedge dz^\ell =0, \label{ccond}
\end{equation}
where we have set to zero the terms in the external directions
$dz^\a$, and the $C$-terms containing both holomorphic and
antiholomorphic indices, which are zero already because of the
$N=2$ special geometry properties \cite{dlv},\cite{abcdffm}.
Again, we see that equation  (\ref{ccond}) is satisfied by
imposing the same condition (\ref{ccond1}) on the special-K\"ahler
manifold.\par From the analysis of the fermionic terms in the
supersymmetry transformation laws of the fermions \cite{corto}, it
is possible to find a further condition on the C-tensor:
\begin{equation}
C_{\a\b\g} |_{\cM_R}=0
\end{equation}
which, together with (\ref{ccond1}), implies
\begin{equation}
R^i_{\ \a\b\g} |_{\cM_R}=0
\end{equation}

\subsection{$N=2$ vector multiplets $\longrightarrow N=1$ matter
multiplets}
 Let us now discuss the precise identification of the $N=1$
matter multiplets obtained by reduction of the $N=2$ vector multiplets.\\
From the above analysis we have found that the indices labelling
$N=1$ chiral and vector multiplets are not related anymore, as it
was instead the case in the $N=2$ theory.
\par
As far as equation  (\ref{chiral}) is concerned, we immediately
see that, after reduction of the index ${\mathcal I}$ and
comparison with the corresponding $N=1$ formula (\ref{trachi1}),
 we can make the following identification:
\begin{eqnarray}\label{comp}
 \lambda^{i1} &=& \chi^i
\label{chi}\\
gW^{{ i}11}& =& N^{ i}={\rm i}g_{(X)} \left(P^1_{X}-{\rm
i}P^2_{X} \right)g^{{
i}{\bar\jmath}}f^{X}_{{\bar\jmath}}\label{shiftN}
\end{eqnarray}
 that is we may interpret the $\lambda^{{ i}1}$ as $n_C$ $N=1$
chiral spinors belonging to $N=1$ left-handed chiral multiplets
$(\chi^{i} , z^{i})$, ${i}=1,\dots,n_C$. It can be easily
verified  that the consistency condition
\begin{equation}
\l^{\a 1} =0 \Rightarrow \delta\l^{\a 1} =0
\end{equation}
gives
\begin{equation}
k^\a_\L =0
\end{equation}
using $f^{X}_\a =0$.
\par
Let us now discuss the $N=1$ vector multiplets coming from the
truncation. \\
The transformation law for the $n_V +1$ vectors of the $N=2$
theory (\ref{gaugtrasf}) becomes, after truncation:
\begin{eqnarray}
\delta A^{ \L }_\mu &=& -{\rm i} f^{ \L }_i \bar \lambda^{i2}
\gamma_\mu \epsilon^1  -{\rm i} f^{ \L }_\a \bar \lambda^{\a 2}
\gamma_\mu \epsilon^1+ h.c. = -{\rm i} f^{ \L }_\a
\bar \lambda^{\a 2} \gamma_\mu \epsilon^1+ h.c.\label{vect1}\\
\delta A^{X }_\mu &=& -{\rm i} f^{X }_i \bar \lambda^{i2}
\gamma_\mu \epsilon^1  -{\rm i} f^{X }_\a \bar \lambda^{\a 2}
\gamma_\mu \epsilon^1+ h.c. =0\label{vect2}
\end{eqnarray}
where in (\ref{vect1}) we have used (\ref{dl0}). Eq.
(\ref{vect2}) is consistently zero if we put
\begin{equation}
\l^{i2}=0\label{li2}
\end{equation}
 since $f^{X}_\a |_{\cM_R } =\nabla_\a L^{X}|_{\cM_R }=
0$ \footnote{This follows by looking at the expression of the
$N=2$ K\"ahler metric \cite{abcdffm}
\begin{equation}
  g_{{\mathcal I} \bar{\mathcal J}} = -2 {\rm Im} \cN_{{\bf \L}{\bf \S}}
  f^{{\bf \L}}_{\mathcal I} f^{{\bf \S}}_{\bar{\mathcal J}}
\end{equation}
by requiring that its mixed component $g_{i\bar\a}$ is zero.
Indeed, after reduction we get
\begin{equation}
  0= g_{ i {\bar\a}} = -2 {\rm Im} \cN_{XY} f^{X}_{
  i} f^{Y}_{{\bar\a}}
\end{equation}
 implying $f^{Y}_{{\bar\a}}=0$}.
\\
We note that while the gauge index ${\bf \L}$ of the $N=2$
gaugino runs over $n_V+1$ values (because of the presence of the
graviphoton) the indices $ \L$ and $\a$ take only $n'_V \leq n_V$
values. In particular, the index of the graviphoton $A^0$ belongs
to the orthogonal subset $X = 0,1,\cdots , n_C$, so that the
graviphoton is automatically projected out.
\par
To match the corresponding $N=1$ formula  (\ref{travec1}) we have
to set:
\begin{equation}
\lambda_\bullet^{ \L } \equiv -2 f^{ \L }_\a \lambda^{\a 2}.
\label{deflambda}
\end{equation}
 Now, we observe that we may trade the gaugino world index ${\mathcal I}
=1,\cdots , n_V$ with a vector index ${\bf \L}$ already at the
level of the $N=2$ theory, by defining
\begin{equation}
\l^{{\bf \L}A} \equiv -2 f^{{\bf \L}}_{{\mathcal I}}
\l^{{\mathcal I} A}. \label{lambdanew}
\end{equation}
Here the gauge index ${\bf \L}$ of the $N=2$ gauginos runs over
$n_V +1$ values (because of the presence of the graviphoton)
while the index ${\mathcal I}$ takes only $n_V$ values. The extra
gaugino, say $\lambda^0 $, is actually spurious, since
$\lambda^{{\bf \L} A}$ satisfies:
\begin{equation}\label{linear}
T_{{\bf \L}} \lambda^{{\bf \L} A}  = -2T_{{\bf \L}} f^{{\bf
\L}}_{{\mathcal I}} \lambda^{{\mathcal I} A}= 0
\end{equation}
where
\begin{equation}\label{defprogr}
 T_{{\bf \L}} \equiv 2{\rm i} {\rm Im }{\cal N}_{{\bf \L}{\bf \S}}
L^{{\bf \S}} ,
\end{equation}
 due to the special geometry
relation
\begin{equation}\label{iden}
{\rm {Im}}{\cal N}_{{\bf \L} {\bf \S}}L^{{\bf \L}} f^{{\bf \S}
}_{{\mathcal I}} \,=\,0. \label{sg}
\end{equation}
Note that $T_{{\bf \L}}$ is the projector on the graviphoton
field-strength, according to equation  (\ref{gravif}) \cite{cdf}.
\\
Using special geometry, one can see that the transformation law
for the $N=2$ gaugini (\ref{gaugintrasf}) can be rewritten in
terms of the $\l^{{\bf \L}A}$, up to 3-fermions terms, as:
\begin{equation}
\delta\lambda^{{\bf \L}A} = P^{{\bf \L}}_{\phantom{{\bf \L}} {\bf
\S}} F^{-{\bf \S}
}_{\mu\nu}\gamma^{\mu\nu}\epsilon^{AB}\epsilon_B - 2{\rm i}
U^{{\bf \L}{\bf \S}}\left(P^0_{{\bf \S}} \epsilon^{AB} +
P^{AB}_{{\bf \S}}\right)
 \epsilon_B
  \label{gaugin2''}
\end{equation}
where $P^{{\bf \L}}_{\phantom{{\bf \L}} {\bf \S}}$ is the
projector on the matter-vector field strengths and $U^{{\bf \L}
{\bf \S}}$ a tensor of special geometry. They are defined below in
equations (\ref{defp}),(\ref{defu}). The derivation of formula
(\ref{gaugin2''}) is given in Appendix C.
\par
The above formulae allow us to perform the reduction of the
gaugino $\l^{{\bf \L}2}=(\l^{{ \L}2},\l^{{X}2})$
straightforwardly. First of all, $\l^{{X}2}=f^X_i \l^{{i}2}=0$ as

 follows from  (\ref{li2}). Then, setting $A=2$ and ${\bf
\L}=\L$, we have
\begin{equation}
\l^\L_\bullet \equiv \l^{\L 2}= -2f^\L_{{\mathcal I}}
\l^{{\mathcal I}2}  = -2 f^\L_\a \l^{\a 2}
\end{equation}
since in the reduced theory $f^\L_i=0$, and then:
\begin{equation}
\delta\lambda^{ \L }_\bullet= P^{ \L}_{\phantom{{\bf \L}} \S}
F^{-\S }_{\mu\nu}\gamma^{\mu\nu}\epsilon_\bullet - 2 {\rm i}
U^{\L \S}\left(P^0_{\S}  + P^3_{ \S}\right)
 \epsilon_\bullet
  \label{gaugin1}
\end{equation}
Let us now apply the following relations of special geometry
\cite{cdf} to the present reduction:
\begin{eqnarray}
\label{defu} U^{{\bf \L}{\bf \S}}&\equiv & f^{{\bf \L}}_{\mathcal
I}g^{{\mathcal I}\bar{\mathcal J}}f^{{\bf \S}}_{\bar{\mathcal J}}
=-\half \left[\left(\rm {\rm Im}{\cal N}\right)^{-1}\right]^{{\bf
\L} {\bf \S}} -\bar
L^{{\bf \L}} L^{{\bf \S}}\\
\label{defp} P^{{\bf \L}}_{\phantom{{\bf \L}}{\bf \S}} & \equiv &
-2 U^{{\bf \L}{\bf \G}} {\rm Im} \cN_{{\bf \S}{\bf \G}} =
\delta^{{\bf \L}}_{{\bf \S}} -{\rm i} T_{{\bf \S}} \bar L^{{\bf
\L}}.
\end{eqnarray}
After decomposing the indices and using $f^{X}_\a|_{\cM_R } =
f^{\L}_i|_{\cM_R }= g_{i\bar\a}|_{\cM_R } =0$ we have
\cite{abcdffm}:
\begin{eqnarray}
\label{defu1} U^{ \L  \S }&\equiv & f^{ \L }_{\a}g^{\a {\b}}f^{
\S }_{ \b } =-\half \left[\left(\rm {Im}{\cal
N}\right)^{-1}\right]^{ \L  \S } ;
\\
\label{defu2} U^{X Y }&\equiv & f^{X }_{i}g^{i{\bar j}}f^{Y
}_{{\bar j}} =-\half \left[\left(\rm {Im}{\cal
N}\right)^{-1}\right]^{X Y } -\bar
L^{X } L^{Y } ; \\
\label{defp1} P^{  \L}_{\phantom{{\bf \L}} \S} & =  & ({\rm Im
}\cN^{-1}) ^{\L\G} {\rm Im} \cN_{ \S \G} =
\delta^{{\bf \L}}_{{\bf \S}};\\
\label{defp2} P^{X}_{\phantom{X}Y} & = &  = \delta^{X}_{Y} -{\rm
i} T_{Y} \bar L^{X}.
\end{eqnarray}
 Eq. (\ref{gaugin1})
can then be rewritten as:
\begin{equation}
\delta\lambda^\L _\bullet =  \left[
F^{-\L}_{\mu\nu}\gamma^{\mu\nu}
 + {\rm i} \left({\rm Im}{\cal
N}^{-1}\right)^{\L\S }\left(P^0_\S + P^3_\S\right)  \right]
\epsilon_\bullet .\label{gaugin2'}
\end{equation}
\par
We observe that the prepotential $P^0_\S$,
 which gives the special-K\"ahler
manifold contribution to the D-term, can be given an explicit form
in terms of $N=2$ objects. Indeed, let us recall that $P^0_{{\bf
\S}}$ has the following general form, as shown in equation
(\ref{killer}) of Appendix C:
\begin{equation}
P^0_{{\bf \S}} =-2{\rm i}{\rm Im}{\cal N}_{{\bf \S}{\bf \G}}
f^{{\bf \G}}_{{\mathcal I}} k^{{\mathcal I}}_{{\bf \Delta}} \bar
L^{{\bf \Delta}}
\end{equation}
which gives, after reduction:
\begin{equation}
P^0_{ \S} =-2{\rm i}{\rm Im}{\cal N}_{ \S \G} f^{ \G}_{\a}
k^{\a}_{W} \bar L^{W}. \label{p0}
\end{equation}
On the other hand, using the following $N=2$ special geometry
property:
\begin{equation}
f^{{\bf \G}}_{{\mathcal I}} k^{{\mathcal I}}_{{\bf \Delta}}  =
{\rm i} P^0_{{\bf \Delta}}L^{{\bf \G} } -{f}^{{\bf \G} }
_{\phantom{{\bf \G}}{\bf \Delta} {\bf \S}} L^{{\bf \S}}
\label{poiss}
\end{equation}
(${f}^{{\bf \G} } _{\phantom{{\bf \G}}{\bf \Delta} {\bf \S}}$ are
the structure constant of the $N=2$ gauge group $G^{(2)}$) by
contracting with $\bar L^{{\bf \Delta}}$ and reducing it to the
submanifold $\cM_R $, we also find \cite{abcdffm}:
\begin{equation}
P^0_\S =2{\rm i}{\rm Im}{\cal N}_{\S\G}{f}^\G_{\phantom{\G}X
Y}\bar L^{X} L^{Y}\label{p02}.
\end{equation}
\par
 In conclusion we get the final form of the gaugino
transformation law for the $N=1$ theory as:
\begin{equation}\label{final}
\left(\delta\lambda^\L_\bullet\right)_{N =1}=
\cF^{-\L}_{\mu\nu}\gamma^{\mu\nu}\epsilon_\bullet +{\rm i}D^\L
\epsilon_\bullet \, \, , \quad (\L = 1,\cdots , n)
\label{gaugin2'''}
\end{equation}
where, in order to retrieve the transformation law
(\ref{tralamb1}) we have set
\begin{equation}
\label{dterm} D^\L  \equiv  ({\rm Im }\cN^{-1})^{\L\S}
\left(P^0_\S + P^3_\S\right).
\end{equation}
\par
In order to show that equation (\ref{final}) is the correct $N=1$
transformation law of the gauginos we have still to prove that
$\cN_{\L\S}$ is an antiholomorphic function of the scalar fields
$z^i$ (as it is the case for an $N=1$ theory), since the
corresponding object of the $N=2$ special geometry $\cN_{{\bf
\L}{\bf \S}}$ is not antiholomorphic. For this purpose we observe
that in $N=2$ special geometry the following identity holds (at
least when a $N=2$ prepotential function exists \footnote{In
Appendix D we will discuss the reduction with special
coordinates.})\cite{cdf}:
\begin{equation}\label{inversusf1}
{\mathcal N}_{{\bf \L}{\bf \S}}\,=\,{\bar F}_{{\bf \L}{\bf
\S}}-2{\rm i}\bar T_{{\bf \L}} \bar T_{{\bf \S}} (L^{{\bf \G}
}{\rm Im}F_{{\bf \G} {\bf \Delta}}L^{{\bf \Delta}})
\end{equation}
where the matrix $F_{{\bf \L}{\bf \S}}$ is holomorphic.   \\
If we now reduce the indices ${\bf \L}{\bf \S}$ we find:
\begin{equation}\label{holn}
{\mathcal N}_{\L\S}\,=\,{\bar F}_{\L\S}-2{\rm i}\bar T_\L \bar
T_\S (L^{X }{\rm Im}F_{XY}L^{Y})\equiv {\bar F}_{\L\S}
\end{equation}
since $T_\L=0$ is precisely the constraint (\ref{nl}). Therefore
${\mathcal N}_{\L\S}$ is antiholomorphic and the $D$-term
(\ref{dterm}) becomes:
\begin{equation}\label{dtrm1}
  D^\L  \equiv 2{\rm i} f^\L_\a W^{\a 21} = - 2({\rm Im }f^{-1}(z^i))^{\L\S}
\left(P^0_\S + P^3_\S\right).
\end{equation}
where we have defined
\begin{equation}\label{norma}
F_{\L\S}(z^i)=\half f_{\L\S}(z^i)
\end{equation}
in order to match the normalization of the holomorphic kinetic
matrix of the $N=1$ theory appearing in equation
(\ref{dlambda}).\\
We observe that for choices of symplectic sections such that the
function $F_{{\bf \L}{\bf \S}}$ does not exist, the relation
(\ref{holn}) does not hold, but still $\cN_{\L\S}$ has to be
antiholomorphic on $\cM_R$. Un explicit example will be given in
section (6.6).
\par
As a final observation, we note that the above reduction on  the
indices of the $N=2$ Killing vectors gives rise to $k^{\mathcal
I}_{{\bf \L}} \Rightarrow (k^i_\L ,k^\a_\L , k^i_{X}, k^\a_{X})$.
The Killing vectors  $k^i_\L$ gauge the isometries  of the
submanifold $\cM_R$. On the other hand, $k^\a_\L$ are zero on the
submanifold, since they correspond to isometries orthogonal to
$\cM_R$; $k^i_{X}$ are also zero because we have projected out
the corresponding vectors. Finally, $k^\a_{X}$ are in general
different form zero, and enter in the definition of $P^0_\S$,
equation  (\ref{p0}). These conclusions can be formally retrieved
by analyzing the reduction of the special geometry identity
\cite{abcdffm}
\begin{equation}
2{\rm i} g_{{\mathcal I} \bar{\mathcal J}} k^{{\mathcal
I}}_{[{\bf \L}} k^{\bar{\mathcal J}}_{{\bf \S}]} = {f}_{{\bf \L}
{\bf \S}}^{\phantom{{\bf \L}{\bf \S}}{\bf \G}} P^0_{{\bf
\G}}.\label{chiusura}
\end{equation}
One can easily verify that if we set ${\bf \L}=\L ;{\bf \S}=\S$
then one retrieves the analogous of relation (\ref{chiusura}) on
$\cM_R$ provided we set
\begin{equation}
k^\a_\L =0  \, ; \quad P^0_{X}=0 .
\end{equation}
For ${\bf \L}=\L ;{\bf \S}=Y$ equation  (\ref{chiusura}) is
identically satisfied provided we add to the previous condition
the further constraint
\begin{equation}
k^i_{X} =0.
\end{equation}
Finally, when ${\bf \L}=X ;{\bf \S}=Y$, equation (\ref{chiusura})
reduces to the relation:
\begin{equation}
2{\rm i} g_{\a {\bar\b}} k^{\a}_{[X} k^{ {\bar\b}}_{Y ]} = {f}_{X
Y}^{\phantom{{\bf \L}{\bf \S}} \G} P^0_{ \G}\label{chiusura1}
\end{equation}
which has to be satisfied all over the manifold $\cM_R$.

\subsection{Reduction of the hypermultiplet sector} Let us analyze
the sector of the hypermultiplets when the reduction is
implemented. The scalars of the hypermultiplets belong to a
quaternionic manifold ${\cal M}^{Q}$.
 A quaternionic
manifold ${\cal M}^{Q}$ has a holonomy group of the following
type \cite{alex}, \cite{wolf}, \cite{gali}:
\begin{eqnarray}
{\rm Hol}({\cal M}^{Q})&=& SU(2)\otimes {\cal H} \quad
(\mbox{quaternionic}) \nonumber \\ {\cal H} & \subseteq & Sp
(2n_H) \label{olonomia}
\end{eqnarray}
Introducing flat indices $\{A,B,C= 1,2\}\, \{\alpha,\beta,\gamma =
1,.., 2n_H\}$  that run, respectively, in the fundamental
representations of $SU(2)$ and $Sp(2n_H)$ ($n_H$ is the number of
hypermultiplets) we introduce the vielbein 1-form \cite{abcdffm}
\begin{equation}
{\mathcal U}^{A\alpha} = {\mathcal U}^{A\alpha}_u (q) d q^u
\label{quatvielbein}
\end{equation}
such that
\begin{equation}
\label{metric} h_{uv} = {\mathcal U}^{A\alpha}_u {\mathcal
U}^{B\beta}_v \IC_{\alpha\beta}\epsilon_{AB} \label{quatmet}
\end{equation}
where $\IC_{\alpha \beta} = - \,\IC_{\beta \alpha}$ and
$\epsilon_{AB} = - \,\epsilon_{BA}$ are, respectively, the flat
$Sp(2n_H)$ and $Sp(2) \sim SU(2)$ invariant metrics. The vielbein
${\mathcal U}^{A\alpha}$ is covariantly closed with respect to the
$SU(2)$-connection $\omega^x (x=1,2,3)$ and to the $Sp(2n_H)$-Lie
Algebra valued connection $\Delta^{\alpha\beta} = \Delta^{\beta
\alpha}$:
\begin{eqnarray}
\nabla {\mathcal U}^{A\alpha}& \equiv & d{\mathcal U}^{A\alpha}
+{i\over 2} \omega^x  \left(\sigma_x\right)^A_{\ B}
\wedge {\mathcal U}^{B\alpha} \nonumber\\
&+& \Delta^{\alpha}_{\ \beta} \wedge {\mathcal U}^{A\beta}
 =0 \label{quattorsion}
\end{eqnarray}
where
$(\sigma^x)^{AB}\,=\,\epsilon^{AC}(\sigma^x)_C^{\phantom{C}B}$ and
$(\sigma^x)_A^{\phantom{A}B}$ are the standard Pauli matrices.
Furthermore ${ \cal U}^{A\alpha}$ satisfies  the reality
condition:
\begin{equation}
{\mathcal U}_{A\alpha} \equiv ({\mathcal U}^{A\alpha})^* =
\epsilon_{AB} \IC_{\alpha\beta} {\mathcal U}^{B\beta}
\label{quatreality}
\end{equation}
 The supersymmetry transformation laws
of the fields in the hypermultiplets are given in equation
(\ref{iperintrasf}) and (\ref{quatertrasf}), that we rewrite here
using  tangent-space indices for the quaternionic variation:
\begin{eqnarray}
{\mathcal U}_u^{\alpha A}\delta\,q^u &=&  \bar {\zeta}^{\alpha}
  \epsilon^A + \IC^{\alpha  \beta}\epsilon^{AB}\bar {\zeta}_{\beta}
  \epsilon _B \label{quatertrasf1}\\
\label{iperleft}
  \delta\,\zeta _{\alpha}&=&{\rm i}\,
{\mathcal U}^{B \beta}_{u}\, \nabla _{\mu}\,q^u \,\gamma^{\mu}
\epsilon^A \epsilon _{AB}\,\IC_{\alpha  \beta} \,+g\,N^A_\alpha
\,\epsilon _A \\
\label{iperright}\delta\,\zeta ^{\alpha}&=&{\rm i}\, {\mathcal
U}^{A\alpha}_{u}\, \nabla _{\mu}\,q^u \,\gamma^{\mu} \epsilon_A
\,+g\, N^\alpha_A \epsilon ^A
\end{eqnarray}
Let us see what happens to equations
(\ref{quatertrasf1}),(\ref{iperleft}),(\ref{iperright}), when the
truncation is implemented.
\\
First of all let us note that the scalars in $N=1$ supergravity
must lie in chiral multiplets, and have in general a
K\"ahler-Hodge structure. It is therefore required that the
holonomy of the quaternionic manifold be reduced:
\begin{equation}\label{redhol}
 {\rm Hol} \left({\cal M}^{Q}\right) \subset SU(2) \times Sp(2n_H) \to
 {\rm Hol} \left({\cal M}^{KH}\right) \subset U(1) \times
 SU(n).
\end{equation}
Therefore the $SU(2)$ index $A =1,2$ and the $Sp(2n_H)$ index have
to be decomposed accordingly. We set $\a \to (I,\dot I )\in U(1)
\times SU(n_H) \subset Sp(2n_H)$. Since the vielbein ${ \cal
U}^{A\alpha}$ satisfy the reality condition (\ref{quatreality}),
we have, in $U(n_H)$ indices :
\begin{eqnarray}
{ \cal U}_{1I} \equiv ({\mathcal U}^{1I})^* &=& \IC_{I\dot I}
{\mathcal U}^{2\dot I}\nonumber\\
 {\mathcal U}_{2I} \equiv ({\mathcal U}^{2I})^* &=&
- \IC_{\dot I I} {\mathcal U}^{1 \dot I} \label{quatrealdec}
\end{eqnarray}
where we have used the decomposition of the symplectic metric
$\IC_{\alpha\beta} =\left(
  \begin{array}{cc}
    0 & \IC_{I\dot J}  \\
   \IC_{\dot J I} & 0
  \end{array}\right) $
with $\IC_{I\dot J} = - \,\IC_{\dot J I}= \delta_{I\dot J}$. \\
From equation  (\ref{quatrealdec}) one finds that it is sufficient
to refer to the $2n_H$ complex vielbein $ { \cal U}^{1I}, { \cal
U}^{2I}$ since the ones with dotted indices are related to them by
complex conjugation.
\\
 Let us first examine the torsion-free
equation obeyed by the quaternionic vielbein written in the
decomposed indices:
\begin{eqnarray}
&&d{\mathcal U}^{1I} +{i\over 2} \omega^3\wedge {\mathcal U}^{1I}+
{i\over 2} (\omega^1-{\rm i}\omega^2) \wedge {\mathcal
U}^{2I}\nonumber
\\&&
 + \Delta^{I}_{\ J} \wedge {\mathcal U}^{1J} +
\Delta^{I}_{\ \dot J} \wedge {\mathcal U}^{1\dot J} =0 \label
{torsion1}
\end{eqnarray}
\begin{eqnarray}
&&d{\mathcal U}^{2I} -{i\over 2}(\omega^1+{\rm i}\omega^2)  \wedge
{\mathcal U}^{1I}-{i\over 2}\omega^3 \wedge {\mathcal U}^{2I} \nonumber\\
&&+ \Delta^{I}_{\ J} \wedge {\mathcal U}^{2J} + \Delta^{I}_{\ \dot
J} \wedge {\mathcal U}^{2\dot J} =0
 \label{tors2}
\end{eqnarray}
\par
For the $N=1$ reduced K\"ahler--Hodge scalar manifold, the
holonomy has to be $U(1) \times SU(n_H)$, with a non trivial
$U(1)$-bundle, whose field-strength has to be identified with the
K\"ahler form. Since in the $N=2$ quaternionic parent theory
there is a similar non trivial $SU(2)$-bundle, whose
field-strength has to be identified with the Hyper-K\"ahler form,
we assume that the $U(1)$ part of the holonomy should be valued in
 the $U(1)$ subgroup of the $SU(2)$ valued connection of $N=2$ quaternionic
holonomy group.
\\
From equations  (\ref{torsion1}), (\ref{tors2}) we see that,
setting
\begin{equation}\label{cond}
\omega^1=\omega^2=\Delta^I_{\dot J}=0
\end{equation}
we get two K\"{a}hler-Hodge manifolds whose respective vielbeins
obey the torsionless equations for each submanifold.
\par
Let us now check the involution property dictated by the Frobenius
theorem. As we know from section 3, this amounts to demand that
the curvatures of the connections set to zero, equation
(\ref{cond}),
 must satisfy the constraints of being also zero on the
 submanifold. That is we must have:
\begin{equation}
\Omega^1=\Omega^2  = \IR ^I_{\ \dot J}=0\label{constr1}
\end{equation}
where the $SU(2)$ curvature $\Omega^x$  is  given by
\footnote{Note that $\Omega^x = \lambda \, K^x_{uv}$
 with $K^x_{uv}$ given in terms of the three
complex structures by:
\begin{eqnarray}
K^x& = &K^x_{u v} d q^u \wedge d q^v \nonumber \\
 K^x_{uv} &=&
h_{uw} (J^x)^w_v .
\end{eqnarray}
The scale $\lambda$ is fixed by supersymmetry of the Lagrangian
and in our conventions is $\l=-1$.}
\begin{equation}
\Omega^x \, \equiv \, d \omega^x + {1\over 2} \epsilon^{x y z}
\omega^y \wedge \omega^z = \\ {\rm i}\, \lambda \IC_{\alpha\beta}
(\sigma ^x)_{AB} {\mathcal U}^{\alpha A} \wedge {\mathcal
U}^{\beta B} \label{curform}
\end{equation}
 while the $Sp(2n_H)$ curvature
$\IR^{\alpha}_{\ \beta }$ is given by:
\begin{eqnarray} &&\IR^{\alpha}_{\ \beta} \equiv d \Delta^{\alpha}_{\
\beta} + \Delta^{\alpha}_{\ \gamma} \wedge \Delta^{\gamma}_{\
\beta}\\ \nonumber &&=\lambda\epsilon_{AB}{\mathcal
U}^{A\alpha}\wedge {\mathcal U}^{B}_{\beta}+{\mathcal
U}^{A\gamma}\wedge {\mathcal U}^{B\delta}
 \epsilon_{AB}
\IC^{\alpha\rho} \Omega_{\rho\beta\gamma\delta}\ ,\label{simpl1}
\end{eqnarray}
where $\Omega_{\a\b\g\delta}$ is a completely symmetric 4-index
tensor \cite{bw1}.\\
 From equation  (\ref{curform}) we see that the constraint
(\ref{constr1}) for involution is satisfied iff
\begin{equation}
 {\mathcal U}^{1I}\wedge {\mathcal U}^{2I}=0
\label{half}
\end{equation}
that is if, say, the subset ${\mathcal U}^{2I}= \left({\mathcal
U}^{1\dot I}\right)^*$
 of the quaternionic vielbein is set to zero on a submanifold $\cM^{KH}\subset \cM^Q$.
\\
When condition (\ref{half}) is imposed, our submanifold has
dimension at most half the dimension of the quaternionic manifold
(in the following we always refer to the maximal case, where
$I=1,\cdots ,n_H$) and the $SU(2)$ connection is
 reduced to a $U(1)$ connection, whose curvature on $\cM^{KH}$ is:
\begin{equation}  \Omega^3|_{\cM^{KH}} = {\rm
i}\lambda{\mathcal U}^{1I}\wedge  {\mathcal U}_{1 I }= {\rm
i}\lambda{\mathcal U}^{1I}\wedge \overline{ {\mathcal U}^{1 I }}
\label{omega3}
\end{equation}
so that the $SU(2)$-bundle of the quaternionic manifold is reduced
to a $U(1)$-Hodge bundle for the $n_H$ dimensional complex
submanifold spanned by the $n_H$ complex vielbein ${\mathcal
U}^{1I}$.
\\
The truncation corresponds therefore to select a $n_H$-complex
dimensional submanifold $\cM^{KH} \subset \cM^{Q}$ spanned by the
vielbein ${\mathcal U}^{1I}$ and to ask that, on the submanifold,
the $2n_H$ extra degrees of freedom are frozen, that is:
\begin{equation}
{\mathcal U}^{2I}|_{\cM^{KH}} =\left({\mathcal U}_{1\dot
I}\right)^*|_{\cM^{KH}} \,=\,0 \label{zero}.
\end{equation}
Calling $w^s $ ($s=1,\cdots n_H$) a set of $n_H$ holomorphic
coordinates on ${\cM^{KH}}$ and $n^t$ ($t=2n_H+1, \cdots , 4n_H$)
a set of $2n_H$ real coordinates for the space orthogonal to
${\cM^{KH}}$, we see that equation  (\ref{zero}), which can be
rewritten as:
\begin{equation}\label{zerobis}
  {\mathcal U}^{2I}|_{\cM^{KH}} =\left({\mathcal U}^{2I}_s dw^s +
  {\mathcal U}^{2I}_{\bar s}
  d\bar w^{\bar s} + {\mathcal U}^{2I}_t dn^t \right) |_{\cM^{KH}}
  =0\,,
\end{equation}
implies:
\begin{equation}\label{viel2cond}
  {\mathcal U}^{2I}_s |_{\cM^{KH}} ={\mathcal U}^{2I}_{\bar s}|_{\cM^{KH}} =0
\end{equation}
since:
\begin{equation}
dn^t |_{\cM^{KH}} =0 .
\end{equation}
On the other hand, we also have:
\begin{equation}\label{viel1cond}
 {\mathcal U}^{1I}_t |_{\cM^{KH}} =0 .
\end{equation}
since the vielbein ${\mathcal U}^{1I}$ is tangent to the
submanifold. \noindent
 Let us note that the conditions (\ref{constr1}) on the
curvatures $\Omega^{1},\Omega^{2}$ imposed on the submanifold do
not imply that all their components are also zero there, and
indeed from (\ref{viel2cond}), (\ref{viel1cond}) and the
definition (\ref{curform}) it follows:
\begin{equation}\label{omega1,2}
 \Omega^1_{s\bar s}|_{\cM^{KH}} = \Omega^1_{t  t'}|_{\cM^{KH}}
  =  \Omega^2_{s\bar s}|_{\cM^{KH}} =
  \Omega^2_{t t'}|_{\cM^{KH}} =0
\end{equation}
while the mixed components $\Omega^1_{s t}|_{\cM^{KH}}$,
$\Omega^2_{s t}|_{\cM^{KH}} $ (together with their com\-plex
con\-ju\-gates $\Omega^1_{\bar st}|_{\cM^{KH}}$, $\Omega^2_{\bar s
t}|_{\cM^{KH}} $) are different from zero. We also observe that,
when the truncation is performed, also the mixed components of the
metric are  zero:
\begin{equation}\label{metric'}
  h_{st}|_{\cM^{KH}} = \left(\cU^{1I}_s \bar\cU_{1I |t} \right)|_{\cM^{KH}} =
  0 .
\end{equation}
From (\ref{viel2cond}), (\ref{viel1cond}) and (\ref{curform}) it
also follows that the only components different from zero of the
2-form $\Omega^3$ are $\Omega^3_{s\bar s}$ and $\Omega^3_{t t'}$.
\par
Let us now analyze in detail whether the involution of the
constraint $\Delta^I_{\ \dot J} =0$ is satisfied:
\begin{eqnarray}\label{simplconstr}
&&\IR^{I}_{\ \dot J}= \lambda\IC_{\dot J K} \left({\mathcal
U}^{1I} \wedge {\mathcal U}^{2K} - {\mathcal U}^{2I}\wedge
{\mathcal U}^{1K}\right)
+\nonumber\\
&&+ 2\IC^{I \dot I} \Bigl[{\mathcal U}^{1K}\wedge {\mathcal
U}^{2L} \Omega_{\dot I\dot J KL}\, + 2{\mathcal U}^{1K}\wedge
{\mathcal U}^{2\dot L}
  \Omega_{ \dot I\dot J K\dot L}\,+
  {\mathcal U}^{1\dot K}\wedge {\mathcal
U}^{2\dot L}
  \Omega_{\dot I\dot J \dot K\dot L}\Bigr]\, =\, 0
\end{eqnarray}
After imposing (\ref{zero}), the first line in equation
(\ref{simplconstr}) is automatically zero, and equation
(\ref{simplconstr}) is reduced to the constraint:
\begin{equation}
\label{constromega} 4\IC^{I \dot I} {\mathcal U}^{1K}\wedge
{\mathcal U}^{2\dot L}
  \Omega_{ \dot I\dot J K\dot L}\, =\, 0
  \end{equation}
  Furthermore, let us note that when the constraint
  (\ref{simplconstr}) is imposed, the $Sp(2n_H)$ holonomy gets reduced
  to  $U(1)\times SU(n_H)$.
  The $U(n_H)$ curvature becomes:
\begin{equation}\label{uncurv}
 \IR^{I \dot J}= \lambda{\mathcal
U}^{1I}\wedge {\mathcal U}^{2 \dot K} + \IC^{I \dot K} \IC^{ \dot
J L} {\mathcal U}^{1M}\wedge {\mathcal U}^{2\dot N} \Omega_{\dot
N\dot K ML}\,
\end{equation}
 Choosing cordinates such that $q^{4s+3}=q^{4s+4}=0,
s=0,\dots,n_H-1$ we may introduce complex coordinates $w^s=
q^{1+4s}+{\rm i}q^{2+4s}$ with K\"{a}hler 2-form $K=\Omega^3$
which is automatically closed. The $U(1)$ connection of the Hodge
bundle is  given by $\omega^3_s$ as can be ascertained from the
reduced form of equation (\ref{torsion1}) expressing the vanishing
of the torsion on the K\"{a}hler-Hodge submanifold $\cM^{KH}$:
\begin{equation}\label{torsion2}
\nabla {\mathcal U}^{1I} \equiv  d{\mathcal U}^{1I} +{i\over 2}
\omega^3
 \wedge {\mathcal U}^{1I} + \Delta^{I}_ J \wedge {\mathcal U}^{1 J}  =0
\end{equation}
\par
  In conclusion, what we have found is that the conditions for the
  truncation of a quaternionic manifold (spanning the scalar sector
  of $n_H$ $N=2$ hypermultiplets) to a K\"ahler--Hodge
  one (spanning the scalar sector of $n_h$ $N=1$  chiral multiplets)
are the following:
\begin{itemize}
\item{
${\mathcal U}^{2I}= \omega^1= \omega^2=\Delta^I_{\dot J}=0$

}
\item{The quaternionic manifold cannot be generic; in particular,
the completely symmetric tensor
$\Omega_{\alpha\beta\gamma\delta}\in Sp(2n_H)$, appearing in the
$Sp(2n_H)$ curvature, must have the following constraint on its
components:
\begin{equation}
 \Omega_{ \dot I\dot J K\dot L}\, =\, 0. \label{omega condition}
  \end{equation}
}
\end{itemize}
The resulting submanifold, denoted by $\cM^{KH}$, has at most
$n_H$ complex dimensions \cite{am} and is of K\"ahler--Hodge type,
with K\"ahlerian  vielbein $P^I,{\bar P}^{\bar I}$ (for its
normalization, see equation (\ref{trans1}) below):
\begin{eqnarray}
{\mathcal U}^{1I}_u dq^u &\longrightarrow &\frac {1} {\sqrt 2}\bar
P^{\bar I}_{\bar s}
d{\bar w}^{\bar s}\\
{\mathcal U}^{2\dot I}_u dq^u &\longrightarrow & \frac {1} {\sqrt
2} P^I_{ s} dw^{s}
\end{eqnarray}
(where $w^s,s=1,\cdots n_H$ are complex coordinates on the reduced
K\"ahler manifold) and $U(1)\times U(n_H)$ curvature given by:
\begin{equation}
{\cal R}^I_{J } \equiv {\cal R}^u_{v}{\mathcal U}^{1I}_u {\mathcal
U}_{1J}^v \longrightarrow \frac{\rm i}{2} \Omega^3 \delta^I_J +
\IR^I_{\ J}.
\end{equation}
\par
Once the dimension of the manifold has been truncated, the only
constraint on the quaternionic manifold is given by equation
(\ref{omega condition}). Let us therefore discuss how general it
is, and which quaternionic manifolds  satisfy it.
\\
First of all we note that the family of symmetric spaces $ Sp(2m,
2)/Sp(2)\times Sp(2m)$ has a vanishing $\Omega$-tensor,
$\Omega_{\alpha\beta\gamma\delta}=0$ \cite{gali}, and hence {\it a
fortiori} satisfies our requirement.
\par
Furthermore we can now show that the special quaternionic
manifolds obtained by c-map \cite{fesa} from special-K\"ahler
manifolds do indeed satisfy the condition:
\begin{equation}
\Omega_{\dot I\dot J K\dot L} =0 \label{Omega}
\end{equation}
Indeed the tensor (\ref{Omega}) appears in equation
(\ref{simplconstr}) multiplied by the product of the vielbein $
{\mathcal U}^{1K} \wedge  {\mathcal U}^{2\dot L}$. The same
sub-block of the $Sp(2n)$ curvature is denoted in \cite{fesa} by
$r'^A_{\phantom{A} \bar B}$. Now, it is easy to recognise that the
set of $n+1$ complex vielbein $(v, e^a)$ of \cite{fesa} have to be
identified with our vielbein ${\mathcal U}^{1 K}$. However, no
wedge product of type $\bar v\wedge e^a$ nor of type $\bar e^a
\wedge e^b$ appear in $r'^A_{\phantom{A} \bar B}$, which means
that the corresponding coefficient $ \Omega_{\dot I\dot J K\dot L}
=0$. Therefore, all the special quaternionic manifolds (including
non symmetric quaternionic spaces) can be reduced to
K\"ahler-Hodge manifolds in a way consistent with our procedure.

There are however other symmetric spaces which do not correspond
to $c$-map of special-K\"{a}hler manifolds, yet they satisfy our
constraints. Indeed consider the following reduction from
quaternionic to K\"{a}hler-Hodge manifolds:

\begin{equation}
\frac {SO(4,n)} {SO(4)\times SO(n)}\longrightarrow \frac
{SO(2,n_1)} {SO(2)\times SO(n_1)}\times \frac {SO(2,n_2)}
{SO(2)\times SO(n_2)} \label{firsta}
\end{equation}
where $(n_1 +n_2 =n)$. We see that they satisfy our constraints.
Indeed, the K\"{a}hler-Hodge manifold on the right of the
correspondence in equation  (\ref{firsta}) is apparently a
submanifold of the corresponding  quaternionic with half
dimension. Therefore the conditions for the validity of the
Frobenius theorem have to be satisfied, in particular equation
(\ref{omega condition}). Indeed, for symmetric spaces we can
compute explicitly the $\Omega$-tensor by comparing the general
formula of the Riemann tensor for symmetric spaces:
\begin{equation}\label{riemann}
{\cal R}^{uv}_{\phantom{uv}{ts}} {\cal U}^{\alpha A}_u {\cal
U}^{\beta B}_v = -\half {f}^{\alpha A|\beta B}_{\phantom{\alpha
A|\beta B}h} {f}_{\gamma C|\delta D}^{\phantom{\alpha A|\beta B}h}
{\cal U}^{\gamma C}_{[t}{\cal U}^{\delta D}_{s]},
\end{equation}
(where we have denoted by $f^{\alpha A|\beta B}_{\phantom{\alpha
A|\beta B}\g C}$ the structure constants of the isometry group of
the symmetric manifold $K=G/H$, the index $h$ running on the Lie
algebra of $H$, the couple of indices $A\alpha$ labelling the
coset generators)
 with its general form in the case of
quaternionic manifolds:
\begin{equation}
{\cal R}^{uv}_{\phantom{uv}{ts}} {\cal U}^{\alpha A}_u {\cal
U}^{\beta B}_v = -\,{{\rm i}\over 2} \Omega^x_{ts}
 (\sigma_x)^{AB} C^{\alpha \beta}+
 \IR^{\alpha\beta}_{ts}\epsilon^{AB}.
\label{2.65}
\end{equation}
One easily obtains
\begin{equation}\label{tensorone}
  \Omega_{\a\b\g\delta}= - \frac \lambda 2 \left( \IC _{\a\g}
  \IC_{\b\delta} +\IC _{\a\delta}
  \IC_{\b\g}
  \right) - \frac{\rm i}{4} \epsilon^{AC} \epsilon^{BD}
  {f}_{C\{\a |\b\} D  | h} {f}^h_{\phantom{h} A\{\g |  \delta \} B}
\end{equation}
where the curly brackets mean symmetrization of the corresponding
indices. \\
Using equation  (\ref{tensorone}),  we have explicitly verified
the validity of (\ref{omega condition}) in the case of the
omega-tensor appearing in (\ref{firsta}). These quaternionic
reductions explicitly appear in some effective lagrangians coming
from superstring theory models \cite{fgkp2}.
 \vskip 5mm
\par
We still have to analyze the effects of the reduction on the
hyperini and on the supersymmetry transformation laws. They
become, after putting $\epsilon_2 =0$:
\begin{eqnarray}
{\mathcal U}_u^{1I}\delta\,q^u &=&  \bar {\zeta}^{I}
  \epsilon^1 \label{quat1}\\
  {\mathcal U}_u^{2I}\delta \,q^u &=& -\IC^{I\dot J}\bar {\zeta}_{\dot J}
  \epsilon _1 \label{quat2}\\
\label{iper1}
  \delta\,\zeta _{I}&=&{\rm i}\,
{\mathcal U}^{2 \dot J}_{u}\, \IC_{I \dot J} \, \nabla _{\mu}\,q^u
\,\gamma^{\mu} \epsilon^1 \,+ g N^1_I \epsilon _1 =\left(\delta\,\zeta ^{I}\right)^*\\
\label{iper2} \delta\,\zeta _{\dot J}&=&{\rm i}\, {\mathcal U}^{2
I}_{u}\, \IC_{\dot J I } \, \nabla _{\mu}\,q^u \,\gamma^{\mu}
\epsilon^1 \,+ g N^1_{\dot J} \epsilon _1 =\left(\delta\,\zeta
^{\dot J}\right)^*
\end{eqnarray}
Choosing the normalization in such a way to match the
normalization of the kinetic terms of the $N=1$ theory, we set:
\begin{equation}\label{trans1}
{\mathcal U}^{2\dot J}C_{I\dot J}=\frac {1} {\sqrt 2}P_{Is}
\end{equation}
\begin{equation}\label{defn1}
  N^1_I =\frac{1} {\sqrt 2}P_{Is}N^s
\end{equation}
\begin{equation}\label{zeta1}
\zeta^s \equiv \sqrt{2} P^{sI} \zeta_I = \sqrt{2} g^{s \bar s}
\bar P^I_{\bar s} \zeta_I ,
\end{equation}
 \begin{equation} {\mathcal U}^{2 \dot J}_u \IC _{I\dot J}
  \nabla_\mu q^u |_{\cM^{KH}}
   =  \frac{1}{\sqrt{2}} P_{Is} \nabla_\mu w^s \end{equation}
which implies:
\begin{equation}
{\mathcal U}_u^{2\dot J} \IC _{I\dot J}\delta\,q^u |_{\cM^{KH}}=
\frac{1}{\sqrt{2}} \bar P_{Is} \delta  w^{s} \label{pincopallo}
\end{equation}
where $\zeta^s$ denote chiral left-handed spinors with holomorphic
world indices, $P_{Is}$ are the vielbein of the K\"ahler-Hodge
manifold ${\cM^{KH}}$ and $w^s$ its holomorphic coordinates. We
observe that due to the definition (\ref{trans1}) the 2-form
$\Omega^3$ defined in equation  (\ref{omega3}) is one half the
K\"ahler 2-form on ${\cal M}^{KH}$.\\
 In that way we obtain the standard formulae for the $N=1$
supersymmetry transformation laws of the chiral multiplets
$(\zeta^s, \w^s)$, that is:
 \begin{eqnarray}
  \delta \zeta^s &
=& {\rm i} \nabla_\mu w^s \gamma^\mu \epsilon^\bullet  \,+\, N^s
\epsilon _\bullet  \label{rediper} \\
\delta w^s &=&  \zeta^s \epsilon_\bullet \end{eqnarray} where
\begin{equation}
N^s \equiv \,\sqrt{2}\, g_{({\bf \L})}P^{ s J} N^1_J =
\,2\sqrt{2}\, g_{({\bf \L})} P^{ s J} \IC_{J\dot J} {\mathcal
U}^{1\dot J}_t k^t_{{\bf \L}} \bar L^{{\bf \L}}. \label{rediper2}
\end{equation}
Note that the shift term $N^s$ is indeed different from zero, but
depends only on the isometries of the projected out part of the
quaternionic manifold \footnote{ Indeed, the request ${\mathcal
U}^{1\dot I}|_{\cal M} = \left( {\mathcal U}^{1\dot I}_s dw^s
+h.c. + {\mathcal U}^{1\dot I}_t dn^t \right)|_{\cal M} = 0$
implies ${\mathcal U}^{1\dot I}_s=0$ but does not impose any
restriction on the components orthogonal to the retained
submanifold.}. The explicit $N=1$ form of the gauging contribution
will be given in the next section. \\
From equation  (\ref{quat2}), however, we see that the condition
${\mathcal U}^{2I} = 0 $ implies that the subset of $n_H$
hyperinos $\zeta_{\dot I}$ have to be truncated out.
\\
Consistency of the truncation in equation  (\ref{iper2}) implies
\begin{equation}
\label{cons} N^1_{\dot J} \equiv 2 g_{({\bf \L})} \IC_{I \dot J}
{\mathcal U}^{1I}_u k^u_{{\bf \L}} \bar L^{{\bf \L}} = 2 g_{({\bf
\L})}\IC_{I \dot J} {\mathcal U}^{1I}_s k^s_{{\bf \L}} \bar
L^{{\bf \L}} =0 \quad\Rightarrow \quad g_{({\bf \L})} k^s_{{\bf
\L}} \bar L^{{\bf \L}} = 0.
\end{equation}
 The restrictions on the theory imposed by this constraint will be
 discussed  in the subsection 6.4.


\subsection{Further consequences of the gauging}
 The truncation $N=2 \to N=1$
implies, as we have seen in the previous subsections, a number of
consequences that we are now going to discuss, and in particular:
\begin{itemize}
\item{
The $D$-term of the $N=1$-reduced gaugino $\lambda^\L = -2 f^\L_i
\l^{i2} $ is:
\begin{equation}
D^\L = W^{i21} =-2g_{(\L)}({\rm Im}f)^{-1 \L\S} \left(P^3_\S(w^s)
+P^0_\S (z^i)\right)\label{d2}
\end{equation}
}
\item{ The $N=1$-reduced superpotential, that is the gravitino mass, is:
\begin{equation}
L(z,w)=\frac{\rm i}{2}g_{(X)}L^{X} \left( P^1_{X} - {\rm
i}P^2_{X} \right) \label{grashift}
\end{equation}
}
 \item{The fermion shifts of the $N=1$ chiral spinors $\chi^i =\l^{i1}$
coming from the $N=2$  gaugini are:
\begin{equation}
N^i =2  g^{i\bar\jmath}\nabla_{\bar\jmath}\bar L
\label{chir1shift}
\end{equation}
}
\item{The fermion shifts of the $N=1$ chiral spinors $\zeta^s$
coming from $N=2$  hypermultiplets are:
\begin{equation}
N^s =-4 g_{(X)}k^t_{X} \bar L^{X} {\mathcal U}^{1\dot I}_t
{\mathcal U}_{2\dot I}^s .\label{chir2shift}
\end{equation}
}
\end{itemize}
In order for the shifts given in eqs.
(\ref{grashift}), (\ref{chir1shift}), (\ref{chir2shift}) to
define the correct transformation laws of the $N=1$ theory, we
still have to show that the superpotential $L$ is covariantly
holomorphic with respect to the $w^s$ coordinates:
\begin{equation}
\nabla _{\bar s} L =0
\end{equation}
and that the
$N^s$ shift for the chiral multiplets coming from the quaternionic
sector can be written with the standard expression for an $N=1$
chiral multiplets shift, that is as:
\begin{equation}
N^s = 2\,g\,g^{s\bar s} \nabla_{\bar s} \bar L \label{ns}
\end{equation}
These features do indeed follow, as a consequence of the reduction
$SU(2) \to U(1)$ in the holonomy group.
 Indeed:
 \begin{equation} \nabla
_{\bar s} L = \frac {{\rm i}} {2} L^{{\bf \L}} \nabla_{\bar s}
P^x_{{\bf \L}} \left(\s^x \right)_1^{\ 2} = {\rm i} k^t_{X} L^{X}
\Omega^x_{\bar s t} \left(\s^x\right)_1^{\ 2}
 \end{equation}
Now, recalling that:
\begin{equation}
\Omega^x \left(\s^x\right)_1^{\ 2}  = 2 {\mathcal U}_1^\a \wedge
{\mathcal U}^{2\beta} \IC_{\a\b} = 4 {\mathcal U}_1^I \wedge
{\mathcal U}^{2\dot J} \IC_{I\dot J} = 4{\mathcal U}_{1t}^I
{\mathcal U}^{2\dot J}_s \IC_{I\dot J}dn^t \wedge dw^s
\end{equation}
we immediatly get: $\Omega^x _{st} \left(\s^x\right)_1^{\ 2} \neq
0$ while $\Omega^x _{\bar st} \left(\s^x\right)_1^{\ 2} = 0$, so
that $\nabla_{\bar s} L =0 $ follows.
\\
Let us now compute $N^s$ explicitly:
\begin{eqnarray}
N^s &=&  \sqrt{2}\, P^{ s J} \,g\,N^1_J = \,4 \,g_{({\bf \L})}\,
\,\IC_{J\dot J} g^{s \bar s} {\mathcal U}^{ J}_{2\bar
s} {\mathcal U}^{1\dot J}_t k^t_{{\bf \L}} \bar L^{{\bf \L}} \nonumber\\
&=& 4\,\,g_{({\bf \L})}\, \IC_{J\dot J} g^{s \bar s} \frac{\rm
i}{2} \Omega^x
_{t\bar s} (\s^x)^1_{\ 2}k^t_{X} \bar L^{X} \nonumber\\
&=& - {\rm i}\,g_{({\bf \L})}\, g^{s \bar s} \nabla_{\bar s}
P^x_{X} (\s^x)^1_{\ 2}\bar L^{X} = - {\rm i} \,g_{({\bf \L})}\,
g^{s \bar s} \nabla_{\bar s}\left(P^1_{X} + {\rm i} P^2_{X}
\right)\bar
L^{X} \nonumber\\
&=&2  \,\,g_{({\bf \L})}\, g^{s \bar s} \nabla_{\bar s}\bar L ,
\end{eqnarray}
that is it has the right expression for an $N=1$ chiral shift,
according to equation (\ref{defn}). \vskip5mm
\par
Let us now discuss the implications of the gauging constraints
(\ref{first}), (\ref{second}) and (\ref{cons}) on the $N=1$ theory
obtained by reduction, that is the consistency of the truncation
of the second gravitino multiplet $\delta \psi_{\mu 2} =0$ and of
the spinors $\zeta_{\dot I}$ in the hypermultiplets sector for the
gauged theory:
\begin{eqnarray}
\hat \omega_1^{\phantom{1} 2}=0 &\Longrightarrow & g_{({\bf \L})}
A^{{\bf \L}}
\left(P^1_{{\bf \L}} -{\rm i} P^2_{{\bf \L}}\right)= 0 \label{orto1}\\
S_{12}=0 &\Longrightarrow & g_{({\bf \L})} L^{{\bf \L}} P^3_{{\bf
\L}} =0
\label{orto2}\\
\delta \zeta_{\dot I} = 0 & \Longrightarrow & g_{({\bf \L})}
k^s_{{\bf \L}} \bar L^{{\bf \L}} = 0\label{orto3}
\end{eqnarray}
Since the vectors of the $N=2$ theory which are not gauged do not
enter in the previous equations we may limit ourselves to
consider the case where the index ${\bf \L}$ runs over the
adjoint representation of the $N=2$ gauge group. If we call
$G^{(2)}$ the gauge group of the $N=2$ theory and
$G^{(1)}\subseteq G^{(2)}$ the gauge group of the corresponding
$N=1$ theory, then we have that the adjoint representation of
$G^{(2)}$ decomposes as
\begin{equation}
Adj(G^{(2)}) \Rightarrow  Adj(G^{(1)}) + R(G^{(1)}),\label{decrep}
\end{equation}
where $R(G^{(1)})$ denotes some representation of $(G^{(1)})$
(the representation $R(G^{(1)})$ is of course absent for $
G^{(1)} = G^{(2)}$). The gauged vectors of the $N=1$ theory are
restricted to the subset $\{A^{ \L}\}$ generating $Adj(G^{(1)})$
(that is the index ${\bf \L}$ is decomposed as ${\bf \L}\to (\L ,
X )$, with $\L\in Adj(G^{(2)}$ and $X \in
R(G^{(1)})$).\\
This decomposition of the indices is of course the same as the
one used in analyzing the consequences of the constraint
(\ref{gravifot}) in section (6.2). In particular,
 the
graviphoton index ${\bf \L}=0$ always belongs to the set $X$
since the graviphoton $A^0$ is projected out.
\\
 The quaternionic Killing vectors of the $N=2$ theory then decompose
as \begin{equation} k^u_{{\bf \L}} \Rightarrow \{k^s_{\L}, k^{\bar
s}_{\L}, k^t_{\L}, k^s_{X}, k^{\bar s}_{X}, k^t_{X}\}.
\end{equation}
 Obviously, we must have that $k^s_{X} =0$ since  the
Killing vectors of the reduced submanifold have to span the
adjoint representation of $G^{(1)}$. Viceversa, the Killing
vectors with world index in the orthogonal complement, $k^t_{{\bf
\L}}$, must obey $k^t_{\L} =0$, while $k^t_{X}$ are in general
different from zero. Indeed, the isometries generated by
$k^t_{{\bf \L}}$ would not leave invariant the hypersurface
describing the submanifold ${\cal M}^{KH} \subset {\cal M}^Q$.
These properties will be in fact confirmed in Appendix E, by a
careful analysis of the reduction of the
quaternionic Ward identities.\\
 Coming back to the implications of the constraints (\ref{orto1}), (\ref{orto2}),
 (\ref{orto3}), they can be rewritten,
 using the results of section (6.2), as
 follows:
\begin{eqnarray}
 g_{(\L)}
A^{\L}
\left(P^1_{\L} -{\rm i} P^2_{\L}\right)&=& 0 \label{porto1}\\
 g_{(X)} L^{X}
P^3_{X} &=&0
\label{porto2}\\
g_{(X )} k^s_{X} \bar L^{X} = 0\label{porto3}
\end{eqnarray}
Since we have found that $k^s_{X}=0$, equation (\ref{porto3}) is
identically satisfied. \\
Eq.s (\ref{porto1}) and (\ref{porto2}) are  satisfied by
requiring:
\begin{equation}\label{portati}
  P^1_\L = P^2_\L =0 \, ; \quad  P^3_{X}=0
\end{equation}
Then the superpotential of the theory is given by \cite{ps} -
\cite{tatar}:
\begin{equation}
L =\frac{\rm i}2 L^{X}(z,\bar z) \left(P^1_{X}(w,\bar w ) - {\rm
i} P^2_{X}(w,\bar w )\right). \label{supfin}
\end{equation}
 We are left with an $N=1$ theory
coupled to $n'_V$ vector multiplets ($ \L =1,\cdots ,n'_V$) and
$n_C +n_H$ chiral multiplets ($X =0,1,\cdots ,n_C$) with
superpotential (\ref{supfin}). All the isometries of the scalar
manifolds are in principle gauged since the D-term of the reduced
$N=1$ theory depends on $P^0_\L (z,\bar z )+ P^3_\L (w,\bar w )$.
\par
In the particular case where the gauge group $G^{(1)}$ of the
$N=1$ reduced theory is the same as the gauge group $G^{(2)}$ of
the $N=2$ parent theory, the index $X$ takes only the value zero
and all the scalars are truncated out $(L^\L=0,L^{\dot 0}=1)$.
The vectors $A^\L$ are retained in the truncation while $A^0$ is
projected out.
 In this case the superpotential reduces to:
 \begin{equation} L =\frac{\rm i}2 L^0
\left(P^1_0 - {\rm i} P^2_0\right). \label{sup1}
\end{equation}
Moreover, from equation  (\ref{p02}) we have that in this case the
prepotential $P^0_\L =0$, and the D-term depends only on $P^3_\L
(w,\bar w)$.
 We
then have an $N=1$ theory coupled to $n_V$ vector multiplets and
$n_H$ chiral multiplets, with gauged isometries and
superpotential (\ref{sup1}). \\
 Note that when $P^1_0 - {\rm i}
P^2_0$ is constant, (\ref{sup1}) gives a constant F-term. This
case can only be obtained in absence of hypermultiplets. Indeed,
from the general quaternionic formula \cite{cere}
\begin{equation}
n_H P^x_{{\bf \L}} = -\half \Omega^x_{uv} \nabla^u k^v_{{\bf \L}}
\label{eq}
\end{equation}
we see that if $n_H \neq 0$ a Fayet-Iliopoulos term, as well as a
constant F-term, is excluded \cite{cere}, since a constant $P^x_0$
is not compatible with the covariance of the r.h.s. under $SU(2)$
and the gauge group. Even when the theory is ungauged ($k^u_{{\bf
\L}}=0$) a constant $P^x_0$ is still excluded for $n_H\neq 0$,
since in this case equation  (\ref{eq}) reduces to $n_H
P^x_{{\bf \L}} =0$, implying $P^x_{{\bf \L}} =0$.\\
 If $n_H =0$, then a constant $P^x_{\bf \L}$ is possible ($N=2$ Fayet--Iliopoulos term) \cite{fevp}
 \footnote{An $N=2$ Fayet--Iliopoulos term coming from $P^0_\L$ is
 excluded by the Ward identity (\ref{poiss}).},
 provided the gauge group is abelian (otherwise it breaks the
gauge group) and provided it satisfies the identity
\begin{equation}\label{ward}
  \epsilon^{xyz} P^y_{{\bf \L} }
  P^z_{\bf \S } =0
\end{equation}
which follows from the general quaternionic Ward identity
\cite{abcdffm}, \cite{gali}
\begin{equation}
 \frac {1} { \l} \Omega^x_{uv} k^u_{\bf\L } k^v_{\bf\S } +
\half \epsilon^{xyz}P^y_{\bf\L }
  P^z_{\bf\S } -\half f_{{\bf\L}\bf\S}^{\phantom{{\bf\L}{\bf\S}}\bf\G} P^x_{\bf\G} =0
  \label{wardquat}
\end{equation}
in absence of hypermultiplets.
\\
When we reduce the theory to $N=1$, a constant value of $P^{\rm
i}_{X} \equiv \xi^{\rm i}_{X}\neq 0$ (${\rm i} =1,2$) or $P^3_{
\L} \equiv \xi^3_{ \L}\neq 0$ are both compatible with all the
constraints (\ref{porto1})-(\ref{porto3}); in particular $L^{{\bf
\L}} \xi^3_{{\bf \L}}=0$ and $A^{{\bf \L}} \xi^{\rm i} _{{\bf
\L}}=0$ implying the presence of a $N=1$ Fayet-Iliopoulos term
corresponding to $\xi^3_{ \L}$, or a constant F-term
corresponding to $\xi^{\rm i}_{X}$.

\subsection{$N=2 \to N=1$  scalar potential}
Let us now compute explicitely the reduction of the scalar
potential of the $N=2$ theory down to $N=1$. The $N=2$ scalar
potential is given by:
\begin{equation}\label{pot2}
  \cV^{N=2}
 =\left( g_{{\mathcal I}\bar{\mathcal J}}k^{{\mathcal I}}_{{\bf \L}} k^{\bar{\mathcal J}}_{{\bf \S}} +
 4h_{uv}k^u_{{\bf \L}}
 k^v_{{\bf \S}}\right)\bar L^{{\bf \L}} L^{{\bf \S}} + \left(- \frac {1} {2} ({\rm
 Im}{\mathcal N^{-1})}^{{\bf \L}{\bf \S}} -\bar L^{{\bf \L}} L^{{\bf \S}}\right)
 P^x_{{\bf \L}}
 P^x_{{\bf \S}}
-3P^x_{{\bf \L}} P^x_{{\bf \S}} {\bar L}^{{\bf \L}} L^{{\bf \S}}
\end{equation}
while the $N=1$ scalar potential can be written in terms of the
covariantly holomorphic superpotential $L$ as:
\begin{equation}\label{pot1}
  \cV^{N=1} = 4\left(\nabla_\ell L \nabla_{\bar\ell}\bar L g^{\ell\bar\ell}
  -3 |L|^2 +\frac{1}{16} {\rm Im}f_{\L\S} D^\L D^\S \right),
\end{equation}
where the holomorphic index $\ell$ runs over all the scalars of
the theory.
\\
 Before performing the reduction it is instructive to
work out in detail the supersymmetry Ward identity involving the
scalar potential \cite{maiafe}, \cite{cegipo}:
\begin{equation}\label{ward2}
  \delta^A_B \cV^{N=2} = -12 \bar S^{AC} S_{CB} + g_{{\mathcal I}\bar{\mathcal J}}
  W^{{\mathcal I} AC} W^{\bar{\mathcal J}}_{BC} + 2 N^A_\a N^\a_B.
\end{equation}
Instead of taking the trace of (\ref{ward2}) on the $SU(2)$
indices $A,B$, thus recovering the potential (\ref{pot2}), one can
alternatively write down the stronger relations:
\begin{equation}\label{pot2bis}
   \delta^1_1  \cV^{N=2} = \cV^{N=2} =
    -12 \bar S^{1C} S_{C1} + g_{{\mathcal I}\bar{\mathcal J}} W^{{\mathcal I} 1C}
  W^{\bar{\mathcal J}}_{1C} + 2 N^1_\a N^\a_1
\end{equation}
\begin{equation}\label{pot2ter}
   \delta^2_2  \cV^{N=2} = \cV^{N=2} =
    -12 \bar S^{2C} S_{2C} + g_{{\mathcal I}\bar{\mathcal J}} W^{{\mathcal I}2C}
  W^{\bar{\mathcal J}}_{C2} + 2 N^2_\a N^\a_2.
\end{equation}
and furthermore:
\begin{equation}\label{zero'}
\delta^2_1 \cV^{N=2} =0= -12 \bar S^{2C} S_{1C} + g_{{\mathcal
I}\bar{\mathcal J}} W^{iC2}
  W^{\bar{\mathcal J}}_{C1} + 2 N^2_\a N^\a_1.
\end{equation}
 When we pass to the truncated theory, the matrix
$S_{AB}$ becomes diagonal ($S_{12}\sim P^3_\L \bar L^\L=0$) and
its eigenvalues are the masses of the 2 gravitini:
\begin{equation}\label{massgrav}
 S_{AB} = \left(
  \begin{array}{cc}
    L & 0 \\
 0& \tilde L
  \end{array}\right).
\end{equation}
where:
\begin{eqnarray}\label{L}
  L&=&\frac {{\rm i}} {2}L^{X}(P^1_{X}-{\rm i}P^2_{X})\\
 \tilde L&=&\frac {{\rm i}} {2}L^{X}(-P^1_{X}-{\rm i}P^2_{X})
\end{eqnarray}
so that:
\begin{eqnarray}\label{massgrav'}
  |L|^2& =& S_{11}S^{11}=\frac {1} {4} L^{X}\bar L^{Y} \left[ P^x_{X} P^x_{Y} +{\rm
  i} \left(P^1_{X} P^2_{Y} - P^2_{X} P^1_{Y}\right)\right]\\
   |\tilde L|^2& =& S_{22}S^{22}=\frac {1} {4} L^{X}\bar L^{Y} \left[ P^x_{X} P^x_{Y} -{\rm
  i} \left(P^1_{X} P^2_{Y} - P^2_{X} P^1_{Y}\right)\right].
\end{eqnarray}
The difference between the 2 gravitino mass eigenvalues can be
written
 in terms of the fermionic shifts as:
\begin{eqnarray}
|L|^2 - |\tilde L|^2 &=&  \frac{\rm
  i}{2} \left(P^1_{X} P^2_{Y} - P^2_{X} P^1_{Y}\right)= \bar S^{1C} S_{C1}-
 \bar S^{2C} S_{C2}\nonumber\\
& =&\frac{1}{12}\left(  g_{i\bar \jmath} W^{i1C}
  W^{\bar\jmath}_{1C} + 2 N^1_\a N^\a_1 -  g_{i\bar \jmath} W^{i2C}
  W^{\bar\jmath}_{2C} - 2 N^2_\a N^\a_2\right) .
\end{eqnarray}
 \par
 Let us now perform the reduction.
 Using,  e.g.,
equation  (\ref{pot2bis}) and recalling that $S_{12}=0$  and
 $N^1_{\dot I} =0$ (see equation (\ref{cons})), we find  :
\begin{equation}\label{potenziale}
{\cal V}^{N=2 \to N=1} =-12 \bar S^{11} S_{11} + g_{{\mathcal I}
\bar{\mathcal J}} \left(W^{{\mathcal I} 11} W^{\bar{\mathcal
J}}_{11}+ W^{{\mathcal I} 12} W^{\bar{\mathcal J}}_{12}\right) +
2N^1_I N^I_1
\end{equation}
Using equations (\ref{w11}), (\ref{w12}), the first two terms of
equation  (\ref{potenziale}) give:
\begin{equation}
\label{SS}
 -12 \bar S^{11}
S_{11} = -3 P^{\rm i}_{X} P^{\rm i}_{Y} L^{X} \bar L^{Y} +3 {\rm
i} \left(P^2_{X} P^1_{Y} - P^1_{X} P^2_{Y} \right)L^{X }\bar
L^{Y} = -12 L \bar L
\end{equation}
\begin{equation}\label{WW}
 g_{{\mathcal I} \bar{\mathcal J}}
W^{{\mathcal I} 11} W^{\bar{\mathcal J}}_{11} = \left(P^1_{X}+{\rm
i}P^2_{X}\right)\left(P^1_{Y}-{\rm i}P^2_{Y}\right) U^{XY} =
4g^{k\bar l}\nabla_k L \nabla_{\bar l}\bar L
\end{equation}
For the term $ g_{{\mathcal I} \bar{\mathcal J}} W^{{\mathcal I}
21} W^{\bar{\mathcal J}}_{21}$ we obtain:
\begin{equation}
g_{{\mathcal I} \bar{\mathcal J}} W^{{\mathcal I} 21}
W^{\bar{\mathcal J}}_{21} = -2 {\rm Im}\cN_{ {\bf \L}{\bf \S}}
f^{{\bf \L}}_{{\mathcal I}} f^{{\bf \S}}_{ \bar{\mathcal
J}}W^{{\mathcal I} 21} W^{\bar{\mathcal J}}_{21}= -\half {\rm
Im}\cN _{ \L \S}D^\L D^\S = \frac14 {\rm Im} f_{\L\S} D^\L D^\S
\end{equation}
where we have reduced the indices according to the results of
subsection (6.3) and used equations  (\ref{dtrm1}), (\ref{holn}),
(\ref{norma}).
\par
To compute the last term in equation (\ref{potenziale}) we use
equation (\ref{defn1}) and (\ref{ns}) and we find
\begin{equation}\label{NN}
2 N^1_I N^I_1 = g_{s \bar s}N^s \bar N^{\bar s}\,=\,4g^{s\bar
s}\nabla_s L \nabla_{\bar s}\bar L
\end{equation}
\par
Collecting all the terms we find that the reduction of the $N=2$
scalar potential gives:
\begin{equation}
\label{21pot} {\cal V}^{N=2 \to N=1} = 4 \left[ -3 L \bar L +
g^{i\bar\jmath} \nabla_i L \nabla_{\bar\jmath} \bar L + g^{s\bar
s} \nabla_s L \nabla_{\bar s} \bar L + \frac{1}{16} {\rm Im}
f_{\L\S} D^\L D^\S \right]\end{equation} which coincides with the
scalar potential (\ref{pot1}) of the $N=1$ theory, where we have
decomposed  the indices according to the fact that the $\s$-model
is a product manifold .
\par
We note that our computation of the reduction of the scalar
potential has been  performed by first reducing the $N=2$
fermionic shifts to $N=1$  and then computing the potential. Of
course, we could also have performed the computation by directly
computing the reduction of each term of the $N=2$ potential. In
the latter case, to obtain the desired results requires some non
trivial computations. In particular, there are some subtleties
related to the observation that the $N=2$ potential does not
contain ``interference'' contributions of the form $P^0_{{\bf
\L}}P^x_{{\bf \S}}$ or $P^x_{[{\bf \L}}P^y_{{\bf \S} ]}$, while
such terms are instead present in the $N=1$ potential, given the
form (\ref{d2}) of the D-term and (\ref{grashift}) of the
superpotential. To solve the puzzle and recover the precise
correspondence between the $N=2$ and $N=1$ theories, one has to
use several times the reduced forms of the Ward identities of
quaternionic and special-K\"ahler geometries, the definition of
the quaternionic Killing vectors \cite{df}, \cite{cere},
\cite{alevan} and the expression that the special geometry
prepotential gets in the reduction, equation  (\ref{p02}). The
explicit computation is given in Appendix F.

\subsection{Examples of truncation to $N=1$ gauged supergravity}
As an application of the
formalism developed in this section, we can now consider reduction
on $N=8$ to $N=1$ or in general of $N=2$ theories down to $N=1$.
\par
The simplest case is to consider $N=2$ special-K\"ahler manifolds
which are also $N=1$ Hodge-K\"ahler, or submanifolds of half the
dimension of quaternionic manifolds which are ``dual'' (under
c-map) to special-K\"ahler.
\par
We first consider ``dual quaternionic manifolds'' which are
symmetric spaces; they were all given in Table 4 of ref.
\cite{cfg}. This immediately gives the $N=2 \to N=1$ reduction of
theories with only hypermultiplets as follows:
\begin{table}
  \centering
  \caption{$N=2 \to N=1$}\label{21}
  \begin{tabular}{|c|c|}
    \hline
    $N=2$ ($n_V =0$ , ${\cal M}_Q$ (dim$_Q = n$) & $N=1$ ($n_V =0, {\cal M}_{KH}$ (dim$_C = n$) \\
    \hline
    $\frac{U(2,n+1)}{U(2)\times U(n+1)} $& $\frac{SU(1,1)}{U(1)}\times
    \frac{U(1,n)}{U(1)\times U(n)} $ \\
    $\frac{SO(4,n+1)}{SO(4)\times SO(n+1)}\quad (n \geq 2) $ &
    $\frac{SU(1,1)}{U(1)}\times \frac{SU(1,1)}{U(1)}\times
    \frac{SO(2,n-1)}{SO(2)\times SO(n-1)} $ \\
     $\frac{G_{2(2)}}{SO(4)} $& $\frac{SU(1,1)}{U(1)}\times
    \frac{SU(1,1)}{U(1)} $ \\
   $\frac{F_{4(4)}}{USp(6)\times USp(2)} $& $\frac{SU(1,1)}{U(1)}\times
    \frac{Sp(6,\IR)}{U(3)} $ \\
    $\frac{E_{6(2)}}{SU(6)\times SU(2)} $& $\frac{SU(1,1)}{U(1)}\times
    \frac{SU(3,3)}{SU(3)\times SU(3)\times U(1)} $ \\
    $\frac{E_{7(-8)}}{SO(12)\times SU(2)} $& $\frac{SU(1,1)}{U(1)}\times
    \frac{SO^*(12)}{U(6)} $ \\
   $\frac{E_{8(-24)}}{E_7\times SU(2)} $& $\frac{SU(1,1)}{U(1)}\times
    \frac{E_{7(-26)}}{E_6\times SO(2)} $ \\ \hline
  \end{tabular}
\end{table}
It is interesting to note that if ${\cal M}_Q = \frac{G_Q}{H_Q}$,
${\cal M}_{SK} = \frac{SU(1,1)\times G}{U(1) \times H}$ then $H_Q
= SU(2) \times G_c $, where $G_c$ is the compact form of $G$!.\\
From the previous table we can immediately obtain $N=1$
truncations of $N=8$ supergravity with $(n_V,n_H)$ replaced by
$(n_V^{(N=1)}, n_C = n_V + n_Q)$.
\\
In all these models (unless $n_Q =0$) the K\"ahler-Hodge manifold
will be of the form
\begin{equation}
SK(n_V) \times SK(n_Q -1) \times \frac{SU(1,1)}{U(1)}.
\end{equation}
\par
As a simple example, motivated by string construction
\cite{fgkp2}, for the application of the results of the previous
sections, we consider a $N=4$, $D=4$ matter coupled supergravity
with gauge group $SO(n)$ ($n$ even). The $\s$-model of the
scalars in presence of gauging is  given by:
\begin{equation}
\frac{SU(1,1)}{U(1)}\times \frac{SO(6,\frac{n(n-1)}{2})}{SO(6)
\times SO(\frac{n(n-1)}{2}) },
\end{equation}
and the content of the scalar sector can be encoded in the
vielbein 1-form $P_{AB I}$ where the antisymmetric couple $AB$
labels the irrep.  ${\bf 6} \in SU(4)$ and $I$ labels the
fundamental representation of $SO(\frac{n(n-1)}{2})$.
\\
 This $N=4$ theory is reduced to $N=2$ through the
action of a $\ZZ_2$ group and to $N=1$ by the action of $\ZZ_2
\times \ZZ '_2$.
  The generators of $\ZZ_2 \times \ZZ
'_2$ in the R-symmetry group $SU(4)$ are given by:
\begin{equation}
 \a = \pmatrix{
   1 & 0&  0 & 0  \cr
   0& 1 & 0 & 0 \cr
    0 & 0 & e^{{\rm i} \pi}  & 0 \cr
   0 & 0& 0& e^{{\rm i} \pi} \cr}
   \, ; \quad \b = \pmatrix{1 & 0&  0 & 0  \cr
   0& e^{{\rm i} \pi}  & 0 & 0 \cr
    0 & 0 & 1 & 0 \cr
   0 & 0& 0& e^{{\rm i} \pi} \cr}
\end{equation}
so that two gravitinos are singlets with respect to $\ZZ_2$ and
one gravitino is invariant with respect to $\ZZ_2 \times \ZZ '_2$.
\par
To obtain charged matter in the $N=4 \to N=2$  reduction, we
implement the action of $\ZZ_2$ on the gauge group. Let us make
the following decomposition
\begin{equation}
  SO(n) \stackrel{\ZZ_2}   \longrightarrow  SO(n_A ) \times
  SO(n_B)
  \end{equation}
  so that,
  under the action of $\ZZ_2$:
\begin{eqnarray}
  n_A  &  \Rightarrow &n_A \nonumber\\
 n_B  & \Rightarrow& \alpha n_B
 \end{eqnarray}
 and then
 \begin{eqnarray}
Adj(SO(n_A))&\stackrel{\ZZ_2}   \longrightarrow & Adj(SO(n_A))\nonumber\\
Adj(SO(n_B))&\stackrel{\ZZ_2}   \longrightarrow & Adj(SO(n_B))\nonumber\\
 (n_A,n_B) &\stackrel{\ZZ_2}   \longrightarrow & \a (n_A , n_B ) .
 \end{eqnarray}
Correspondingly, for the group $SU(4)$ we have:
\begin{eqnarray}
{\bf 4}& \stackrel{\ZZ_2}   \longrightarrow& \alpha {\bf 4}\nonumber\\
 {\bf 2}_1& \stackrel{\ZZ_2}   \longrightarrow& {\bf 2}_1
\end{eqnarray}
The scalars transforming non trivially under $\ZZ_2$ are
projected out and we are left with the coset manifold:
\begin{equation}
\frac{SU(1,1)}{U(1)}\times
\frac{SO(2,\frac{n_A(n_A-1)}{2}+\frac{n_B(n_B-1)}{2} )}{SO(2)
\times SO(\frac{n_A(n_A-1)}{2}+\frac{n_B(n_B-1)}{2}) }\times
\frac{SO(4,n_A n_B )}{SO(4) \times SO(n_An_B) } \label{cinquina}
\end{equation}
where the first two factors define an $N=2$ special-K\"ahler
manifold and the last factor is a quaternionic manifold.
\par
  In order to obtain an $N=1$ supergravity theory, the gauge groups $SO(n_A )$
and $SO(n_B)$ are  further decomposed as follows:
\begin{eqnarray}
 SO(n_A) &\to& SO(n_1 ) \times SO(n_2) \\
SO(n_B) &\to& SO(n_3 ) \times SO(n_4)
\end{eqnarray}
and we define the action of  $\ZZ '_2$ as:
\begin{eqnarray}
n_1 \Rightarrow n_1 \, ,& \quad & n_2 \Rightarrow \b n_2 \nonumber\\
n_3 \Rightarrow n_3 \, ,& \quad & n_4 \Rightarrow \b n_4 .
\end{eqnarray}
 \\
This induces an action of $\ZZ_2 \times \ZZ '_2$ on the
decomposition of the gauge group:
\begin{eqnarray}
Adj(SO(n)) &\stackrel{\ZZ_2}   \longrightarrow & Adj(SO(n_A )) +
Adj(SO(n_B)) + (n_A , n_B )_\a
\nonumber\\
&\stackrel{\ZZ_2 \times \ZZ '_2}   \longrightarrow & Adj(SO(n_1
))_1 + Adj(SO(n_2))_1  + Adj(SO(n_3 ))_1 +
Adj(SO(n_4))_1  + \nonumber\\
&&+ (n_1 , n_2,1,1)_\b+ (1,1, n_3, n_4)_\b + (n_1,1,  n_3,1)_\a
+\nonumber\\
&&+ (n_1, 1,1, n_4)_{\a\b} + ( 1, n_2, n_3,1 )_{\a\b}  + (1,
n_2,1, n_4)_\a \label{casino}
\end{eqnarray}
In equation  (\ref{casino}) we have labelled each representation
with indices $1,\a , \b , \a\b $ whose meaning is that the
corresponding representation  is invariant or transforms under
$\a$, $\b$ or $\a\b$ respectively. That is the representations
$Adj(SO(n_I))$ are invariant under $\ZZ_2\times \ZZ '_2$, while
the remaining bifundamental representations $(n_I , n_J)$
transform as follows:
\begin{eqnarray}
(n_1 , n_3 ) ; (n_2 , n_4) & \mbox{transform under} & \a
\nonumber\\
(n_1 , n_2 ) ; (n_3 , n_4) & \mbox{transform under} & \b
\nonumber\\
(n_2 , n_3 ) ; (n_1 , n_4) & \mbox{transform under} & \a\b
\end{eqnarray}
With the same notation, let us now consider the $\ZZ_2\times \ZZ
'_2$ action on the ${\bf 6}$ of $SU(4)$:
\begin{eqnarray}
 {\bf 6} &\stackrel{\ZZ_2  }   \longrightarrow & 4_\a + 2_1
 \nonumber\\
 &\stackrel{\ZZ_2 \times \ZZ '_2}   \longrightarrow & (2_\a
 + 2_{\a\b}) + 2_\b
\end{eqnarray}
Joining the information coming from the the decomposition of
$SU(4)$ and $SO(\frac{n(n-1)}{2})$ we see that the scalars which
remain invariant under the action of $\ZZ_2 \times \ZZ '_2$ are
given by the  vielbein in the following representations:  $P_{2_\a
(n_1 , n_3 )}$;$P_{2_\a (n_2 , n_4 )}$; $P_{2_\b (n_1 , n_2
)}$;$P_{2_\b (n_3 , n_4 )}$; $P_{2_{\a\b} (n_1 , n_4
)}$;$P_{2_{\a\b} (n_2 , n_3 )}$. This means that the
special-K\"ahler manifold reduces to:
\begin{equation}
\frac{SU(1,1)}{U(1)}\times
\frac{SO(2,\frac{n_A(n_A-1)}{2}+\frac{n_B(n_B-1)}{2} )}{SO(2)
\times SO(\frac{n_A(n_A-1)}{2}+\frac{n_B(n_B-1)}{2}) } \to
\frac{SU(1,1)}{U(1)}\times \frac{SO(2,n_1 n_2 + n_3 n_4)}{SO(2)
\times SO(n_1 n_2 + n_3 n_4)
 }.\label{speciale}
\end{equation}
while the quaternionic manifold splits as
follows:
\begin{equation}
\frac{SO(4,n_A n_B )}{SO(4) \times SO(n_An_B) } \to
\frac{SO(2,n_1 n_3 + n_2 n_4)}{SO(2) \times SO(n_1 n_3 + n_2 n_4)
 }\times \frac{SO(2,n_1 n_4 + n_2 n_3)}{SO(2) \times SO(n_1 n_4 + n_2 n_3)
  }.\label{quaterna}
\end{equation}
\par
Let us now comment this result.
\\
From the analysis in section (6,2) we have learnt that when we
reduce a gauged $N=2$ theory to $N=1$ (with $G^{(2)} \to G^{(1)}$)
the surviving scalars from the vector multiplets sector are those
which are in the representation  $R(G^{(1)})$ according to
equation (\ref{decrep}), while all the scalars in the adjoint
representation of $G^{(1)}$ are truncated out. Precisely this
happens in our case. Indeed, from equation  (\ref{speciale}) the
irreps $(n_1, n_2)$ and $( n_3, n_4)$ belong to the left over
representations in equation (\ref{casino}).
 Furthermore, all the other bifundamental
rep. belong to the scalars  coming from the quaternionic sector,
according to equation  (\ref{quaterna}). Note that the total
dimensional of the product manifold of equation  (\ref{quaterna})
is exactly half the dimension of the parent quaternionic manifold,
according to the general result found in section (6.1). It is
interesting to observe that the same kind of result appears in the
decomposition $N=4 \to N=2$ described by equation
(\ref{cinquina}). In fact, the reduced product manifold in
equation (\ref{cinquina}) has a $\s$-model whose scalars belong
again to the representation $R=(n_A,n_B)$ left over in the
reduction of the adjoint representation  of the $N=4$ gauge group.

Other examples can be obtained \cite{fp} from heterotic strings
compactified on $\ZZ_N$ orbifolds with reduced non abelian gauge
group $E_6$.
\par
We finally observe that the $N=2$ special-K\"ahler manifold in the
l.h.s. of (\ref{speciale}) can be parametrized with the symplectic
section $(L^{{\bf \L}}, M_{{\bf \L}} = \eta_{{\bf \L}{\bf \S}} S
L^{{\bf \S}})$ (with $L^{{\bf \L}}L^{{\bf \S}}\eta_{{\bf \L}{\bf
\S}} =0$ and $\eta_{{\bf \L}{\bf \S}} =(1,1,-1,\cdots ,-1)$)
where a prepotential $F$ does not exist \cite{cdfv}. In this case
the $N=2$ vector kinetic matrix has the form:
\begin{equation}
\cN_{{\bf \L}{\bf \S}} = (S-\bar S) \left(\Phi_{{\bf \L}}\bar
\Phi_{{\bf \S}}+\bar\Phi_{{\bf \L}}\Phi_{{\bf \S}}\right) + \bar S
\eta_{{\bf \L}{\bf \S}} \, ; \quad \Phi_{{\bf \L}} \equiv
\frac{L^{{\bf \L}}}{\sqrt{L^{{\bf \L}}\bar L_{{\bf \L}}}}.
\end{equation}
When we perform the truncation to $N=1$, the sections $L^\L$
become zero, and the $N=1$ vector kinetic matrix takes the form:
\begin{equation}
\cN_{ \L \S} =   \bar S \eta_{ \L\S},
\end{equation}
that is it becomes antiholomorphic in the complex scalar $S$
parametrizing the manifold $\frac{SU(1,1)}{U(1)}$.

\section*{Acknowledgements}
We acknowledge interesting discussions with I. Antoniadis, A.
Ceresole, G. Dall'Agata, M.A. Lled\'o, J. Louis, P. Mayr, E.
Sokatchev  and R. Varadarajan. One of the authors (S.F.) warmly
thanks the Department of Physics of the Politecnico di Torino,
where part of this work has been done.
\\
Work supported in part by the European Community's Human
Potential Programme under contract HPRN-CT-2000-00131 Quantum
Spacetime, in which L. Andrianopoli and R. D'Auria are associated
to Torino University.
 The work of S. Ferrara has been supported in part by the D.O.E. grant
DE-FG03-91ER40662, Task C.

\appendix
\section*{Appendix A: Supersymmetry
re\-duc\-tion from su\-per\-space Bianchi
identities} \setcounter{equation}{0} \addtocounter{section}{1}
 We want now to show that the constraints found at the
level of supersymmetry transformation laws are actually
sufficient to guarantee the closure of the supersymmetry algebra
of the reduced
theory.\\
We prove this statement by considering the reduction of the
superspace Bianchi identities of the $N=8$ theory (which, as is
well known, is equivalent to the ``on-shell'' closure of the
supersymmetry algebra). The $N=8$ Bianchi identities are
\cite{cadafre},\cite{adf} (we omit the wedge product symbols
among the products of forms):
\begin{eqnarray}
&& R^{pq} \wedge V_q + {\rm i} \bar \psi _A \g^p \rho^A - {\rm i}
\bar \rho_A \g^p \psi^A =0\\
&& \nabla \rho_A + \frac 14 R^{pq} \g_{pq} \psi_A
-R_A^{\phantom{A}B} \psi_B =0 \\
&& \nabla F^{\L\S} -2 f^{\L\S}_{AB} \bar\rho^A \psi^B -2 \bar
f^{\L\S AB} \bar \rho_A\psi_B + \nonumber\\
&&- \half \bar f^{\L\S CD} P_{ABCD} \bar\psi^A \psi^B  - \half
f^{\L\S}_
{CD} \bar P^{ABCD} \bar\psi_A \psi_B =0 \\
&& \nabla \left(\nabla \chi_{ABC}\right) - 3
R_{[A}^{\phantom{[A}L}\chi_{BC]L} + \frac 14 R^{pq} \g_{pq}
\chi_{ABC} =0 \\
&& \nabla P_{ABCD} =0
\end{eqnarray}
in terms of the supercovariant field-strengths:
\begin{eqnarray}
&T^{p}&\equiv \mathcal{D}V^{p}-\frac{i}{2}\ \
\overline{\psi}_{A}\gamma^p\psi^A =0\\
&R^{pq}&\equiv d \omega^{pq} - \omega^p_{\ r} \omega^{rq}\\
&F^{\L \S}&\equiv dA^{\L \S}+f^{\L \S}_{AB}
\overline{\psi}^{A}\psi^{B}+ \bar f^{\L \S |AB}\, \bar
{\psi}_{A}\psi_{B}\\
&\rho_{A}&\equiv \mathcal{D}\psi_{A}+\omega_A^{\phantom{A}B} \psi_B\\
&\nabla\chi_{ABC}&\equiv \mathcal{D}\chi_{ABC}+3\omega_{[A}^{\phantom{[A}L} \chi_{BC]L}\\
&R_A^{\phantom{A}B} \equiv & d\omega_A^{\phantom{A}B} +
\omega_A^{\phantom{A}C}  \omega_C^{\phantom{A}B}.
\end{eqnarray}
Note that all the fields are actually superfield 1-forms whose
restriction at $\theta =d\theta =0$ gives the ordinary
space--time fields.\\
To show how the Bianchi identities of the $N=8$ theory reduce to
the Bianchi identities of the $N=N'$ theory, we just work out the
example of the $N=8 \to N=6$ reduction. The other cases can be
analyzed in analogous way.\\
First of all we see that, by decomposing the R-symmetry indices
as in section 2  and setting $\psi_i =0$ ($i=7,8$), the
supercovariant field-strengths get reduced as follows: the
superspace bosonic vielbein $V^p$ and the spin connection
$\omega^{pq}$ ($p,q$ denote space--time flat indices) remain
untouched by the reduction, and the same happens of course for
the Lorentz curvature $R^{pq}$ and supertorsion $T^p$. \\As far
as the gravitinos are concerned, we find:
\begin{eqnarray}
&& \rho_{a}\equiv \mathcal{D}\psi_{a}+\omega_a^{\phantom{a}b} \psi_b \\
&& 0=\rho_{i} =\omega_i^{\phantom{i}a} \psi_a
\end{eqnarray}
which implies $\omega_i^{\phantom{i}a}=0$, consistently with what
we found in section 4. As a consequenc, the gravitinos Bianchi
identities reduce to:
\begin{equation}
\nabla \rho_{a}+\frac 14 R^{pq} \g_{pq}\rho_a + R_a^{\phantom{a}b}
\psi_b =0
\end{equation}
which is the correct Bianchi identity for an $N=6$ gravitino,
while consistency of the truncation implies:
\begin{equation}
\nabla \rho_{i}= R_i^{\phantom{i}a} \psi_a =0 \, \quad \to
R_i^{\phantom{i}a} =0
\end{equation}
again in agreement with the $\sigma$-model results.\\
Let us analyze the spin one-half sector. It gives
\begin{eqnarray}
\nabla \chi_{abc} &=& {\cal D} \chi_{abc} +
3\omega_{[a}^{\phantom{[a}d}\chi_{bc]d}+
\omega_{[a}^{\phantom{[a}i}\chi_{bc]i} \\
\nabla \chi_{abi} &=& {\cal D} \chi_{abi} -
2\omega_{[a}^{\phantom{[a}d}\chi_{b]di}
+2\omega_{[a}^{\phantom{[a}j}\chi_{b]ij}+
\omega_{i}^{\phantom{i}d}\chi_{dab}
+\omega_{i}^{\phantom{i}j}\chi_{jab}\\
\nabla \chi_{aij} &=& {\cal D} \chi_{aij} -
\omega_{a}^{\phantom{a}d}\chi_{dij} +
2\omega_{[i}^{\phantom{i}d}\chi_{j]ad}
+\omega_{i}^{\phantom{i}dj}\chi_{jab}.
\end{eqnarray}
Since $\omega_{a}^{\phantom{[a}i}=0$, we see that the last
equation  is satisfied only setting $\chi_{iab} =0$, (as already
known from section 2, since they belong to the gravitino
multiplets truncated out). What is left is the spin one-half
sector of the $N=6$ theory. It is now straightforward to see that
the Bianchi identities for $\chi_{abc}$ and $\chi_{aij}$ reduce,
after imposing again the constraint $\omega_{a}^{\phantom{a}i}=0$,
to the corresponding $N=6$ Bianchi identities, while the Bianchi
identity for $\chi_{abi}$
is, consistently, identically satisfied.\\
The analysis of the scalar sector $P_{ABCD}$ and its Bianchi
identity  is identical to what has been already discussed in
section 4, and does not deserve further analysis.\\
Finally, the Bianchi identity for the vector field strengths,
with $\psi_i = \rho_i =0$, reduces to:
\begin{equation}
\nabla F^{\L\S} = -2 f^{\L\S}_{ab} \bar\rho^a \psi^b - \half \bar
f^{\L\S |cd} P_{abcd} \bar\psi^a\psi^b - \half \bar f^{\L\S |ij}
P_{abij} \bar\psi^a\psi^b - \half \bar f^{\L\S |ci} P_{abci}
\bar\psi^a\psi^b.
\end{equation}
Here, the scalar vielbein $P_{abci}=0$ according to the discussion
of section 2 and 4. Furthermore, the reduction of the couple of
indices $\L\S$ goes according to what we have discussed in section
5. Since the duality group acts now on the electric and magnetic
field-strengths in the representation  ${\bf 32}$ of $SO^*(12)$,
we simply substitute the couple $\L\S$ with an index $r$ running
from 1 to 16. Note that the corresponding quantities
$f^r_{ab},f^r_{ij} $ are $16 \times 16$ sub-blocks of the $32
\times 32$ matrix $U$, which has exactly the same form of equation
(\ref{u}), but valued in $Sp(32,\IR )$, which gives the embedded
coset representative.

\section*{Appendix B: Consistency of the Bianchi identities
for $N=2 \to N=1$ gauged theory in $D=4$.}
\setcounter{equation}{0} \addtocounter{section}{1}
 In the same spirit of the analysis of section (5.1), it is easy
 to show that
 the closure of Bianchi identities of the $N=2$ theory
implies the consistent closure of the reduced $N=1$ theory.\\
The definition of the supercurvatures and superspace Bianchi
identities for the $N=2$ theory have been given in ref
\cite{abcdffm} (Appendix A).\\
We have to reduce these objects to their $N=1$ expressions, and to
show that they coincide with the definitions of the
supercurvatures and superspace Bianchi identities for the $N=1$
theory. We quote in the following their standard expression.
\vskip 3mm \par
 {\bf Curvatures of $N=1$ gauged theory}
\begin{eqnarray}
T^a & \equiv & {\mathcal D} V^a - {\rm i} \bar\psi^\bullet \g^a
\psi_\bullet \equiv 0 \\
R^{ab} &=& d\omega^{ab} - \omega^a_{\phantom{a} c} \omega^{cb} \\
\rho_\bullet &=& \nabla \psi_\bullet = {\mathcal D}
\psi_\bullet + \frac{\rm i}{2} \hat { Q} \psi_\bullet \\
R\left(\chi^i\right) &=& \hat \nabla \chi^i = {\mathcal D}\chi^i +
\hat \G^i_{\ j}\chi^j - \frac{\rm i}{2} \hat { Q} \chi^i\\
F^\L &=& dA^\L + \half C^\L_{\phantom{\L}\S\G}A^\S A^\G \\
\nabla \lambda^\L &=& {\mathcal D} \lambda^\L + \frac{\rm i}{2}
\hat { Q} \lambda^\L  + C^\L_{\phantom{\L}\S\G}A^\S \lambda^\G\\
\nabla z^i &=& d z^i + g_{(\L)} k^i_\L A^\L
\end{eqnarray}
where the gauged connections are defined as:
\begin{eqnarray}
\hat \G^i_{\ j} &=& \G^i_{\ j} + g_{(\L)} \nabla _j k^i_\L A^\L\\
\hat {Q} &=& {Q} + g_{(\L)} P_\L A^\L.
\end{eqnarray}
The ungauged connection ${Q}$ is given by
\begin{equation}
Q=Q_i \nabla z^i +Q_{\bar\imath}\nabla \bar z^{\bar\imath}.
\end{equation}
\vskip 3mm \par {\bf Bianchi identities of $N=1$ gauged theory}
\begin{eqnarray}
&&R^{ab} V_b -{\rm i} \bar \psi^\bullet \g^a \rho_\bullet + {\rm
i} \bar \psi_\bullet \g^a \rho^\bullet  = 0
\\
&&{\mathcal D}R^{ab} =0
\\
&&\nabla^2 \psi_\bullet + \frac 14 \g_{ab} R^{ab} \psi_\bullet
- \frac{\rm i}{2} \hat {\mathcal K} \psi_\bullet = 0\\
&&  \nabla^2 \chi^i +\frac 14 \g_{ab} R^{ab}\chi^i +
\hat R^i_{\ j}\chi^j + \frac{\rm i}{2} \hat {\mathcal K} \chi^i = 0\\
&& \nabla F^\L = 0 \\
&& \nabla^2 \lambda^\L + \frac 14 \g_{ab} R^{ab} \lambda^\L -
\frac{\rm i}{2}
\hat {\mathcal K} \lambda^\L  - C^\L_{\phantom{\L}\S\G}A^\S \lambda^\G = 0\\
&& \nabla^2 z^i - g_{(\L)} k^i_\L F^\L = 0
\end{eqnarray}
In the ungauged case, it is straightforward to see that the
conditions found in the text from the analysis of the reduction
of the quaternionic sector and of  supersymmetry transformation
laws are indeed necessary and sufficient, after setting $\psi_2
=\rho_2 =0$, for reducing the $N=2$ supercurvatures and Bianchi
identities to the corresponding $N=1$ expressions.\\
We only observe that in the covariant differential of $\zeta_\a$
and its Bianchi identity, after decomposition of the index $\a
=(I, \dot I)$, we get, as integrability condition:
\begin{equation}
\nabla^2 \zeta_I + \frac 14 R^{ab}\g_{ab} \zeta_I + \frac{\rm
i}{2} {K} \zeta_I + \IR_I^{\phantom{I} J} \zeta_J =0 \, , \quad
\mbox{(since } \IR_I^{\phantom{I} \dot J}=0 \mbox{).}
\end{equation}
This equation can be converted in world indices on ${\cal M}^{KH}$
using equation  (\ref{trans1}). Using further the reduction of
equation (\ref{2.65}) one then recovers the correct $N=1$ result,
in terms of the Riemann curvature of the ${\cal M}^{KH}$ manifold,
with
\begin{equation}
{\mathcal  K}^{(N=1)} = {\mathcal K}^{(N=2)} + \Omega^3.
\end{equation}
Note that $\Omega^3$ is one half of the K\"{a}hler form of the
K\"{a}hler-Hodge manifold ${\mathcal M}^{KH}$.
\par
As far as the gauged theory is concerned, we observe that the
ungauged conditions
\begin{equation}
\G^\a_{\phantom{\a}i} = R^\a_{\phantom{\a}i}=\omega^1 = \omega^2
=\Omega^1 =\Omega^2 =\Delta_I^{\phantom{I}\dot J} =
R_I^{\phantom{I}\dot J}=0
\end{equation}
become the corresponding ones for the gauged quantities
\begin{equation}
\hat\G^\a_{\phantom{\a}i} = \hat R^\a_{\phantom{\a}i}=\hat\omega^1
= \hat\omega^2 =\hat\Omega^1 =\hat\Omega^2
=\hat\Delta_I^{\phantom{I}\dot J} = \hat R_I^{\phantom{I}\dot
J}=0.
\end{equation}
Recalling the definition of the hatted quantities, we find that
the following objects must be zero:
\begin{eqnarray}
g_{({\bf \L})} A^{{\bf \L}} {\mathcal D}_{  j} k^{\a }_{{\bf \L}}
&=&
g_{({\bf \L})} F^{{\bf \L}} {\mathcal D}_{{\mathcal J}} k^{\a}_{{\bf \L}} = 0\\
g_{({\bf \L})} A^{{\bf \L}} (P^1_{{\bf \L}} -{\rm i} P^2_{{\bf
\L}} ) &=& g_{({\bf \L})} F^{{\bf \L}}
(P^1_{{\bf \L}} -{\rm i} P^2_{{\bf \L}} ) = 0\\
g_{({\bf \L})} A^{{\bf \L}} \partial_u k^v_{{\bf \L}} {\mathcal
U}^{u | AI} {\mathcal U}_{v| A\dot J} &=& 0
\end{eqnarray}
The previous conditions can be analyzed in the light of the
results obtained in section 6, and it is straightforward to see
that they are actually satisfied.
  Thus the reduced theory has the correct
integrability conditions.
\section*{Appendix C: A useful formula for the $N=2$ gaugino
transformation law} \setcounter{equation}{0}
\addtocounter{section}{1} In this Appendix we show how to retrieve
equation  (\ref{gaugin2''}) from (\ref{gaugintrasf}). To avoid a
too heavy notation, we write in this Appendix the world indices
and gauge indices without hat and tilde, since we are not going to
perform any reduction.
 We are interested in trading the world index $ i$ of the gauginos $\l^{ i A}$ with a gauge index
 $\L $, through the definition:
 \begin{equation}
 \l^{\L A} \equiv -2 f^{\L}_{ i} \l^{i A}
 \end{equation}
However, the gauge index of the $N=2$ theory runs over $n_V +1$
values (because of the presence of the graviphoton) while the
index $  i$ takes only $n_V$ values. The extra gaugino, say
$\lambda^0 $, is actually spurious, since, as discussed in
section (6.2), $\lambda^{ \L  A} $ satisfies:
\begin{equation}\label{linear'}
T_\L  \lambda^{ \L  A}  =  0
\end{equation}
where
 $T_\L$ is the projector on the graviphoton
field-strength, according to equation  (\ref{gravif}) \cite{cdf}.
 In order to show that the $n_V$ gauginos $\lambda^\L$ do appropriately transform into
 the $n_V$ matter-vector field strengths, let us now calculate the susy transformation law of the new
fermions $\lambda^{ \L  A}$, which, up to 3-fermions terms, is:
\begin{equation}
\delta\lambda^{ \L  A}= -2 f^{ \L  }_{ i} \delta\lambda^{ i A}
 = -2 f^{ \L  }_{ i} \left[-g^{ i {\bar \jmath}} f^{ \S }_{ {\bar\jmath}}
 {\rm
Im}{\cal N}_{ \G  \S } F^{- \G
}_{\mu\nu}\gamma^{\mu\nu}\varepsilon^{AB} + W^{  i AB}\right]
\epsilon_B \label{gaugapp}
\end{equation}
Now we use the following relations of special geometry \cite{cdf}:
\begin{eqnarray}\label{gij}
g_{i\bar j}&=&-2f^\L _i {\rm Im}{\cal N}_{\L \S }f^\S _{\bar j}\\
\nonumber \delta^i_\ell& =& g^{i\bar j} g_{\ell\bar j} = -2
g^{i\bar j} f^\L _\ell {\rm {Im }}{\cal N}_{\L \S }f^\S _{\bar j}
\end{eqnarray}
\begin{equation}\label{defu'}
U^{\L \S }\equiv f^\L _ig^{i\bar j}f^\S _{\bar j} =-\half
\left[\left(\rm {Im}{\cal N}\right)^{-1}\right]^{\L \S } -\bar
L^\L  L^\S .
\end{equation}
\begin{equation}\label{ide}
{\rm {Im}}{\cal N}_{\L  \S }L^{\L } \bar L^{\S }_i\,=\,- \half
\end{equation}
Eq. (\ref{gaugapp}) can then be rewritten as:
\begin{equation}
\delta\lambda^{\L  A}  = \left[2U^{\L \S }{\rm Im}{\cal N}_{\S \G
} F^{-\G }_{\mu\nu} \varepsilon^{AB}- 2 f^\L_\ell k^\ell_\Delta
\bar L^\Delta \varepsilon^{AB} -2{\rm i}U^{\L\S} P^x_\S
(\s^x)^{AB}\right] \epsilon_B \label{gaugapp2}
\end{equation}
Now, using the definition of the special geometry Killing vectors
\begin{equation}
k^i_\L = {\rm i} g^{i\bar\jmath} \partial_{\bar\jmath} P^0_\L
\label{killing}
\end{equation} we have:
\begin{equation}
2f^\L_\ell k^\ell_\Delta \bar L^\Delta  = 2{\rm i} f^\L_\ell
g^{\ell\bar \ell}\partial_{\bar\ell}P^0_\Delta \bar L^\Delta = -
2{\rm i} f^\L_\ell g^{\ell\bar \ell} f^\Delta
_{\bar\ell}P^0_\Delta = - 2{\rm i}U^{\L\S} P^0_\S\label{killer}
\end{equation}
where we have used the special geometry formulae \cite{abcdffm}:
\begin{equation}
P^0_\L L^\L = P^0_\L \bar L^\L =0 \,, \quad f^\L_i \equiv\nabla_i
L^\L.
\end{equation}
Therefore equation  (\ref{gaugapp2}) becomes:
\begin{equation}
\delta\lambda^{\L  A}  = -2U^{\L \S }\left[{\rm Im}{\cal N}_{\S
\G } F^{-\G }_{\mu\nu} \g^{\mu\nu}\varepsilon^{AB} +{\rm i}
\left(- P^0_\S\varepsilon^{AB} + P^x_\S (\s^x)^{AB}\right)\right]
\epsilon_B \label{gaugapp3}
\end{equation}
\par
 Let us now set
\begin{equation}\label{defp'}
P^\L _{\phantom{\G  }\G } \equiv -2U^{\L \S }{\rm Im}{\cal N}_{\S
\G } = \delta^\L _\G  + 2{\rm Im}{\cal N}_{\G \S }\bar L^\L L^\S
=\delta^\L _\G  -{\rm i} T_\G  \bar L^\L
\end{equation}
\begin{equation}\label{trans}
 \bar P_\S^{\phantom{\L}\L}=\delta^\L _\S
+{\rm i} \bar T_\S  L^\L  =(P^t)^\L_{\phantom {\S}\S}
\end{equation}
where $T_\L$ is defined  by equation  (\ref{defprogr})
 and satisfies \cite{cdf} :
 \begin{equation}\label{pollo}
T_\L  \bar L^\L  =-{\rm i};\,\,\quad T_\L  f^\L _i=0
\end{equation}
Then we have:
\begin{equation}\label{orto}
P^\L_{\phantom{\L}\S}P^\S_{\phantom{\S}\G}= P^\L_{\phantom{\L}\G
\, ; \quad} T_\L  P^\L _\G  =0
\end{equation}
Therefore $P^\L _{\phantom{\G }\G }$ is the projector orthogonal
to the
  graviphoton, that is it projects the $n_v +1$ vector
  field-strengths onto the $n_v$ field-strengths of the vector
  multiplets.
\par
We can then rewrite equation  (\ref{gaugapp3}) as:
\begin{equation}
\delta\lambda^{\L  A} = P^\L_{\phantom{\L}\S}F^{-\S}_{\mu\nu}
\g^{\mu\nu}\varepsilon^{AB} +{\rm i}U^{\L\G}  \left(-
P^0_\G\varepsilon^{AB} + P^x_\G (\s^x)^{AB}\right) \epsilon_B
\label{gaugappb}
\end{equation}
which is the equation given in the text.
 Note that
\begin{equation}\label{lambda}
\lambda^{\L A } \,=\,P^{\L}_{\phantom{\S}\S}\lambda^{\S A }
\,\,;\quad \,\,f^{\L}_i\,=\,P^{\L}_{\phantom{\S}\S}f^{\S}_i.
\end{equation}
  \\
It is useful to write down the explicit decomposition of the
field strength $F^\L$ into the graviphoton and matter vectors
part, that is:
\begin{equation}
F^{-\L}_{\mu\nu} =   {\rm i} \bar L^\L T_\S F^{-\S}_{\mu\nu} +
P^\L_{\phantom{\L}\S}F^{-\S}_{\mu\nu}.
\end{equation}
Equation (\ref{gaugappb}) becomes:
\begin{equation}
\delta\lambda^\L_\bullet = P^\L_{\phantom{\L}\S}\left[F^{-\G
}_{\mu\nu} \g^{\mu\nu}\varepsilon^{AB} +{\rm i}\left({\rm Im}{\cal
N}\right)^{-1 \S \G }  \left(- P^0_\G\varepsilon^{AB} + P^x_\G
(\s^x)^{AB}\right)\right] \epsilon_B \label{gaugapp4}
 \end{equation}
where we see, as expected, that the gauginos $\l^{\L A}$ do
transform only into the matter-vector field strengths
$P^\L_{\phantom{\L}\S}F^{-\S }_{\mu\nu}$.
 Hence, equation  (\ref{gaugapp4}) intrinsically defines only $n_V$ independent
 gauginos transforming into the $N=2$ field
 strengths $(P^\L_{\phantom{\L}\S}
F^{-\S }_{\mu\nu})$.
\section*{Appendix D: Reduction of special geometry in special coordinates}
\setcounter{equation}{0} \addtocounter{section}{1} If we choose
special coordinates for special geometry
\cite{dlv},\cite{cadf},\cite{cdf},\cite{cdfv}, then the indices
${\bf \L}=(\L , Y)$ and ${\mathcal I}=(i,\a )$ are identified by
the fact that
\begin{equation}
t^{{\mathcal I}} = \frac{X^{{\bf \L}}}{X^0} \, , \quad \left( \L
=\a \, ,\, Y = i  \right)
\end{equation}
and a prepotential $F(X)$ exists such that $f(t) =
\frac{1}{(X^0)^2} F(X)$, with:
\begin{eqnarray}
X^0 F_0 &=& 2f - t^{{\mathcal I}} f_{{\mathcal I}} \, \,,
 \quad \left(f_{{\mathcal I}} = \frac{\partial f}{\partial t^{{\mathcal I}}} \right)  \\
X^0 F_{{\mathcal I}} &=& \partial_{{\mathcal I}}f .
\end{eqnarray}
Furthermore,
\begin{equation}
e^{-\cK} = {\rm i} \left[ 2f - 2 \bar f + (\bar t^{{\mathcal I}} -
t^{{\mathcal I}}) (f_{{\mathcal I}} + \bar f_{{\mathcal
I}})\right] .
\end{equation}
The constraints that define the submanifold $\cM_R$ become:
\begin{eqnarray}
&&W_{ij\a} = \partial_i \partial_j \partial_\a f =0\,,\quad
W_{\a_1\a_2\a_3} = \partial_{\a_1} \partial_{\a_2} \partial_{\a_3} f =0\\
&&X^\L =\frac{\partial F}{\partial X^\L}=\partial_i X^\L =
\partial _\a X^{X}=
\partial_i f_\L =
\partial _\a f_{X}=0
\end{eqnarray}
where we used the fact that $\cK_\a |_{\cM_R} =0$.\par In this
basis $\bar \cN_{{\bf \L}\S} = \partial_{{\bf \L}} \partial _{\S}
f$ and the K\"ahler potential on $\cM_R$ is:
\begin{equation}
e^{-\cK_R} = {\rm i} \left[ 2f - 2 \bar f + (\bar t^{ i} - t^{
i}) (f_{  i} + \bar f_{  i})\right] .
\end{equation}
Note that $F_\L |_{\cM_R}=0$ implies $\partial _\a f|_{\cM_R}=0$
which in turn implies $\partial _\a \partial_i f|_{\cM_R}=0$,
$W_{\a ij} = \partial _\a \partial_i\partial_j f|_{\cM_R}=0$.
Therefore the most general form for $f$ is ($t^{{\mathcal I}}
\Rightarrow (t^i , z^\a )$):
\begin{equation}
f(t^i,z^\a ) = f(t) + \sum_{n\geq 2} z^{\a_1} \cdots
z^{\a_n}f_{\a_1\cdots \a_n} (t)\,, \quad f_{\a_1\a_2 \a_3} (t)=0
.\label{f(t)}
\end{equation}
\par
For the manifold $SU(1,1)/U(1) \times SO(2,n)/[SO(2)\times SO(n)]$
used in section (6.6), with coordinates $(t^0,t^1,\cdots ,
t^{n'},z^1,\cdots ,z^{n-n'})$, the reduced manifold $(z^1,\cdots
,z^{n-n'})=0$ is parametrized with coordinates $(t^0,t^1,\cdots ,
t^{n'})$ and the holomorphic prepotential is
\cite{fevp},\cite{cdf}:
\begin{equation}
f(t^i,z^\a ) = {\rm i} t^0 \left( \sum_{i=1}^{n'} \eta_{ij} t^i
t^j - \sum_{i=1}^{n-n'} \delta_{\a\b}z^\a z^\b \right) \,, \quad
[\eta_{ij} = (1,-1,\cdots , -1)]
\end{equation}
in accordance to equation  (\ref{f(t)}).

\section*{Appendix E: Re\-duc\-tion of the qua\-ter\-nionic Ward iden\-tity}
\setcounter{equation}{0} \addtocounter{section}{1} We derive here
the conditions on the quaternionic prepotentials and Killing
vectors, discussed in Section (6.5), from the
 reduction of the quaternionic Ward identity (\ref{wardquat}),
which is essential for the validity of the $N=2$ supersymmetric
Ward identity involving the scalar potential, that is the
relation \cite{abcdffm}:
\begin{equation}
 \frac {1} { \l} \Omega^x_{uv} k^u_{\bf \L } k^v_{\bf \S } +
\half \epsilon^{xyz}P^y_{\bf \L }
  P^z_{\bf\S } -\half f_{{\bf\L}\bf\S}^{\phantom{\L\S}\bf\G} P^x_{\bf\G} =0
  \label{wardquat'}
\end{equation}
 After projection, and using the just found results
$P^{\rm i}_{\L}=0$; $P^3_{X}=0$,
 it decomposes in a set of equations
 \begin{itemize}
\item{${\bf \L}=\L$; ${\bf \S}=\S$
\begin{eqnarray}
  &&\frac {1} { \l} \Omega^{\rm i}_{uv} k^u_{ \L } k^v_{ \S } +
 \half \epsilon^{{\rm i}{\rm j}}P^{\rm j}_{ \L }
  P^3_{ \S } -\half f_{{ \L} \S}^{\phantom{\L \S}{\bf \G}}
  P^{\rm i}_{{\bf \G}}
  =0
    \label{award1}\\
  && \frac {1} { \l} \Omega^3_{uv} k^u_{\L } k^v_{\S } +
 \half \epsilon^{{\rm i}{\rm j}}P^{\rm i}_{  \L }
  P^{\rm j}_{ \S } -\half f_{{ \L} \S}^{\phantom{\L\S}{\bf \G}} P^3_{{\bf \G}} =0
  \label{award2}
\end{eqnarray}
Since $P^{\rm j}_{ \L}=0$, and since $f_{{ \L} \S}^{\phantom{\L
\S}Z}=0$ (because $G^{(1)} \subset G^{(2)}$), then equation
(\ref{award1}) gives
\begin{equation}
k^t_{ \L}=0, \label{kt}
\end{equation}
 as indeed was expected
from geometrical considerations.\\
Then, equation  (\ref{award2}) becomes
\begin{equation}
 \frac {1} { \l} \Omega^3_{s\bar s} k^s_{  \L } k^{\bar s}_{ \S }
  -\half f_{{ \L} \S}^{\phantom{\L \S} \G}
  P^3_{ \G}
  =0.\label{pesce}
  \end{equation}
 Setting $\l =-1$ and recalling that $- \Omega_{s\bar s} $
 is half of the K\"ahler form of
 the reduced submanifold, we recognize that equation  (\ref{pesce})
 expresses the Poissonian realization of the Lie algebra of the
 prepotentials $P^3_{ \L}$ on the K\"ahler-Hodge submanifold
 ${\cal M}^{KH}$, namely:
 \begin{equation}
 \{P^3_{ \L} ,  P^3_{ \S}\}_P = f_{{ \L} \S}^{\phantom{\L \S} \G}
 P^3_{ \G}.
 \end{equation}
 }
 \item{
${\bf \L}=\L$; ${\bf \S}=Y$
\begin{eqnarray}
  &&\frac {1} { \l} \Omega^{\rm i}_{st} k^s_{ \L } k^t_{Y } -
\half  \epsilon^{{\rm i}{\rm j}}P^3_{ \L }
  P^{\rm j}_{Y } -\half f_{{ \L}Y}^{\phantom{\L Y}Z}
  P^{\rm i}_{Z}
  =0
    \label{bward1}\\
  && \frac {1} { \l} \Omega^3_{uv} k^u_{ \L } k^v_{Y } +
 \half \epsilon^{{\rm i}{\rm j}}P^{\rm i}_{ \L }
  P^{\rm j}_{Y } -\half f_{{ \L}Y}^{\phantom{\L \S}{\bf \G}}
  P^3_{{\bf \G}} =0
  \label{bward2}
\end{eqnarray}
Eq. (\ref{bward1}) gives a relation which has to be valid
everywhere on the submanifold. \\
Since $\Omega^3_{st} =0$,  $P^3_{X} =P^{\rm i}_{ \L}=0$, and
considering (\ref{kt}), then equation  (\ref{bward2}) is
identically satisfied. }
\item{
${\bf \L}=X$; ${\bf \S}=Y$
\begin{eqnarray}
  &&\frac {1} { \l} \Omega^{\rm i}_{st} k^s_{X } k^t_{Y } -
\half  \epsilon^{\rm ij}P^3_{X }
  P^{\rm j}_{Y } -\half f_{{X}Y}^{\phantom{\L Y}Z}
  P^{\rm i}_{Z}
  =0
    \label{cward1}\\
  && \frac {1} { \l} \Omega^3_{tt'} k^t_{X } k^{t'}_{Y } +
\half  \epsilon^{{\rm i}{\rm j}}P^{\rm i}_{X }
  P^{\rm j}_{Y } -\half f_{{X}Y}^{\phantom{\L{\bf \S}} \G}
  P^3_{ \G} =0
  \label{cward2}
\end{eqnarray}
Eq. (\ref{cward1}) is identically satisfied for
$f_{{X}Y}^{\phantom{\L Y}Z}$, while equation (\ref{cward2}) is a
relation to be satisfied all over the submanifold. }
\end{itemize}
\section*{Appendix F: Computation of the $N=2 \to N=1$
scalar potential} \setcounter{equation}{0}
\addtocounter{section}{1} We want to solve here a puzzle raised in
the text about the scalar potential. In the $N=2$ theory, the
scalar potential has the form \cite{abcdffm}:
\begin{eqnarray}\label{pot2'}
  \cV^{N=2}
  &=&-12 S^{11}\bar S_{11} +  g_{{\mathcal I}\bar{\mathcal J}}
 \left(W^{{\mathcal I}11}\bar W^{\bar{\mathcal J}}_{11}+ W^{{\mathcal I}21}\bar W^{\bar{\mathcal J}}_{21}\right)
 + 2N^1_I N^I_1
 \nonumber\\
 &=&\left( g_{{\mathcal I}\bar{\mathcal J}}k^{{\mathcal I}}_{{\bf \L}}
 k^{\bar{\mathcal J}}_{{\bf \S}} + 4h_{uv}k^u_{{\bf \L}}
 k^v_{{\bf \S}}\right)\bar L^{{\bf \L}} L^{{\bf \S}} +
 U^{{\bf \L}{\bf \S}}   P^x_{{\bf \L}}
 P^x_{{\bf \S}}
-3P^x_{{\bf \L}} P^x_{{\bf \S}} {\bar L}^{{\bf \L}} L^{{\bf \S}} .
\end{eqnarray}
which manifestly does not contain contributions antisymmetric in
the quaternionic prepotentials or Killing vectors, nor it has
interference terms $P^0_{{\bf \L}} P^x_{{\bf \S}}$ between
quaternionic and special-K\"ahler isometries.\\
On the other hand, the $N=1$ scalar potential:
\begin{equation}\label{pot1'}
  \cV^{N=1} = 4\left(\nabla_\ell L \nabla_{\bar\ell}\bar L g^{\ell\bar\ell}
  -3 |L|^2 +\frac{1}{16} {\rm Im}f_{\L\S} D^\L D^\S \right),
\end{equation}
does instead contain both kinds of interference contributions,
given the form of the superpotential $L=\frac{\rm i}2
L^{X}(P^1_{X}-{\rm i}P^2_{X})$ and of the D-term $D^\L =-2 ({\rm
Im} f)^{-1\L\S}(P^0_\S + P^3_\S )$ which appear quadratically in
(\ref{pot1'}).
\par
The interference contributions in (\ref{pot1'}) have therefore to
cancel each other. As we are going to show, this does indeed
happen, in a way which involves non trivially the properties
obeyed by the special-K\"ahler and quaternionic Killing vectors.
Let us analyze and reduce separately the various contributions to
the $N=2$ potential, using all the constraint relations found in
Section 6.
\begin{eqnarray}
-12 S^{11}\bar S_{11}&\Rightarrow& -12 |L|^2 \nonumber\\
&=&  -3P^{\rm i}_{X}P^{\rm i}_{Y}{\bar L}^{X} L^{Y} -6{\rm i}
P^1_{[X}P^2_{Y]}L^{X} {\bar L}^{Y}\label{ss}
\end{eqnarray}
 \begin{eqnarray}
  g_{{\mathcal I}\bar{\mathcal J}}
 W^{{\mathcal I}11}\bar W^{\bar{\mathcal J}}_{11}U^{{\bf \L}{\bf \S}}&\Rightarrow&
4g^{i\bar\jmath}
 \nabla_i L \nabla_{\bar\jmath} \bar L \nonumber\\
 &=& g^{i\bar\jmath}
 \nabla_i L^{X}\nabla_{\bar\jmath} \bar L^{Y} P^{\rm i}_{X}
 P^{\rm i}_{Y}  - 2 {\rm i}
P^{1}_{[X}
 P^{2}_{Y]}U^{X Y} \nonumber\\
 &=& g^{i\bar\jmath}
 \nabla_i L^{X}\nabla_{\bar\jmath} \bar L^{Y} P^{\rm i}_{X}
 P^{\rm i}_{Y} +
 2 {\rm i}
P^{1}_{[X}
 P^{2}_{Y]}L^{X} {\bar
L}^{Y}\label{ww1}
\end{eqnarray}
\begin{equation}
 g_{{\mathcal I}\bar{\mathcal J}}
 W^{{\mathcal I}21}\bar W^{\bar{\mathcal J}}_{21}\Rightarrow
({\rm Im} f)^{-1 \L\S} \left(P^0_\L P^0_\S +P^3_\L P^3_\S\right)+
2U^{\L\S} P^0_\L P^3_\S \label{ww2}
\end{equation}
where we have used equation  (\ref{killer}) of Appendix C,  the
identity of special geometry (\ref{chiusura}) and the definition
of the prepotential $P^0_\G$, equation  (\ref{p02}).
\begin{eqnarray}
2 N^1_I N^I_1 &\Rightarrow& 4 g^{s\bar s} \nabla_s L \nabla_{\bar
s} \bar L \nonumber\\
&=& g^{s\bar s} \nabla_s P^{\rm i}_{X} \nabla_{\bar s} P^{\rm
i}_{Y} +2{\rm i} g^{s\bar s} \nabla_s P^{\rm i}_{[X} \nabla_{\bar
s} P^{\rm i}_{Y]}\label{nn}
\end{eqnarray}
The last term in equation  (\ref{nn}) is transformed using the
definition of quaternionic Killing vectors:
\begin{equation}\label{prepo}
2k^v_\L  \Omega_{uv}^x\,=\,\nabla_u P^x_\L
\end{equation}
the realization of the $SU(2)$ algebra on the curvatures
$\Omega^x$:
\begin{equation}
h^{st} \Omega^x_{us} \Omega^y_{tw} = - \lambda^2 \delta^{xy}
h_{uw} + \lambda \epsilon^{xyz} \Omega^z_{uw} \label{2.67}
\end{equation}
and the normalization chosen for the metric on $\cM^{KH}$:
\begin{equation}
h_{s\bar s} =\half g_{s\bar s}.
\end{equation}
After some calculations we get:
\begin{eqnarray}
2{\rm i} g^{s\bar s} \nabla_s P^{\rm i}_{[X} \nabla_{\bar s}
P^{\rm i}_{Y]} &=& 4 {\rm i} \Omega^3_{t \bar t}k^{t}_{[X}
h^{\bar t}_{Y]}L^{X} \bar L^{Y} \nonumber\\ &=&\left( 4{\rm
i}P^{1}_{[X} P^{2}_{Y]} - \frac{\rm i}2 f^\G_{\phantom{\G}X \dot
\S} P^3_\G \right) L^{\dot
\L} \bar L^{Y}\nonumber\\
&=& 4{\rm i}P^{1}_{[X} P^{2}_{Y]}L^{X} \bar L^{Y} - 2 P^3_\L
P^0_\S ({\rm Im}f)^{-1 \L\S}\label{nn'}
\end{eqnarray}
 where we have applied the quaternionic
Ward identity (\ref{wardquat}) discussed in Appendix D to the
present case and the definition of the prepotential $P^0_\L$,
equation (\ref{p02}).
 Collecting
together all the terms in eqs. (\ref{ss}), (\ref{ww1}),
(\ref{ww2}), (\ref{nn'}), we find that the antisymmetric  parts in
$\L ,\S$ and the two terms in $P^0_\L P^3_\S$ cancel against each
other identically.

\section*{Appendix G: The  $N=2$ and $ N=1$ Lagrangians in $D=4$}
\setcounter{equation}{0} \addtocounter{section}{1} For reference
of the reader we give here the Lagrangian of the $N=2$ theory and
of the $N=1$ theory as given in reference \cite{df}\footnote{
Some misprints of ref. \cite{df} have been corrected}.
\par
The $N=1$ lagrangian is, up to four-fermions terms:
\begin{eqnarray}
\label{lag1}(\rm {det}V)^{-1}{\mathcal L^{N=1}}&=&
-\frac{1}{2}\mathcal{R} + {\rm i}\left( f_{\L \S }{\cal F}^{-\L
}_{\mu\nu}{\cal F}^{-\S  \mu\nu} - { {\bar f}}_{\L \S } {\cal
F}^{+\L }_{\mu\nu}{\cal F}^{+ \S  {\mu\nu}}\right )  +
g_{i\bar\jmath}\nabla_{\mu}z^i \nabla^{\mu}z^{\bar\jmath}
\nonumber
\\&+& {{\epsilon^{\mu\nu\lambda\sigma}}\over{\sqrt{-g}}} \left(
\bar\psi^{\bullet}_\mu\gamma_\sigma D_\nu\psi_{\bullet \lambda} -
\bar\psi_{\bullet \mu} \gamma_\sigma D_\nu\psi^{\bullet}_{\lambda}
\right ) +\nonumber\\
&+& \frac {1}{8} \left({ {\bar f}}_{\L \S }\bar\lambda^{\bullet
\L }\gamma^\mu \nabla_\mu\lambda_{\bullet}^{\S } -{ { f}}_{\L \S
}\bar\lambda_{\bullet}^{ \L } \gamma^\mu
\nabla_\mu\lambda^{\bullet \S } \right) \nonumber \\&-& {\rm{i}}
\frac{1}{2} g_{i\bar\jmath} \left(\bar\chi^{i}\gamma^{\mu}
\nabla_{\mu} \chi^{\bar\jmath} +\bar \chi^{\bar\jmath}
\gamma^{\mu} \nabla_{\mu} \chi^{i} \right )- g_{i\bar\jmath}
\left(
 \bar \psi_{\bullet \nu} \gamma^{\mu}\gamma^{\nu} \chi^i \nabla^{\mu}\bar z^{\bar j}
+ \bar \psi^{\bullet}_{ \nu} \gamma^{\mu} \gamma^{\nu} \chi^{\bar
 j} \nabla_{\mu}z^{i} \right) \nonumber \\
 &-&  \,{\rm{i}}\, {\rm Im} f_{\L \S } \left ( {\cal
 F}^{+\L }_{\mu\nu}\bar \lambda_{\bullet}^{ \S }
 \gamma^{\mu} \psi ^{\bullet \nu} + {\cal
 F}^{-\L }_{\mu\nu}\bar \lambda^{\bullet \S }
 \gamma^{\mu} \psi _{\bullet}^{ \nu} \right)\nonumber \\
&-&\frac {\rm {i}}{8} \left (\partial_i {f}_{\L  \S } {\cal
F}^{-\L }_{\mu\nu} \bar \chi^i
 \gamma^{\mu \nu} \lambda _{\bullet}^{\S } - \partial_{\bar\imath} {\bar f}_{\L  \S } {\cal
 F}^{+\L }_{\mu\nu} \bar \chi^{\bar\imath}
 \gamma^{\mu \nu} \lambda ^{\bullet \S } \right) \nonumber
 \\&+&2L\bar\psi_{\mu}^\bullet\gamma^{\mu\nu}\psi_{\nu}^\bullet+2\bar L
\bar\psi_{\mu \bullet}\gamma^{\mu\nu}\psi_{\nu \bullet}\nonumber \\
&+&{\rm{i}}
 g_{i\bar\jmath} \left ( {\overline N}^{\bar j} \bar\chi^i
\gamma^{\mu}\psi_{\mu}^{\bullet}+N^i \bar\chi^{\bar\jmath}
\gamma^{\mu}\psi_{\bullet \mu}\right) + \frac{1}{2} P_{\L }
\left( \bar \lambda^{\bullet \L } \gamma^{\mu} \psi _{\bullet
\mu}- \bar \lambda_{\bullet} ^{\L } \gamma^{\mu} \psi
_{\mu}^{\bullet|} \right)\nonumber \\ &+&\mathcal M_{ij}
\bar\chi^i\chi^j + {\overline \mathcal M}_{\bar\imath \bar\jmath}
\bar\chi^{\bar\imath} \chi^{\bar\jmath} +\mathcal{M}_{\L \S }
\bar\lambda^{\L }_{\bullet}\lambda^{\S }_{\bullet}+{\overline
{\mathcal{M}}}_{\L  \S } \bar\lambda^{\L \bullet}\lambda^{\S
\bullet} \nonumber \\ &+& \mathcal{M}_{\L  i} \bar\lambda^{\L
}_{\bullet} \chi ^i +{\overline{\mathcal{M}}}_{\L  \bar\imath}
\bar\lambda^{\L \bullet} \chi ^{\bar\imath} \nonumber - {\mathcal
{V}}(z,\bar z, q)
\end{eqnarray}
where the kinetic matrix ${f}_{\L  \S }$ is a holomorphic
function of $z^i$, and the mass matrices ${\cal M}_{ij},{\cal
M}_{\L\S},{\cal M}_{\L i}$ are given by:
\begin{eqnarray}
\label{def4} {\mathcal {M}}_{ij} &=& \nabla_i \nabla_j L \\
{\mathcal {M}}_{\L  \S } &=& \frac {{\rm i}} {8} N^i
\partial_i \overline{{\mathcal{N}}}_{\L
\S } \\
 M_{\L  i}&=& - {\rm {i}} \frac {1}{4} {\rm {Im}}
{\mathcal{N}}_{\L  \S } \partial_i D^{\S } -\frac {1} {2}
k^{\bar\jmath}_{\L } g_{i\bar\jmath} \label{mista}
\end{eqnarray}
where we have set  ${\mathcal F}^{\pm \L }_{\mu\nu}=\frac {1}{2}
({\mathcal F}^\L _{\mu\nu}\pm {{\rm
i}\over2}\epsilon^{\mu\nu\rho\sigma} {\mathcal F}_{\rho\sigma}^\L
)$, ${\mathcal F}^\L _{\mu\nu}$ being the field-strengths of the
vectors $A^\L _\mu$.
\\
 Note that, since the scalar
manifold is a K\"ahler-Hodge manifold, all the fields and the
bosonic sections have a definite $U(1)$ weight $p$ under $U(1)$.
We have
\begin{eqnarray}\label{weight}
  p(V^a_{\mu})&=& p(A^{\L })= p(z^i)= p(g_{i\bar\jmath})=p(
  {\mathcal {N}}_{\L  \S }) = p( D^{\L }) = p(P_{\L }) =p({\mathcal {V}}) =0 \nonumber \\
  p(\psi_{\bullet})&=& p(\chi^{\bar\imath})=
  p(\lambda^{\L }_{\bullet})=p(\varepsilon_{\bullet}) = \frac {1} {2} \nonumber \\
p(\psi^{\bullet})&=& p(\chi^{ i})=
  p(\lambda^{\L  \bullet})=p(\varepsilon^{\bullet}) = - \frac {1} {2} \nonumber \\
p(L)&=& p( {\mathcal {M}}_{ij}) = p( \overline {\mathcal {M}}_{\L
\S }) =1 \nonumber \\
p(\bar L)&=& p({\overline {\mathcal M}}_{\bar\imath \bar\jmath})
= p( {\mathcal {M}}_{\L  \S }) =- 1
\end{eqnarray}
Accordingly, when a covariant derivative acts on a field  $\Phi$
of weight $p$ it is  also $U(1)$ covariant (besides possibly
Lorentz, gauge and scalar manifold coordinate symmetries)
according to the following definitions:
\begin{equation}
\begin{array}{ccccccc}
\nabla_i \Phi &=&
 (\partial_i + {1\over 2} p \partial_i {\cal K}) \Phi &; &
\nabla_{i^*}\Phi &=&(\partial_{i^*}-{1\over 2} p \partial_{i^*}
{\cal K}) \Phi \cr
\end{array}
\end{equation}
where $ {\mathcal {K}(z,\bar z)}$ is the K\"{a}hler potential.
\\
On the other hand, the $N=2$ Lagrangian, up to four-fermions
terms, is:
\begin{eqnarray} \label{lagn2}
({\rm det}V)^{-1}\, {\mathcal L^{N=2}} &=& -\frac{1}{2} R +
g_{i\bar\jmath}\nabla^\mu z^i \nabla_\mu \bar z^{\bar\jmath}+
h_{uv}\nabla_\mu q^u \nabla^\mu q^v +
{{\epsilon^{\mu\nu\lambda\sigma}}\over{\sqrt{-g}}   } \left(
\bar\psi^A_\mu\gamma_\sigma \rho_{A\nu\lambda} -  \bar\psi_{A\mu}
\gamma_\sigma \rho^A_{\nu\lambda} \right )
\nonumber \\
&-& {{\rm i}\over2}g_{i\bar\jmath}
\left(\bar\lambda^{iA}\gamma^\mu \nabla_\mu\lambda^{\bar\jmath}_A
+\bar\lambda^{\bar\jmath}_A \gamma^\mu
\nabla_\mu\lambda^{iA}\right ) -{\rm i}\left
(\bar\zeta^\alpha\gamma^\mu\nabla_\mu\zeta_\alpha
+\bar\zeta_\alpha\gamma^\mu \nabla_\mu \zeta^\alpha \right) \nonumber \\
&+& 2{\rm i}\left( \bar {\cal N}_{\L \S }{\cal F}^{-\L
}_{\mu\nu}{\cal F}^{-\S  \mu\nu} - {\cal N}_{\L \S } {\cal
F}^{+\L }_{\mu\nu}{\cal F}^{+ \S  {\mu\nu}}\right )+ \Big\{
-g_{i\bar\jmath}
 \nabla_\mu \bar z^{\bar\jmath} \bar \psi^\mu_A \lambda^{i A} \nonumber\\
&-& 2 {\mathcal U}^{A\alpha}_u \nabla_\mu q^u \bar \psi_A^\mu
\zeta _\alpha + g_{i\bar\jmath}  \nabla _\mu \bar z^{\bar\jmath}
\bar \lambda^{iA} \gamma^{\mu\nu} \psi_{A\nu} + 2{\mathcal
U}^{\alpha A}_u \nabla_\mu q^u \bar \zeta_\alpha \gamma^{\mu\nu}
\psi_{A\nu}+{\rm h.c.}\Big\} \nonumber \\
&+& \{ {\cal F}^{-\L }_{\mu\nu} {\rm Im }\,{\cal N}_{\L \S }\,
{\lbrack} 4 L^\S   \bar \psi^{A\mu} \psi^{B\nu}
\epsilon_{AB}-4{\rm i} {\bar f}^\S _{\bar\imath}\bar
\lambda^{\bar\imath}_A \gamma^\nu
\psi_B^\mu \epsilon^{AB}  \nonumber \\
&+& \frac{1}{2} \nabla_i f^\S _j \bar \lambda^{iA}
\gamma^{\mu\nu} \lambda^{jB}\epsilon_{AB}- L^\S
\bar\zeta_\alpha\gamma^{\mu\nu} \zeta_\beta C^{\alpha\beta}
{\rbrack}+{\rm h.c.}\}
\nonumber  \\
&+& + {\rm i} g_{i\bar\jmath} W^{iAB}
\bar\lambda^{\bar\jmath}_A\gamma_\mu \psi_B^\mu+
 2{\rm i} N^A_\alpha\bar\zeta^\alpha\gamma_\mu \psi_A^\mu \nonumber \\
&+& {\cal M}^{\alpha\beta}{\bar \zeta}_\alpha \zeta_\beta +{\cal
M}^{\alpha}_{\phantom{\alpha}iB} {\bar\zeta}_\alpha \lambda^{iB}
+ {\cal M}_{ij\,AB} {\bar \lambda}^{iA} \lambda^{j B} +
\mbox{h.c.}] -{\mathcal V}\bigl ( z, {\bar z}, q \bigr ).
\end{eqnarray}
Furthermore $L^\L (z,\,\bar z)$ are the covariantly holomorphic
sections of the special geometry, $f^\L _{i}\equiv \nabla_i L^\L
$ and the kinetic matrix ${\cal N}_{\L \S }$ is constructed in
terms of $ L^\L $ and its magnetic dual according to reference
\cite{abcdffm}. The normalization of the kinetic term for the
quaternions depends on the scale $\lambda$ of the quaternionic
manifold
 for which we have chosen the value $\lambda=-1$.
Finally, the  mass matrices of the spin $ \frac {1}{2}$ fermions
${\mathcal M}^{\alpha \,\beta}$, ${\mathcal
 M}_{AB \,ij}$, ${\mathcal M}^{\alpha}_{iA}$ (and
 their hermitian conjugates) and the scalar potential
 $\mathcal V$ are given by:
\begin{eqnarray}\label{mass2}
{\cal M}^{\alpha\beta}  &=&- \, {\mathcal U}^{\alpha A}_u \,
{\mathcal U}^{\beta B}_v \, \varepsilon_{AB}
\, \nabla^{[u}   k^{v]}_{\L }  \, L^{\L } \\
{\cal M}^{\alpha }_{\phantom{\alpha} iB} &=& -4 \, {\mathcal
U}^{\alpha}_{B  u} \, k^u_{\L } \,
 f^{\L }_i \\
{\cal M}_{AB \,\, ik} &=&  \,  \epsilon_{AB} \, g_{l^{\star} [i}
 f_{k]}^\L   k^{l^{\star}}_ \L  \,+ \frac {1}{2}
{\rm {i}}  P_ {\L  AB} \, \nabla_i f^\L  _k \label{pesamatrice}
\end{eqnarray}

\begin{equation}\label{potn2}
 {\mathcal V}(z,\bar z, q)\,=\,g^2 \Bigl[\left(g_{i\bar\jmath} k^i_\L  k^{\bar\jmath}_\S +4 h_{uv}
k^u_\L  k^v_\S \right) \bar L^\L  L^\S
+ g^{i\bar\jmath} f^\L _i f^\S _{\bar\jmath} {\cal P}^x_\L {\cal
P}^x_\S  -3\bar L^\L  L^\S {\cal P}^x_\L  {\cal P}^x_\S \Bigr]\ .
\end{equation}
\par
The $U(1)$-K\"ahler weight of the Fermi fields is
\begin{equation} P(\psi_A ) = P(\l^{\bar \imath}_A) = P(\zeta_\a )
=\half .
\end{equation}

\end{document}